%
%
%
\def\unredoffs{} \def\redoffs{\voffset=-.31truein\hoffset=-.48truein}
\def\speclscape{}
%
%
%
%
%
\newbox\leftpage \newdimen\fullhsize \newdimen\hstitle \newdimen\hsbody
\tolerance=1000\hfuzz=2pt
\catcode`\@=11 
\ifx\hyperdef\UNd@FiNeD\def\hyperdef#1#2#3#4{#4}\def\hyperref#1#2#3#4{#4}\fi
\def\bigans{b }
\def\answ{b }
%
\ifx\answ\bigans\message{(This will come out unreduced.}
\magnification=1200\unredoffs\baselineskip=16pt plus 2pt minus 1pt
\hsbody=\hsize \hstitle=\hsize 
\else\message{(This will be reduced.} \let\l@r=L
\magnification=1000\baselineskip=16pt plus 2pt minus 1pt
\vsize=7truein \redoffs
\hstitle=8truein\hsbody=4.75truein\fullhsize=10truein\hsize=\hsbody
\output={\ifnum\pageno=0 
  \shipout\vbox{\speclscape{\hsize\fullhsize\makeheadline}
    \hbox to \fullhsize{\hfill\pagebody\hfill}}\advancepageno
  \else
  \almostshipout{\leftline{\vbox{\pagebody\makefootline}}}\advancepageno
  \fi}
\def\almostshipout#1{\if L\l@r \count1=1 \message{[\the\count0.\the\count1]}
      \global\setbox\leftpage=#1 \global\let\l@r=R
 \else \count1=2
  \shipout\vbox{\speclscape{\hsize\fullhsize\makeheadline}
      \hbox to\fullhsize{\box\leftpage\hfil#1}}  \global\let\l@r=L\fi}
\fi
%
\newcount\yearltd\yearltd=\year\advance\yearltd by -1900

\def\Title#1#2{\nopagenumbers\abstractfont\hsize=\hstitle\rightline{#1}%
\vskip 1in\centerline{\titlefont #2}\abstractfont\vskip
.5in\pageno=0}
\def\Date#1{\vfill\leftline{#1}\tenpoint\supereject\global\hsize=\hsbody%
\footline={\hss\tenrm\hyperdef\hypernoname{page}\folio\folio\hss}}%
%

\def\draftmode{\message{ DRAFTMODE }\def\draftdate{{\rm preliminary draft:
\number\month/\number\day/\number\yearltd\ \ \hourmin}}%
\headline={\hfil\draftdate}\writelabels\baselineskip=20pt plus
2pt minus 2pt
 {\count255=\time\divide\count255 by 60 \xdef\hourmin{\number\count255}
  \multiply\count255 by-60\advance\count255 by\time
  \xdef\hourmin{\hourmin:\ifnum\count255<10 0\fi\the\count255}}}
\def\nolabels{\def\wrlabeL##1{}\def\eqlabeL##1{}\def\reflabeL##1{}}
\def\writelabels{\def\wrlabeL##1{\leavevmode\vadjust{\rlap{\smash%
{\line{{\escapechar=` \hfill\rlap{\sevenrm\hskip.03in\string##1}}}}}}}%
\def\eqlabeL##1{{\escapechar-1\rlap{\sevenrm\hskip.05in\string##1}}}%
\def\reflabeL##1{\noexpand\llap{\noexpand\sevenrm\string\string\string##1}}}
\nolabels
%
\global\newcount\secno \global\secno=0 \global\newcount\meqno
\global\meqno=1
\def\s@csym{}
\def\newsec#1{\global\advance\secno by1%
{\toks0{#1}\message{(\the\secno. \the\toks0)}}%
\global\subsecno=0\eqnres@t\let\s@csym\secsym\xdef\secn@m{\the\secno}\noindent
{\bf\hyperdef\hypernoname{section}{\the\secno}{\the\secno.} #1}%
\writetoca{{\string\hyperref{}{section}{\the\secno}{\the\secno.}} {#1}}%
\par\nobreak\medskip\nobreak}
\def\eqnres@t{\xdef\secsym{\the\secno.}\global\meqno=1\bigbreak\bigskip}
\def\sequentialequations{\def\eqnres@t{\bigbreak}}\xdef\secsym{}
\global\newcount\subsecno \global\subsecno=0
\def\subsec#1{\global\advance\subsecno by1%
{\toks0{#1}\message{(\s@csym\the\subsecno. \the\toks0)}}%
\ifnum\lastpenalty>9000\else\bigbreak\fi
\noindent{\it\hyperdef\hypernoname{subsection}{\secn@m.\the\subsecno}%
{\secn@m.\the\subsecno.} #1}\writetoca{\string\quad
{\string\hyperref{}{subsection}{\secn@m.\the\subsecno}{\secn@m.\the\subsecno.}}
{#1}}\par\nobreak\medskip\nobreak}
\def\appendix#1#2{\global\meqno=1\global\subsecno=0\xdef\secsym{\hbox{#1.}}%
\bigbreak\bigskip\noindent{\bf Appendix \hyperdef\hypernoname{appendix}{#1}%
{#1.} #2}{\toks0{(#1. #2)}\message{\the\toks0}}%
\xdef\s@csym{#1.}\xdef\secn@m{#1}%
\writetoca{\string\hyperref{}{appendix}{#1}{Appendix {#1.}} {#2}}%
\par\nobreak\medskip\nobreak}
%
%
\def\checkm@de#1#2{\ifmmode{\def\f@rst##1{##1}\hyperdef\hypernoname{equation}%
{#1}{#2}}\else\hyperref{}{equation}{#1}{#2}\fi}
\def\eqnn#1{\DefWarn#1\xdef #1{(\noexpand\relax\noexpand\checkm@de%
{\s@csym\the\meqno}{\secsym\the\meqno})}%
\wrlabeL#1\writedef{#1\leftbracket#1}\global\advance\meqno by1}
\def\f@rst#1{\c@t#1a\em@ark}\def\c@t#1#2\em@ark{#1}
\def\eqna#1{\DefWarn#1\wrlabeL{#1$\{\}$}%
\xdef #1##1{(\noexpand\relax\noexpand\checkm@de%
{\s@csym\the\meqno\noexpand\f@rst{##1}}{\hbox{$\secsym\the\meqno##1$}})}
\writedef{#1\numbersign1\leftbracket#1{\numbersign1}}\global\advance\meqno
by1}
\def\eqn#1#2{\DefWarn#1%
\xdef #1{(\noexpand\hyperref{}{equation}{\s@csym\the\meqno}%
{\secsym\the\meqno})}$$#2\eqno(\hyperdef\hypernoname{equation}%
{\s@csym\the\meqno}{\secsym\the\meqno})\eqlabeL#1$$%
\writedef{#1\leftbracket#1}\global\advance\meqno by1}
\def\xeqn{\expandafter\xe@n}\def\xe@n(#1){#1}
\def\xeqna#1{\expandafter\xe@n#1}
\def\eqns#1{(\e@ns #1{\hbox{}})}
\def\e@ns#1{\ifx\UNd@FiNeD#1\message{eqnlabel \string#1 is undefined.}%
\xdef#1{(?.?)}\fi{\let\hyperref=\relax\xdef\next{#1}}%
\ifx\next\em@rk\def\next{}\else%
\ifx\next#1\xeqn#1\else\def\n@xt{#1}\ifx\n@xt\next#1\else\xeqna#1\fi
\fi\let\next=\e@ns\fi\next}

\def\DefWarn#1{\ifx\UNd@FiNeD#1\else
\immediate\write16{*** WARNING: the label \string#1 is already
defined ***}\fi}
%
\newskip\footskip\footskip14pt plus 1pt minus 1pt 
\def\footnotefont{\ninepoint}\def\f@t#1{\footnotefont #1\@foot}
\def\f@@t{\baselineskip\footskip\bgroup\footnotefont\aftergroup\@foot\let\next}
\setbox\strutbox=\hbox{\vrule height9.5pt depth4.5pt width0pt}
\global\newcount\ftno \global\ftno=0
\def\foot{\global\advance\ftno by1\def\foot@rg{\hyperref{}{footnote}%
{\the\ftno}{\the\ftno}\xdef\foot@rg{\noexpand\hyperdef\noexpand\hypernoname%
{footnote}{\the\ftno}{\the\ftno}}}\footnote{$^{\foot@rg}$}}
%
\newwrite\ftfile
\def\footend{\def\foot{\global\advance\ftno by1\chardef\wfile=\ftfile
\hyperref{}{footnote}{\the\ftno}{$^{\the\ftno}$}%
\ifnum\ftno=1\immediate\openout\ftfile=\jobname.fts\fi%
\immediate\write\ftfile{\noexpand\smallskip%
\noexpand\item{\noexpand\hyperdef\noexpand\hypernoname{footnote}
{\the\ftno}{f\the\ftno}:\ }\pctsign}\findarg}%
\def\footatend{\vfill\eject\immediate\closeout\ftfile{\parindent=20pt
\centerline{\bf Footnotes}\nobreak\bigskip\input \jobname.fts }}}
\def\footatend{}
%
%
\global\newcount\refno \global\refno=1
\newwrite\rfile
\def\ref{[\hyperref{}{reference}{\the\refno}{\the\refno}]\nref}
\def\nref#1{\DefWarn#1%
\xdef#1{[\noexpand\hyperref{}{reference}{\the\refno}{\the\refno}]}%
\writedef{#1\leftbracket#1}%
\ifnum\refno=1\immediate\openout\rfile=\jobname.refs\fi
\chardef\wfile=\rfile\immediate\write\rfile{\noexpand\item{[\noexpand\hyperdef%
\noexpand\hypernoname{reference}{\the\refno}{\the\refno}]\ }%
\reflabeL{#1\hskip.31in}\pctsign}\global\advance\refno
by1\findarg}
\def\findarg#1#{\begingroup\obeylines\newlinechar=`\^^M\pass@rg}
{\obeylines\gdef\pass@rg#1{\writ@line\relax #1^^M\hbox{}^^M}%
\gdef\writ@line#1^^M{\expandafter\toks0\expandafter{\striprel@x #1}%
\edef\next{\the\toks0}\ifx\next\em@rk\let\next=\endgroup\else\ifx\next\empty%
\else\immediate\write\wfile{\the\toks0}\fi\let\next=\writ@line\fi\next\relax}}
\def\striprel@x#1{} \def\em@rk{\hbox{}}
\def\lref{\begingroup\obeylines\lr@f}
\def\lr@f#1#2{\DefWarn#1\gdef#1{\let#1=\UNd@FiNeD\ref#1{#2}}\endgroup\unskip}

\def\addref#1{\immediate\write\rfile{\noexpand\item{}#1}} 
\def\listrefs{\footatend\vfill\supereject\immediate\closeout\rfile\writestoppt
\baselineskip=\footskip\centerline{{\bf References}}\bigskip{\parindent=20pt%
\frenchspacing\escapechar=` \input
\jobname.refs\vfill\eject}\nonfrenchspacing}
\def\startrefs#1{\immediate\openout\rfile=\jobname.refs\refno=#1}
\def\xref{\expandafter\xr@f}\def\xr@f[#1]{#1}
\def\refs#1{\count255=1[\r@fs #1{\hbox{}}]}
\def\r@fs#1{\ifx\UNd@FiNeD#1\message{reflabel \string#1 is undefined.}%
\nref#1{need to supply reference \string#1.}\fi%
\vphantom{\hphantom{#1}}{\let\hyperref=\relax\xdef\next{#1}}%
\ifx\next\em@rk\def\next{}%
\else\ifx\next#1\ifodd\count255\relax\xref#1\count255=0\fi%
\else#1\count255=1\fi\let\next=\r@fs\fi\next}
%

%
\newwrite\ffile\global\newcount\figno \global\figno=1
\def\fig{fig.~\hyperref{}{figure}{\the\figno}{\the\figno}\nfig}
\def\nfig#1{\DefWarn#1%
\xdef#1{fig.~\noexpand\hyperref{}{figure}{\the\figno}{\the\figno}}%
\writedef{#1\leftbracket fig.\noexpand~\xfig#1}%
\ifnum\figno=1\immediate\openout\ffile=\jobname.figs\fi\chardef\wfile=\ffile%
{\let\hyperref=\relax
\immediate\write\ffile{\noexpand\medskip\noexpand\item{Fig.\ %
\noexpand\hyperdef\noexpand\hypernoname{figure}{\the\figno}{\the\figno}.
} \reflabeL{#1\hskip.55in}\pctsign}}\global\advance\figno
by1\findarg}
\def\listfigs{\vfill\eject\immediate\closeout\ffile{\parindent40pt
\baselineskip14pt\centerline{{\bf Figure
Captions}}\nobreak\medskip \escapechar=` \input
\jobname.figs\vfill\eject}}
\def\xfig{\expandafter\xf@g}\def\xf@g fig.\penalty\@M\ {}
\def\figs#1{figs.~\f@gs #1{\hbox{}}}
\def\f@gs#1{{\let\hyperref=\relax\xdef\next{#1}}\ifx\next\em@rk\def\next{}\else
\ifx\next#1\xfig #1\else#1\fi\let\next=\f@gs\fi\next}
\def\figin{\epsfcheck\figin}\def\figins{\epsfcheck\figins}
\def\epsfcheck{\ifx\epsfbox\UNd@FiNeD
\message{(NO epsf.tex, FIGURES WILL BE IGNORED)}
\gdef\figin##1{\vskip2in}\gdef\figins##1{\hskip.5in}
\else\message{(FIGURES WILL BE INCLUDED)}%
\gdef\figin##1{##1}\gdef\figins##1{##1}\fi}
\def\DefWarn#1{}
\def\figinsert{\goodbreak\midinsert}
\def\ifig#1#2#3{\DefWarn#1\xdef#1{fig.~\noexpand\hyperref{}{figure}%
{\the\figno}{\the\figno}}\writedef{#1\leftbracket fig.\noexpand~\xfig#1}%
\figinsert\figin{\centerline{#3}}\medskip\centerline{\vbox{\baselineskip12pt
\advance\hsize by -1truein\noindent\wrlabeL{#1=#1}\footnotefont%
{\bf Fig.~\hyperdef\hypernoname{figure}{\the\figno}{\the\figno}:}
#2}}
\bigskip\endinsert\global\advance\figno by1}
\newwrite\lfile
{\escapechar-1\xdef\pctsign{\string\%}\xdef\leftbracket{\string\{}
\xdef\rightbracket{\string\}}\xdef\numbersign{\string\#}}
\def\writedefs{\immediate\openout\lfile=\jobname.defs \def\writedef##1{%
{\let\hyperref=\relax\let\hyperdef=\relax\let\hypernoname=\relax
 \immediate\write\lfile{\string\def\string##1\rightbracket}}}}%
\def\writestop{\def\writestoppt{\immediate\write\lfile{\string\pageno
 \the\pageno\string\startrefs\leftbracket\the\refno\rightbracket
 \string\def\string\secsym\leftbracket\secsym\rightbracket
 \string\secno\the\secno\string\meqno\the\meqno}\immediate\closeout\lfile}}
\def\writestoppt{}\def\writedef#1{}
\def\seclab#1{\DefWarn#1%
\xdef #1{\noexpand\hyperref{}{section}{\the\secno}{\the\secno}}%
\writedef{#1\leftbracket#1}\wrlabeL{#1=#1}}
\def\subseclab#1{\DefWarn#1%
\xdef #1{\noexpand\hyperref{}{subsection}{\secn@m.\the\subsecno}%
{\secn@m.\the\subsecno}}\writedef{#1\leftbracket#1}\wrlabeL{#1=#1}}
\def\applab#1{\DefWarn#1%
\xdef #1{\noexpand\hyperref{}{appendix}{\secn@m}{\secn@m}}%
\writedef{#1\leftbracket#1}\wrlabeL{#1=#1}}
\newwrite\tfile \def\writetoca#1{}
\def\leaderfill{\leaders\hbox to 1em{\hss.\hss}\hfill}
\def\writetoc{\immediate\openout\tfile=\jobname.toc
   \def\writetoca##1{{\edef\next{\write\tfile{\noindent ##1
   \string\leaderfill {\string\hyperref{}{page}{\noexpand\number\pageno}%
                       {\noexpand\number\pageno}} \par}}\next}}}
\newread\ch@ckfile
\def\listtoc{\immediate\closeout\tfile\immediate\openin\ch@ckfile=\jobname.toc
\ifeof\ch@ckfile\message{no file \jobname.toc, no table of contents this pass}%
\else\closein\ch@ckfile\centerline{\bf Contents}\nobreak\medskip%
{\baselineskip=12pt\footnotefont\parskip=0pt\catcode`\@=11\input\jobname.toc
\catcode`\@=12\bigbreak\bigskip}\fi}
\catcode`\@=12 
%
\edef\tfontsize{\ifx\answ\bigans scaled\magstep3\else
scaled\magstep4\fi} \font\titlerm=cmr10 \tfontsize
\font\titlerms=cmr7 \tfontsize \font\titlermss=cmr5 \tfontsize
\font\titlei=cmmi10 \tfontsize \font\titleis=cmmi7 \tfontsize
\font\titleiss=cmmi5 \tfontsize \font\titlesy=cmsy10 \tfontsize
\font\titlesys=cmsy7 \tfontsize \font\titlesyss=cmsy5 \tfontsize
\font\titleit=cmti10 \tfontsize \skewchar\titlei='177
\skewchar\titleis='177 \skewchar\titleiss='177
\skewchar\titlesy='60 \skewchar\titlesys='60
\skewchar\titlesyss='60
\def\titlefont{\def\rm{\fam0\titlerm}
\textfont0=\titlerm \scriptfont0=\titlerms
\scriptscriptfont0=\titlermss \textfont1=\titlei
\scriptfont1=\titleis \scriptscriptfont1=\titleiss
\textfont2=\titlesy \scriptfont2=\titlesys
\scriptscriptfont2=\titlesyss \textfont\itfam=\titleit
\def\it{\fam\itfam\titleit}\rm}
 \ifx\answ\bigans\else scaled\magstep1\fi
\ifx\answ\bigans\def\abstractfont{\tenpoint}\else
\font\absit=cmti10 scaled \magstep1 \font\abssl=cmsl10 scaled
\magstep1 \font\absrm=cmr10 scaled\magstep1 \font\absrms=cmr7
scaled\magstep1 \font\absrmss=cmr5 scaled\magstep1
\font\absi=cmmi10 scaled\magstep1 \font\absis=cmmi7
scaled\magstep1 \font\absiss=cmmi5 scaled\magstep1
\font\abssy=cmsy10 scaled\magstep1 \font\abssys=cmsy7
scaled\magstep1 \font\abssyss=cmsy5 scaled\magstep1
\font\absbf=cmbx10 scaled\magstep1 \skewchar\absi='177
\skewchar\absis='177 \skewchar\absiss='177 \skewchar\abssy='60
\skewchar\abssys='60 \skewchar\abssyss='60
\def\abstractfont{\def\rm{\fam0\absrm}
\textfont0=\absrm \scriptfont0=\absrms \scriptscriptfont0=\absrmss
\textfont1=\absi \scriptfont1=\absis \scriptscriptfont1=\absiss
\textfont2=\abssy \scriptfont2=\abssys \scriptscriptfont2=\abssyss
\textfont\itfam=\absit \def\it{\fam\itfam\absit}\def\footnotefont{\tenpoint}%
\textfont\slfam=\abssl \def\sl{\fam\slfam\abssl}%
\textfont\bffam=\absbf \def\bf{\fam\bffam\absbf}\rm}\fi
\def\tenpoint{\def\rm{\fam0\tenrm}
\textfont0=\tenrm \scriptfont0=\sevenrm \scriptscriptfont0=\fiverm
\textfont1=\teni  \scriptfont1=\seveni  \scriptscriptfont1=\fivei
\textfont2=\tensy \scriptfont2=\sevensy \scriptscriptfont2=\fivesy
\textfont\itfam=\tenit \def\it{\fam\itfam\tenit}\def\footnotefont{\ninepoint}%
\textfont\bffam=\tenbf
\def\bf{\fam\bffam\tenbf}\def\sl{\fam\slfam\tensl}\rm}
\font\ninerm=cmr9 \font\sixrm=cmr6 \font\ninei=cmmi9
\font\sixi=cmmi6 \font\ninesy=cmsy9 \font\sixsy=cmsy6
\font\ninebf=cmbx9 \font\nineit=cmti9 \font\ninesl=cmsl9
\skewchar\ninei='177 \skewchar\sixi='177 \skewchar\ninesy='60
\skewchar\sixsy='60
\def\ninepoint{\def\rm{\fam0\ninerm}
\textfont0=\ninerm \scriptfont0=\sixrm \scriptscriptfont0=\fiverm
\textfont1=\ninei \scriptfont1=\sixi \scriptscriptfont1=\fivei
\textfont2=\ninesy \scriptfont2=\sixsy \scriptscriptfont2=\fivesy
\textfont\itfam=\ninei \def\it{\fam\itfam\nineit}\def\sl{\fam\slfam\ninesl}%
\textfont\bffam=\ninebf \def\bf{\fam\bffam\ninebf}\rm}
%
%

\hyphenation{anom-aly anom-alies coun-ter-term coun-ter-terms}
\def\inv{^{\raise.15ex\hbox{${\scriptscriptstyle -}$}\kern-.05em 1}}

\def\Dsl{\,\raise.15ex\hbox{/}\mkern-13.5mu D} 
\def\dsl{\raise.15ex\hbox{/}\kern-.57em\partial}

 \def\Tr{{\rm Tr}}
\def\lspace{\ifx\answ\bigans{}\else\qquad\fi}
\def\lbspace{\ifx\answ\bigans{}\else\hskip-.2in\fi} 
\def\boxeqn#1{\vcenter{\vbox{\hrule\hbox{\vrule\kern3pt\vbox{\kern3pt
    \hbox{${\displaystyle #1}$}\kern3pt}\kern3pt\vrule}\hrule}}}
\def\mbox#1#2{\vcenter{\hrule \hbox{\vrule height#2in
        \kern#1in \vrule} \hrule}}  
%

\def\darr#1{\raise1.5ex\hbox{$\leftrightarrow$}\mkern-16.5mu #1}

\def\half{{\textstyle{1\over2}}} 
\def\roughly#1{\raise.3ex\hbox{$#1$\kern-.75em\lower1ex\hbox{$\sim$}}}

\input psfig
\newcount\figno
\figno=0
\def\fig#1#2#3{
\par\begingroup\parindent=0pt\leftskip=1cm\rightskip=1cm\parindent=0pt
\global\advance\figno by 1
\midinsert
\epsfxsize=#3
\centerline{\epsfbox{#2}}
\vskip 12pt
{\bf Fig. \the\figno:} #1\par
\endinsert\endgroup\par
}
\def\figlabel#1{\xdef#1{\the\figno}}
\def\hat{\widehat}
\def\encadremath#1{\vbox{\hrule\hbox{\vrule\kern8pt\vbox{\kern8pt
\hbox{$\displaystyle #1$}\kern8pt}
\kern8pt\vrule}\hrule}}
\def\underarrow#1{\vbox{\ialign{##\crcr$\hfil\displaystyle
 {#1}\hfil$\crcr\noalign{\kern1pt\nointerlineskip}$\longrightarrow$\crcr}}}
%
\overfullrule=0pt

%
\def\tilde{\widetilde}
\def\bar{\overline}

\font\zfont = cmss10 
\font\litfont = cmr6

\def\bigone{\hbox{1\kern -.23em {\rm l}}}
\def\ZZ{\hbox{\zfont Z\kern-.4emZ}}
\def\half{{\litfont {1 \over 2}}}

\Title{hep-th/0107177}
{\vbox{\centerline{$M$-Theory Dynamics}
\bigskip
\centerline{ On A Manifold of $G_2$ Holonomy}}}
\smallskip
\centerline{Michael  Atiyah}
\smallskip
\centerline{\it Mathematics Department, University of Edinburgh, Edinburgh
Scotland UK}
\smallskip
\centerline{and}
\smallskip
\centerline {\it Dept. of Physics, Caltech, Pasadena CA, USA}
\smallskip\smallskip
\centerline{Edward Witten}
\smallskip
\centerline{\it Dept. of Physics, Caltech, Pasadena CA, USA}
\centerline{and}

\centerline{\it CIT-USC Center For Theoretical Physics, USC, Los
Angeles CA} \centerline{and} \centerline{\it Institute for
Advanced Study, Princeton NJ 08540 USA}

\def\RP{{\bf RP}}
\def\HP{{\bf HP}}
\def\H{{\bf H}}
\def\N{{\bf N}}
\def\C{{\bf C}}
\def\CP{{\bf CP}}
\def\RP{{\bf RP}}
\def\Z{{\bf Z}}
\def\hat{\widehat}
\def\tilde{\widetilde}
\def\T{{\bf T}}
\def\S{{\bf S}}
\def\R{{\bf R}}
\def\B{{\bf B}}

\medskip

\noindent
\bigskip
We analyze the dynamics of $M$-theory on a manifold of $G_2$
holonomy that is developing a conical singularity.  The known
cases involve a cone on ${\CP}^3$, where we argue that the
dynamics involves restoration of a global symmetry,
$SU(3)/U(1)^2$, where we argue that there are phase transitions
among three possible branches corresponding to three classical
spacetimes, and  $\S^3\times \S^3$ and its quotients, where we
recover and extend previous results about smooth continuations
between different spacetimes and relations to four-dimensional
gauge theory.

 \Date{June, 2001}
\newsec{Introduction}

In studying supersymmetric compactifications of string theory, one
of the important issues is the behavior at a singularity. For example,
in compactifications of $M$-theory and of Type II superstring theory, an
important role is played by
the $A-D-E$ singularities of a K3 surface and by various singularities
of a Calabi-Yau threefold.
Singularities of heterotic string or $D$-brane gauge fields are
also important, though in the present paper we focus on metric singularities.

\nref\agm{P. S. Aspinwall, B. R. Greene, and D. R. Morrison,
``Multiple Mirror Manifolds And Topology Change In String Theory,''
Phys. Lett. {\bf B303} (1993) 249, hep-th/9301043.}%
\nref\ew{E. Witten, ``Phases Of ${\cal N}=2$ Models In Two Dimensions,''
Nucl. Phys. {\bf B403} (1993) 159, hep-th/9301042.}%
\nref\strom{A. Strominger, ``Massless Black Holes And Conifolds In
String Theory,'' Nucl. Phys. {\bf B451} (1995) 96, hep-th/9504090.}%
\nref\gms{B. Greene, D. R. Morrison, and A. Strominger,  ``Black Hole
Condensation And The Unification Of String Vacua,'' Nucl. Phys. {\bf B451}
(1995) 109, hep-th/9504145.}%
The basic questions about string theory and $M$-theory dynamics
at a classical singularity are familiar.  What happens in the quantum
theory when a classical singularity develops?  Does extra gauge symmetry
appear?  Are there new massless particles?  Does the quantum theory flow
to a nontrivial infrared fixed point?  Is it possible to make a transition
to a different classical spacetime by following the behavior of the
quantum theory through a classical singularity?  If such transitions are
possible, do they occur smoothly (as in the case of the classical
``flop'' of Type II superstring theory  \refs{\agm,\ew}) or via a phase transition
to a different branch of the moduli space of vacua
(as in the case of the Type II conifold transition \refs{\strom,\gms})?

Usually, the essential phenomena occurring at a singularity
are local in nature,
 independent of the details of a global spacetime in which the
singularity is embedded.  The most basic singularities from which more
elaborate examples are built are generally {\it conical} in nature.
In $n$ dimensions, a conical metric takes the general form
\eqn\retro{ds^2= dr^2+r^2d\Omega^2,}
where $r$ is the ``radial'' coordinate, and $d\Omega^2$ is a metric
on some compact $(n-1)$-manifold $Y$. An $n$-manifold $X$ with such a metric
 is said to be a cone on $Y$; $X$ has a singularity at the origin
unless $Y={\S}^{n-1}$ and $d\Omega^2$ is the standard round
metric.  For example, the $A-D-E$ singularity in real dimension
four is a cone on $\S^3/\Gamma$, with $\S^3$ a three-sphere and
$\Gamma$ a finite subgroup of $SU(2)$.

In this example, as in many others, the $A-
D-E$ singularity can be resolved (or deformed)
to make a smooth four-manifold $\hat X$ that is only asymptotically conical.
$\hat X$ carries a hyper-Kahler metric that depends on a number of parameters
or moduli.
$\hat X$ is smooth for generic values of the moduli, but becomes singular
 when one varies
the moduli  so that $\hat X$ is exactly, not just asymptotically, conical.

The present paper is devoted to analyzing  $M$-theory dynamics on
a seven-manifold  of $G_2$ holonomy that develops an isolated conical singularity.  The motivation for studying $G_2$-manifolds
is of course that $G_2$ holonomy is the condition for unbroken supersymmetry
in four dimensions.  This is also the reason that it is possible to get
interesting results about this case.

\nref\bryant{R. Bryant and S. Salamon, ``On The Construction Of Some
Complete Metrics With Exceptional Holonomy,'' Duke Math. J. {\bf 58}
(1989) 829.}
\nref\gibbons{G. W. Gibbons, D. N. Page, and C. N. Pope, ``Einstein
Metrics On $S^3$, $R^3$, and $R^4$ Bundles,''  Commun. Math. Phys.
{\bf 127} (1990) 529.}
\nref\acharya{B. Acharya, ``On Realising ${\cal N}=1$ Super Yang-Mills
In $M$ Theory,'' hep-th/0011089.}
\nref\amv{M. F. Atiyah, J. Maldacena, and C. Vafa, ``An $M$-Theory
Flop As A Large $\N$ Duality,'' hep-th/0011256.}
\nref\achtwo{B. Acharya, ``Confining Strings From $G_2$ Holonomy
Spacetimes,'' hep-th/0101206.}
\nref\achvafa{B. Acharya and C. Vafa,  ``On Domain Walls Of ${\cal N}=1$
Supersymmetric Yang-Mills In Four Dimensions,'' hep-th/0103011.}

There are probably many possibilities for an isolated conical singularity
of a $G_2$-manifold, but apparently
only three simply-connected
cases are known  \refs{\bryant,\gibbons}:
 a cone on ${\CP}^3$, $SU(3)/U(1)\times U(1)$, or $\S^3\times \S^3$
 can carry a metric of $G_2$ holonomy;
 each of these cones can be deformed
to make a smooth, complete, and asymptotically
conical  manifold $X$ of $G_2$ holonomy.  We will study the behavior
of $M$-theory on these manifolds, as well as on additional examples
obtained by dividing by a finite group.

The case of a cone on $\S^3\times \S^3$ or a quotient thereof
has been studied previously \refs{\acharya - \achvafa} and found to
be rather interesting.  This example, in fact, is related to earlier
investigations of dualities involving fluxes and branes in  topological
\ref\gopa{R. Gopakumar and C. Vafa, ``On The Gauge Theory/Geometry
Correspondence,'' hep-th/9811131.} and ordinary
\nref\kleb{I. R. Klebanov and M. J. Strassler, ``Supergravity And A
Confining Gauge Theory: Duality Cascades And $\chi$SB-Resolution
Of Naked Singularities,'' JHEP {\bf 0008:052} (2000), hep-th/0007191.}%
\nref\oldvafa{C. Vafa, ``Superstrings And Topological Strings At Large $\N$,'' hep-th/0008142.}%
\nref\nun{J. M. Maldacena and C. Nunez, ``Towards The Large $N$
Limit Of Pure $\N=1$ Super Yang Mills,'' Phys. Rev. Lett. {\bf 86} (2001)
588, hep-th/0008001.}%
\nref\sinha{S. Sinha and C. Vafa, ``$SO$ and $Sp$ Chern-Simons At
Large
$N$,'' hep-th/0012136.}%
\nref\agva{M. Aganagic and C. Vafa, ``Mirror Symmetry And A $G_2$
Flop,'' hep-th/0105225.}%
\refs{\kleb - \nun} strings. We will reexamine it in more detail,
and also investigate the other examples, which turn out to be
easier to understand.  Our results in sections 4 and 6 are
closely related to results in \refs{\oldvafa, \sinha, \agva }.

In section two, we introduce the examples, describe some of their
basic properties, and make a proposal for the dynamics of the first
two examples.
According to our proposal, the dynamics involves in one case the restoration
of a global symmetry in the strong coupling region, and in the second
case a phase transition between three different branches that represent
three different classical spacetimes.
 In section three, we give evidence for this proposal
by relating the manifolds of $G_2$ holonomy to certain
configurations of branes in ${\C}^3$ that have been studied as
examples of singularities of special Lagrangian threefolds
\ref\joycesp{D. Joyce, ``On Counting Special Lagrangian Homology
3-Spheres,'' hep-th/9907013.}.  (For somewhat analogous quotients,
see \ref\arnold{V. I. Arnold, ``Topological Content Of The
Maxwell Theorem On Multipole Representations Of Spherical
Functions,'' {\it Topological Methods In Nonlinear Analysis},
Journal of the Jliusz Schauder Center, {\bf 7} (1996) 205,
``Relatives Of The Quotient Of The Complex Projective Plane By The
Complex Conjugation,'' Proc. Steklov Mathematical Institute
(1998).} and \ref\ggpt{J. P. Gauntlett, G. W. Gibbons, G.
Papadopoulos, and P. K. Townsend,  ``Hyper-K\"ahler Manifolds And
Multiplet Intersecting Branes,'' hep-th/9702202.}.)  By a slight
extension of these arguments, we also give simple examples of
four-dimensional chiral fermions arising from models of $G_2$
holonomy.

  In section four, we analyze the more
challenging example, involving the cone on $\S^3\times \S^3$.  In
this example, refining the reasoning in \amv, we argue that there is
a moduli space of theories of complex dimension one that interpolates
smoothly, without a phase transition, between three different classical
spacetimes.  To describe the interpolation precisely, we introduce
some natural physical observables associated with the deviation of the
geometry at infinity from being precisely conical.  Using these observables
together with familiar ideas of applying holomorphy to supersymmetric
dynamics \ref\seiberg{N. Seiberg, ``The Power
Of Holomorphy -- Exact Results in 4D SUSY Field Theories,''
hep-th/9408013.  }, we give a precise description of the moduli space.

In section five, we compare details of the solution found in section
four to topological subtleties in the membrane effective action.

In section six, following \refs{\acharya,\amv}, we consider
further examples obtained by dividing
by a finite group $\Gamma$. In one limit of the models considered, there
is an effective four-dimensional gauge theory with a gauge group
of type $A$, $D$, or $E$.  We again give a precise description of the
moduli space in these examples, in terms of the natural observables.
For the $A$ series, this enables us to put on a more precise basis
some observations made in \amv\ about the relation of the classical
geometry to chiral symmetry breaking.  For the $D$ and $E$ series, there
is a further surprise: the model interpolates between different possible
classical gauge groups, for example $SO(8+2n)$ and $Sp(n)$ in the case of the
$D$ series, with generalizations of this statement for the $E$ series.
In the case of the $E$ series, the analysis makes contact
with recent developments involving commuting triples and $M$-theory
singularities \ref\sethietal{J. de Boer, R. Dijkgraaf, K. Hori,
A. Keurentjes, J. Morgan, D. R. Morrison,
and S. Sethi,  ``Triples, Fluxes, and Strings,'' hep-th/0103170.}.

Our discussion will be relevant to $M$-theory on a compact manifold of
$G_2$ holonomy if (as is likely but not yet known)  such a compact manifold
can develop a conical singularity of the types we consider.
  The known techniques of construction of compact manifolds
of $G_2$ holonomy are explained in detail in a recent book
\ref\joyce{D. Joyce, {\it Compact Manifolds With Special
Holonomy} (Oxford University Press, 2000).}.
\nref\cvetica{M. Cvetic, H. Lu, and C. N. Pope, ``Brane Resolution
Through Transgression,'' hep-th/0011023.}%
\nref\cveticb{M. Cvetic, G. W. Gibbons, H. Lu, and C. N. Pope,
``Supersymmetric Non-Singular Fractional $D2$-Branes And
NS-NS 2-Branes,'' hep-th/0101096.}%
\nref\cveticc{M. Cvetic, G. W. Gibbons, H. Lu, and C. N. Pope,
``Ricci-Flat Metrics, Harmonic Forms, and Brane Resolutions,''
hep-th/0012011; ``Hyper-Kahler Calabi Metrics, ${\bf L}^2$ Harmonic
Forms, Resolved $M2$-Branes, and ${\rm AdS}_4/{\rm CFT}_3$ Correspondence,
hep-th/0102185; ``New Complete Non-Compact ${\rm Spin}(7)$ Manifolds,''
hep-th/0103155.}%

The asymptotically conical manifolds of $G_2$ holonomy that we study have
been reexamined in recent papers \refs{\cvetica,\cveticb},
along with asymptotically conical metrics with other reduced
holonomy groups \cveticc.
Some features explored there will be relevant below, and other aspects
are likely to be important in generalizations.
\nref\papad{P. Townsend and G. Papadopoulos, ``Compactification
Of $D=11$ Supergravity On Spaces Of Exceptional Holonomy,''
hep-th/9506150, Phys. Lett. {\bf B357} (1995) 472.}%
\nref\harvey{J. A. Harvey and G. Moore, ``Superpotentials And Membrane
Instantons,'' hep-th/9907026.}%
\nref\manyach{B. S. Acharya, J. M. Figueroa-O'Farrill, C. M. Hull,
and B. Spence, ``Branes At Conical Singularities And
Holography,'' Adv. Theor. Math. Phys. {\bf 2} (1999) 1249, hep-th/9808014.}%
\nref\otheracha{B. Acharya, ``$M$-Theory, Joyce Orbifolds, And Super
Yang-Mills,'' Adv. Theor. Math. Phys. {\bf 3} (1999) 227, hep-th/9812205.}%
\nref\fivewrap{B. S. Acharya, J. P. Gauntlett, and N. Kim,
``Fivebranes Wrapped On Associative Three-Cycles,'' hep-th/0011190.}%
\nref\cachazo{F. Cachazo, K. Intriligator, and C. Vafa,
``A Large $N$ Duality Via  A Geometric Transition,''
hep-th/0103067.}%
\nref\edel{J. Edelstein and C. Nunez, ``$D6$ Branes And $M$ Theory Geometrical
Transitions From Gauged Supergravity,'' hep-th/0103167.}%
\nref\gomis{J. Gomis, ``$D$-Branes, Holonomy and $M$-Theory,'' hep-th/0103115.}%
\nref\partouche{H. Partouche and B. Pioline,  ``Rolling Among
$G_2$ Vacua,'' hep-th/0011130, JHEP
{\bf 0103} (2001) 005.}%
\nref\kachru{S. Kachru and J. McGreevy, ``$M$-Theory On Manifolds of
$G_2$ Holonomy and Type IIA Orientifolds,'' hep-th/0103223.}%
\nref\part{P. Kaste, A. Kehagias, and H. Partouche,
``Phases of Supersymmetric Gauge Theories From $M$-Theory On
$G_2$ Manifolds,'' hep-th/0104124.}%
\nref\kon{Y. Konishi and M. Naka, ``Coset Construction Of
Spin(7), $G_2$ Gravitational Instantons,'' hep-th/0104208.}%
\nref\brand{A. Brandhuber, J. Gomis, S. S. Gubser, and S. Gukov,
``Gauge Theory At Large $N$ And New $G_2$ Holonomy Metrics,''
hep-th/0106034.}%
\nref\clp{M. Cvetic, H. Lu, and C. N. Pope, ``Massless 3-Brane In
$M$-Theory,'' hep-th/0105096.}%
\nref\cvglp{M. Cvetic, G. W. Gibbons, H. Lu, and C. N. Pope,
``Supersymmetric $M3$-Branes And $G_2$ Manifolds,''
hep-th/0106026.}%
\nref\cmany{M. Cvetic, G. W. Gibbons, H. Lu, and C. N. Pope,
``Resolved Branes And
$M$-Theory On Special Holonomy Spaces,'' hep-th/0106177.}%
\nref\cvmore{M. Cvetic, G. W. Gibbons, J. T. Liu, H. Lu, and C.
N. Pope, ``A New Fractional $D2$-Brane, $G_2$ Holonomy, and
$T$-Duality,'' hep-th/0106162.}%
 For additional  work on $M$-theory on manifolds of $G_2$
holonomy, see \refs{\harvey - \cvmore}.

\newsec{Known Examples And Their Basic Properties}

In studying $M$-theory on a manifold $X$ that is asymptotic
to a cone on $Y$, the problem is defined by specifying
the fields at infinity, and in particular by specifying $Y$.
The fields are then free to fluctuate in the
interior.  Here the phrase ``in the interior'' means that one should
look at those fluctuations that decay fast enough at infinity to
have finite kinetic energy.  Variations of the fields that would
have infinite kinetic energy are nondynamical; their values are specified
at infinity as part of the definition of the problem.  They are analogous
to coupling constants in ordinary four-dimensional field theory.
The problem of dynamics is to understand the behavior of the fluctuations.
Quantum mechanically, one
type of fluctuation is that there might be different possible
$X$'s once $Y$ is given.

A similar dichotomy holds for
symmetries.  The symmetries of the problem are the symmetries of the fields
at infinity, that is, on $Y$.  An unbroken symmetry is a symmetry that leaves
fixed the fields in the interior.

Much of our work in the present section will be devoted to identifying
the fluctuating fields and the couplings, the symmetries and the unbroken
symmetries, for the various known examples.
In describing the examples, we largely
 follow the notation of \gibbons.

\subsec{$\R^3$ Bundles Over Four-Manifolds}

The starting point for the first two examples is a four-manifold
$M$ with self-dual Weyl curvature and a positive curvature
Einstein metric, normalized so that $R_{\alpha\beta}
=3g_{\alpha\beta}$.  We write the line element of $M$ as
$h_{\alpha\beta}dx^\alpha dx^\beta$, where $h$ is the metric and
$x^\alpha$ are local coordinates. In practice, the known
possibilities for $M$ are $\S^4$ and ${\CP}^2$.  (In the case of
${\CP}^2$, the orientation is taken so that the Kahler form of
${\CP}^2$ is considered self-dual.) At a point in $M$, the space
of anti-self-dual two-forms is three-dimensional. The bundle $X$
of anti-self-dual two-forms is accordingly a rank three real
vector bundle over $M$; it carries an $SO(3)$ connection $A$ that
is simply the positive chirality part of the spin connection of
$M$ written in the spin one representation.

$X$ admits a complete metric of $G_2$ holonomy:
\eqn\gtmet{ds^2={dr^2\over 1-\left({r_0/ r}\right)^4}+{r^2\over 4}
(1-(r_0/r)^4)|d_Au|^2+{r^2\over 2}\sum_{\alpha,\beta=1}^4
 h_{\alpha\beta}dx^\alpha dx^\beta.}
Here $u^i$ are fiber coordinates for the bundle $X\to M$,
and $d_Au$ is the covariant derivative  $d_Au_i=du_i+\epsilon_{ijk}
A_ju_k$.  Also, $r_0$ is an arbitrary positive parameter with dimensions
of length, and $r$ is a ``radial'' coordinate with $r_0\leq r \leq\ \infty$.

This metric is asymptotic to a cone on a six-manifold $Y$ that is
a two-sphere bundle over $M$. Indeed, for $r\to\infty$, we can
drop the $r_0/r$ terms, and then the metric takes the general
form $dr^2+r^2d\Omega^2$, where here $d\Omega^2$ is a metric on
$Y$. $Y$ is the subspace of $X$ with $\sum_iu_i^2=1$ and is known
as the twistor space of $M$.

In fact, we can be more specific: the metric on $X$ differs from a
conical metric by terms of order $(r_0/r)^4$, for $r\to\infty$.
The exponent 4 is greater than half of the dimension of $X$, and
this has the following important consequence.  Let $g$ be the
metric of $X$, and $\delta g$ the variation of $g$ with respect
to a change in $r_0$. Define the ${\bf L}^2$ norm of $\delta g$ by
\eqn\onon{|\delta g|^2=\int_X d^7x \sqrt g \,g^{ii'}g^{jj'}\delta
g_{ij} \delta g_{i'j'}.} Since $\delta g/g \sim r^{-4}$, and we
are in seven dimensions, we have \eqn\impop{|\delta g|^2<\infty.}
This means that in $M$-theory on $\R^4\times X$, the kinetic
energy associated  with a fluctuation in $r_0$ is finite; this
fluctuation gives rise to a massless scalar field $a$ in four
dimensions.

Supersymmetrically, this massless scalar field must be completed
to a massless chiral multiplet $\Phi$. Apart from zero modes of
the metric, massless scalars in four dimensions can arise as zero
modes of the three-form field $C$ of eleven-dimensional
supergravity.  In the present example, an additional massless
scalar arises because there is on $X$ an  ${\bf L}^2$ harmonic
three-form $\omega$, constructed in \cveticb.  This additional
scalar combines with $a$ to make a complex scalar field that is
the bosonic part of $\Phi$.  The superpotential of $\Phi$
vanishes identically for a reason that will be explained later,
so the expectation value of $\Phi$ parameterizes a family of
supersymmetric vacua.

The topological interpretation of $\omega$ will be of some
interest. $\omega$ is zero as an element of $H^3(X;\R)$, since in
fact ($X$ being contractible to $\S^4$ or ${\CP}^2$) that group
vanishes. The key feature of $\omega$ topologically is that
\eqn\fibeq{\int_{\R^3}\omega\not= 0,} where the integral is taken
over any fiber of the fibration $Y\to M$. Since $\R^3$ is
noncompact, this means that $\omega$ can be associated with an
element of the compactly supported cohomology $H_{cpct}^3(X;\R)$.
The physical meaning of this will become clear presently.

\bigskip\noindent{\it Geometrical  Symmetries}

Now we want to discuss the symmetries of these models.
Some symmetries arise from geometrical symmetries of the manifold $Y$.
They will be examined in detail when we consider specific examples.

For now, we merely make some general observations about
geometrical symmetries.
  A symmetry of $X$ may either preserve its orientation or
reverse it.  When $X$ is asymptotic to a cone on $Y$, a symmetry
reverses the orientation of $X$ if and only if it reverses the orientation
of $Y$.  In $M$-theory on $\R^4\times X$, a symmetry of $X$ that reverses
its orientation  is observed in the effective four-dimensional
physics as an ``$R$-symmetry'' that changes the sign of the superpotential.
\foot{For example,
a symmetry of $X$ of order two that preserves the $G_2$ structure
and reverses the orientation of $X$
squares to $-1$ on spinors, so its eigenvalues are $\pm i$.
In $M$-theory, the gravitino field  on $\R^4\times X$ is real,
so such  a symmetry acts as $\pm i$ on positive chirality gravitinos
 on $\R^4$ and as $\mp i$ on their complex conjugates,
the negative chirality gravitinos.  Since it transforms the two chiralities
oppositely, it is an $R$-symmetry.}
Orientation-preserving symmetries of $X$ are not $R$-symmetries;
they leave the superpotential invariant.

For example, the symmetry $\tau:u\to -u$ of \gtmet\ reverses the orientation
of $X$, so it is an $R$-symmetry.
Since it preserves the modulus $|\Phi|$, it transforms $\Phi\to e^{i\alpha}
\Phi$ for some constant $\alpha$.  We can take $\alpha$ to be zero, because
as we will presently see, these models also have a $U(1)$ symmetry (coming
from gauge transformations of the $C$-field) that
rotates the phase of $\Phi$ without acting as an $R$-symmetry.
Existence of a symmetry that leaves $\Phi$ fixed and reverses the
sign of the superpotential means
that the superpotential is zero for a model of this type.
This is consistent with the fact that, as the third homology of $X$
vanishes in these examples, there are no membrane instantons that
would generate a superpotential.

We should also consider geometrical symmetries of the first
factor of $\R^4\times X$.  Apart from the connected part of the
Poincar\'e group, we must consider a ``parity'' symmetry, reflecting
one of the $\R^4$ directions in $\R^4\times X$.  This exchanges chiral
and anti-chiral superfields, changes the sign of the $C$-field, and
maps the chiral superfield $\Phi$ to its complex conjugate $\bar\Phi$.
By contrast, symmetries of $\R^4\times X$ that act only on $X$, preserving
the orientation of $\R^4$, give holomorphic mappings of
chiral superfields to chiral superfields.

\bigskip\noindent{\it Symmetries  From $C$-Field}

Important additional symmetries arise from the $M$-theory
three-form $C$. As always in gauge theories, global symmetries
come from gauge symmetries whose generators do not vanish at
infinity but that leave the fields fixed at infinity.  (Gauge
symmetries whose generators vanish at infinity can be neglected
as they act trivially on all physical excitations. Gauge
transformations that do not leave the fields fixed at infinity
are  not really symmetries of the physics.)

For the $M$-theory three-form $C$, the basic gauge transformation
law is $\delta C = d\Lambda$, where $\Lambda$ is a two-form.  So
a symmetry generator obeys $d\Lambda=0$ at infinity.
In $M$-theory on $\R^4\times X$, with $X$ being asymptotic to a cone
on a six-manifold $Y$, the global symmetry group coming from the $C$-field
is therefore $K=H^2(Y;U(1))$.
What subgroup of $K$ leaves the vacuum invariant?  A two-form $\Lambda$
generates an unbroken symmetry if, in a gauge transformation generated
by $\Lambda$, $0=\delta C=d\Lambda$ everywhere, not just at infinity.
So an element of $G$ is unbroken if it is obtained by restricting to $Y$
an element of $H^2(X;U(1))$.   Such elements of $K$ form the subgroup $L$
of unbroken symmetries.

In practice, for $X$ an $\R^3$ bundle over a four-manifold $M$,
a spontaneously broken symmetry arises for $\Lambda$ such that
\eqn\plok{\int_{\S^2}\Lambda\not= 0,}
where here $\S^2$ is the ``sphere at infinity'' in one of the $\R^3$
fibers.  Indeed,
when this integral is nonzero, $\Lambda$ cannot be extended over
$X$ as a closed form, since $\S^2$ is a boundary in $X$ -- it is the boundary
of a fiber.  But $\Lambda$ can be extended over $X$ somehow, for example,
by multiplying it by a function that is 1 at infinity on $X$ and zero
in the ``interior.''  After picking such an extension of $\Lambda$,
we transform $C$ by $\delta C=d\Lambda$, and we have
$\int_{\R^3}\delta C = \int_{\S^2}\Lambda$, so
\eqn\plook{ \int_{\R^3}\delta C \not= 0. }

Under favorable conditions, and in particular in the examples
considered here, we can pick
 $\Lambda$ so that $\delta C$ is harmonic.  Then
the massless scalar in four dimensions associated with the harmonic three-form
$\delta C$ is a Goldstone boson for the symmetry generated by $\Lambda$.
That in turn  gives us the physical meaning of \fibeq: the
massless scalar in four dimensions associated with the harmonic
three-form $\omega$ is the Goldstone boson of a spontaneously broken
symmetry whose generator obeys \plok.

\subsec{First Example And Proposal For Dynamics}

There are two known examples of this type.  In the first
example,  $M={\S}^4$.  The three-plane bundle $X$ over $M$ is
asymptotic to a cone on $Y={\CP}^3$.  This arises as follows.

${\bf CP}^3$ admits two different homogeneous Einstein metrics.
The usual Fubini-Study metric has $SU(4)$ symmetry, and the second
one, which is relevant here, is invariant under the subgroup
$Sp(2)$ of $SU(4)$.\foot{Our notation is such that $Sp(1)=SU(2)$.
The preceding statements about the symmetry groups can be refined
slightly;  the groups that act faithfully on ${\CP}^3$ are,
respectively, $SU(4)/\Z_4$ and $Sp(2)/\Z_2$.} Thus, ${\CP}^3$ can
be viewed as the homogeneous space $SU(4)/U(3)$, but for the
present purposes, it can be more usefully viewed as the
homogeneous space $Sp(2)/ SU(2)\times U(1) $, where $ SU(2)\times
U(1)\subset SU(2)\times SU(2)=Sp(1)\times Sp(1) \subset Sp(2)$.

If we divide
$Sp(2)$ by $SU(2)\times SU(2)$, we are imposing a stronger equivalence
relation than if we divide by $ SU(2)\times U(1)$.  So
defining a six-manifold and a four-manifold $Y$ and $M$ by
\eqn\defman{Y=Sp(2)/U(1)\times SU(2),~~M=Sp(2)/SU(2)\times SU(2),}
$Y$ fibers over $M$ with fibers that are copies of
$SU(2)/U(1)=\S^2$.  In fact,
as $SO(5)=Sp(2)/\Z_2$ and $SO(4)=SU(2)\times SU(2)/\Z_2$, $M$ is
the same as $SO(5)/SO(4)=\S^4$.
If we replace the $\S^2$ fibers of $Y\to \S^4$ by $\R^3$'s, we get
an asymptotically conical seven-manifold $X$.

$X$ is the bundle of anti-self-dual two-forms over $\S^4$.  Indeed, an
anti-self-dual
two-form at a point on $\S^4$ breaks the isotropy group of that point
from $SU(2)\times SU(2)$ to $SU(2)\times U(1).$  If we restrict
to unit anti-self-dual two-forms, we get the six-manifold $Y$.  $M=\S^4$
admits the standard ``round'' Einstein metric, and an asymptotically
conical metric of $G_2$ holonomy on $X$ is given in \gtmet.

\bigskip\noindent{\it Geometrical Symmetries}

Now let us work out the symmetries of $Y$, and of $X$.  Geometrical
symmetries of $Y$ can be interpreted rather like the symmetries of the
$C$-field that were discussed in section 2.1.
Symmetries of $Y$ are symmetries of $M$-theory on $\R^4\times X$.
A symmetry of $Y$
that extends over $X$ is an unbroken symmetry.  A symmetry of $Y$
that does not extend over $X$ is spontaneously broken; it maps one
$X$ to another possible $X$.  In the particular example at hand, we will
see that all of the symmetries of $Y$ extend over $X$, but that will not
be so in our other examples.

We can represent $Y$ as the space of all $g\in Sp(2)$, with the
equivalence relation $g\cong gh$ for $h\in SU(2)\times U(1)$.
In this description, it is clear that $Y$ is invariant under the left
action of $Sp(2)$ on $g$ (as noted in a footnote above, it is the
quotient $Sp(2)/\Z_2=SO(5)$ that acts faithfully). These symmetries
also act on $M$ and $X$.   Additional symmetries of $Y$  come
from right action by elements $w\in Sp(2)$ that ``centralize''
$H=SU(2)\times U(1)$, that is, for any $h\in H$,
$w^{-1}hw\in H$.
(We should also consider outer automorphisms of $Sp(2)$ that centralize
$H$, but $Sp(2)$ has no outer automorphisms.)
Any $w\in H$
centralizes $H$, but acts trivially on $Y$.  There is only one nontrivial
element $w$ of $Sp(2)$ that centralizes $H$.  It is
contained in the second factor of $SU(2)\times SU(2)\subset Sp(2)$.
If we represent the $U(1)$ factor in $H=SU(2)\times U(1)\subset SU(2)
\times SU(2) $ by the diagonal elements
\eqn\bigerg{\left(\matrix{ e^{i\theta} & 0 \cr 0 & e^{-i\theta}\cr}\right)}
of $SU(2)$, then we can represent $w$ by the $SU(2)$ element
\eqn\jumbop{\left(\matrix{ 0 & 1 \cr -1 & 0 \cr}\right)}
that anticommutes with the generator of $U(1)$.

$w$ acts trivially on the four-manifold $M=Sp(2)/SU(2)\times SU(2)$, since it is contained in $SU(2)\times SU(2)$.
To show that all symmetries of $Y$ extend as symmetries of $X$, we must,
as $X$ is manifestly $Sp(2)$-invariant, find a $\Z_2$ symmetry
of $X$ that acts trivially on $M$.  It is simply the $R$-symmetry $\tau$
that
we discussed earlier:  multiplication by $-1$ on the
$\R^3$ fibers of $X\to M$, or in other words the transformation $u\to -u$.

\bigskip\noindent{\it Dynamics}

Now let us try to guess the dynamics of the chiral superfield $\Phi$.
$\Phi$ is essentially
\eqn\unic{\Phi=V e^{i\int_{\R^3}C},}
where $V\sim r_0^4$ is the volume of the $\S^4$ at the ``center''
of $X$, and the integral in the exponent is taken over a fiber of $X\to M$.

The geometrical symmetry $\tau$ of $Y$ extends over $X$, as we
have  seen, regardless of $r_0$.  So it acts trivially
on the modulus of $\Phi$.   It also leaves fixed the argument
of $\Phi$. (Concretely, $\tau$ reverses the orientation
of $\R^3$, but also, since it reverses the orientation of the overall
spacetime $\R^4\times X$, transforms $C$ with an extra factor of $-1$.
The net effect is to leave fixed $\int_{\R^3}C$.)
Hence, this symmetry plays no role in the low energy dynamics.

We also have the ``parity'' symmetry of $\R^4\times X$, reflecting the first factor.  This symmetry acts by $\Phi\to
\overline\Phi$.  (Indeed, such a transformation maps $C\to -C$ with
no action on $\R^3$.)

The important symmetries for understanding this problem come from
the symmetries of the $C$-field.  In the present example, with
$Y={\CP}^3$, we have $H^2(Y;U(1))=U(1)$, so the symmetry group is
$K=U(1)$. On the other hand, $X$ is contractible to $M=\S^4$, and
$H^2(\S^4;U(1))=0$, so $K$ is spontaneously broken to nothing.

Also, since the second homology group of $M$ is trivial, the
second homology group of $Y$ is generated by a fiber of the
$\S^2$ fibration $Y\to M$. The generator of $K$ is accordingly
derived from a two-form $\Lambda$ with
\eqn\hurgo{\int_{\S^2}\Lambda\not= 0.} As explained in section
2.1, a symmetry generated by such a $\Lambda$ shifts the value of
$\int_{\R^3}C$, and hence acts on $\Phi$ by \eqn\rigno{\Phi\to
e^{i\alpha}\Phi.} This shows explicitly the spontaneous breaking
of $K$.

Supergravity gives a reliable account of the dynamics for large
$|\Phi|$, that, is, for large $V$. We want to guess, using the
symmetries and holomorphy, what happens in the quantum regime of
small $|\Phi|$.  In the present case, there is a perfectly
obvious guess.  Since it has an action of the global symmetry
$K=U(1)$, acting by ``rotations'' near infinity,   the moduli
space must have genus zero.  Its only known ``end'' is
$\Phi\to\infty$ and -- as ends of the moduli space should be
visible semiclassically -- it is reasonable to guess that this is
the only end.  If so, the moduli space must simply be  the complex
$\Phi$-plane, with $\Phi=0$ as a point at which the global
symmetry $K$ is restored.  Our minimal conjecture for the
dynamics is that there are no extra massless particles for strong
coupling; the dynamics remains infrared-free; and  the only
qualitative phenomenon that occurs for small volume is that at a
certain point in the moduli space of vacua, the global $U(1)$
symmetry is restored.

In section 3, we will give supporting evidence for this proposal
by comparing the model to a Type IIA model with $D$-branes.

\subsec{A Second Model}

In the second example, $Y=SU(3)/T$, where $T=U(1)\times
U(1)=U(1)^2$ is the maximal torus of $SU(3)$. Moreover,
$M=SU(3)/U(2)=\CP^2$, and $X$ is the bundle of anti-self-dual
two-forms over $\CP^2$.

Again we must be careful in discussing the metric and symmetries of $Y$.
First let us look at the geometrical symmetries.  The nontrivial symmetries of interest are outer automorphisms
 of $SU(3)$ that centralize the maximal torus $T$, and right action by
elements of $SU(3)$ that centralize $T$.  Actually, the
centralizer of the maximal torus in any compact connected Lie
group
 is the Weyl group $W$, which for
$SU(3)$ is the group  $\Sigma_3$ of permutations of three elements.  If one thinks
of an element of $SU(3)$ as a $3\times 3$ matrix, then elements of $W$ are $3\times 3$
permutation matrices (times $\pm 1$ to make the determinant 1).

$SU(3)$ also has an outer automorphism of complex conjugation,
which centralizes $T$  (it acts as $-1$ on the Lie algebra
of $T$).  It commutes with $\Sigma_3$ (since the permutation
matrices generating $\Sigma_3$ are real).  This gives a symmetry
of $Y$ that extends as a symmetry
of $X$ (acting by complex conjugation on $M=\CP^2$) regardless of the value
of $\Phi$.  Because it is a symmetry for any value of the chiral
superfield $\Phi$, it decouples from the low energy dynamics and will
not be important.
By contrast, the $\Sigma_3$ or ``triality'' symmetry is very important,
as we will see.

$Y$ admits a well-known homogeneous Kahler-Einstein metric, but
that is not the relevant one for the metric \gtmet\ of $G_2$
holonomy on $X$. The reason for this is as follows. The notion of
a Kahler metric on $Y$ depends on a choice of complex structure.
The (complexified) tangent space of $Y$ has a basis in one-to-one
correspondence with the nonzero roots of $SU(3)$.  Picking a
complex structure on $Y$ is equivalent to picking a set of
positive roots.  The Weyl group permutes the possible choices of
what we mean by positive roots, so it is not natural to expect a
Kahler metric on $Y$ to be $\Sigma_3$-invariant. In fact, the
subgroup of $\Sigma_3$ that leaves fixed a Kahler metric is
$\Sigma_2=\Z_2$, the group of permutations of {\it two} elements.
Indeed, if we exchange the positive and negative roots of $SU(3)$
(by making a Weyl transformation that is a reflection with
respect to the highest root), this will reverse the complex
structure of $Y$; but a metric that is Kahler for one complex
structure is also Kahler for the opposite complex structure.
Accordingly, the standard Kahler metric of $Y$ is invariant under
the subgroup $\Sigma_2$ of $\Sigma_3$.

$Y$ also admits $SU(3)$-invariant metrics that are invariant under the full
$\Sigma_3$.  In fact, to give an $SU(3)$-invariant metric on $Y$,
we first give a $T$-invariant
metric on the tangent space at a point, and then transport it by $SU(3)$.
There is up to a scalar multiple
a unique $T$-invariant metric on the tangent space
that is also $\Sigma_3$-invariant: it assigns the same length to each
nonzero root of $SU(3)$.  So an $SU(3)\times \Sigma_3$-invariant metric
$g_{IJ}$ on $Y$ is uniquely determined up to a scale.  This metric
is hermitian but not Kahler (for each of the complex structures of $Y$).

Uniqueness up to scale of the $SU(3)\times \Sigma_3$-invariant
metric implies that this metric is  an Einstein metric.
Indeed, the Ricci tensor $R_{IJ}$ derived from $g_{IJ}$ is again
a symmetric tensor with the same $SU(3)\times \Sigma_3$ symmetry,
so it must be a multiple of $g_{IJ}$.
The $G_2$ metric on $X$ is asymptotic to
a cone on $Y$, where $Y$ is endowed with
this $\Sigma_3$-invariant Einstein metric.

Now, let us ask what subgroup of $\Sigma_3$ is preserved by $X$.
$X$ is an $\R^3$ bundle over $M=\CP^2= SU(3)/SU(2)\times U(1)$.
$M$ in fact has no symmetries that commute with $SU(3)$, since $SU(2)\times
U(1)$ has no nontrivial centralizer in $SU(3)$.  $X$ has a $\Z_2$
symmetry that commutes with $SU(3)$ and acts trivially on $M$: it is the transformation $u\to -u$
in \gtmet.  So picking a particular $X$ that is bounded by $Y$ breaks
the $\Sigma_3$ symmetry of $Y$ down to $\Sigma_2=\Z_2$.  The $\Z_2$ in
question is an antiholomorphic transformation of $Y$ that reverses the
complex structure.

In essence, though $Y$ admits various choices of complex structure and
an Einstein metric on $Y$ can be defined without making a choice,
to fiber $Y$ over $\CP^2$ one must pick (up to sign, that is, up to an
overall reversal of the complex structure) a distinguished complex structure
on $Y$.  So there are three different choices of $X$, determined by
the choice of complex structure on $Y$.

Next, let us consider the symmetries that originate from the
$C$-field. Consider the fibration $SU(3)\to Y$ with fibers
$T=U(1)\times U(1)$. Because the first and second cohomology
groups of $SU(3)$ with $U(1)$ coefficients are zero, the spectral
sequence for this fibration gives us
\eqn\ikop{H^2(Y;U(1))=H^1(T;U(1))} In fact,
\eqn\nikop{H^1(T;U(1))=T^*=U(1)\times U(1),} where $T^*$ is the
dual torus of $T$. On the other hand, $X$ is contractible to
$M=\CP^2$, and $H^2(\CP^2;U(1)) =U(1)$.  So the global symmetry
group coming from the $C$-field is $K=T^*=U(1)\times U(1)$,
spontaneously broken to $L=U(1)$.

Here, we need to be more precise, because there are really, as we have
seen above, three possible choices for $X$, and each choice will give
a different unbroken $U(1)$.  We need to know how the unbroken $U(1)$'s
are related.  We can be more precise using \nikop. Thus, the maximal
torus $T$  is the quotient of $\R^2 $ by the
root lattice $\Lambda$ of $SU(3)$, and $T^*$ is the quotient by the weight lattice $\Lambda^*$.
The unbroken subgroup $L$ is a one-parameter subgroup of $T^*$; such
a subgroup is determined by a choice (up to sign) of a primitive
weight $w\in \Lambda^*$, which one can associate with the generator of $L$.

The three possible $X$'s -- call them $X_1,X_2$, and $X_3$ -- are permuted
by an element of order three in $\Sigma_3$.  The corresponding $w$'s are likewise (if their signs are chosen correctly) permuted by the element of
order three.  Since an element of the Weyl group of order three rotates
the weight lattice of $SU(3)$ by a $2\pi/3$ angle, the weights $w_1,$ $w_2$,
and $w_3$ are permuted by such a rotation, and hence
\eqn\imion{w_1+w_2+w_3=0.}

On the $i^{th}$ branch, there is an unbroken subgroup of $\Sigma_3$
generated by an element $\tau_i$ of order 2.
An element of order 2 in $\Sigma_3$ acts by a reflection on the weight
lattice, so its eigenvalues are $+1$ and $-1$.  We can identify which
is which.  $\tau_i$ (which acts on $X_i$ by $u\to -u$ in the notation of
\gtmet) acts trivially on the chiral superfield $\Phi_i$, by arguments
that we have already seen.
So $\tau_i$ acts trivially on the Goldstone boson field, which is the
argument of $\Phi_i$.  Hence $\tau_i$ leaves fixed the broken symmetry
on the $i^{th}$ branch, and acts by
\eqn\inmo{\tau_i(w_i)=-w_i}
on the generator $w_i$ of the unbroken symmetry.

\bigskip\noindent{\it Proposal For Dynamics}

Now we will make a proposal for the dynamics of this model.

The fact that the unbroken symmetries are different
for the three classical spacetimes $X_i$
implies that one cannot, in this model,
smoothly interpolate without a phase transition
between the three  different limits.  Instead, we claim
one can continuously interpolate between the different
classical spacetimes by passing through a phase transition.

We need a theory with three branches of vacua.  On the $i^{th}$ branch,
for $i=1,2,3$, there must
be a single massless chiral superfield $\Phi_i$.
The three branches are permuted by a $\Sigma_3$ symmetry.  Moreover,
there is a global $U(1)\times U(1)$ symmetry, spontaneously broken on each
branch to $U(1)$.  $\Sigma_3$ acts on $U(1)\times U(1)$ like the Weyl
group of $SU(3)$ acting on the maximal torus, and
 the generators $w_i$ of the unbroken symmetries of the three branches
add up to zero.

We will reproduce this via an effective theory that contains
all three chiral superfields $\Phi_i$, in such a way that there are three branches of the moduli space of vacua on each of which two of the $\Phi_i$
are massive.  We take $\Sigma_3$ to act by permutation of the $\Phi_i$,
and we take $K=U(1)\times U(1)$ to act by
$\Phi_i\to e^{i\theta_i}\Phi_i$ with $\theta_1+\theta_2+\theta_3=0$.
The minimal nonzero  superpotential invariant under $K\times \Sigma_3$ is
\eqn\juggo{W(\Phi_1,\Phi_2,\Phi_3)=\lambda \Phi_1\Phi_2\Phi_3}
with $\lambda$ a constant.  In the classical approximation, a vacuum
is just a critical point of $W$.  The equations for
a critical point are $0=\Phi_2\Phi_3=\Phi_3\Phi_1
=\Phi_1\Phi_2$, and there are indeed three branches of vacua permuted
by $\Sigma_3$.  On
the $i^{th}$ branch, for $i=1,2,3$, $\Phi_i$ is nonzero and the other
$\Phi$'s are zero and massive.   The three branches meet at a singular
point at the origin, where $\Phi_1=\Phi_2=\Phi_3=0$.
Thus, one can pass from one branch to another by going through a phase
transition  at the origin.

On the branch with, say, $\Phi_1\not=0$, the unbroken $U(1)$ is
$\Phi_1\to\Phi_1$, $\Phi_2\to e^{i\theta}\Phi_2$, $\Phi_3\to e^{-i\theta}\Phi_3$, generated by a diagonal matrix $w_1={\rm diag}(0,1,-1)$
acting on the $\Phi_i$.  The generators $w_2$ and $w_3$ of the unbroken
symmetries on the other branches are obtained from $w_1$ by a cyclic permutation
of the eigenvalues, and indeed obey $w_1+w_2+w_3=0$.  Moreover, the nontrivial
element of $\Sigma_3$ that leaves fixed, say, the first branch is the
element
$\tau_1$ that exchanges $\Phi_2$ and $\Phi_3$; it maps $w_1$ to $-w_1$,
as expected according to \inmo.

We will give additional evidence for this proposal in section 3, by comparing
to a construction involving Type IIA $D$-branes.  For now, we conclude
with a further observation about the low energy physics.
The theory with superpotential $W$ has an additional $U(1)$ $R$-symmetry
that rotates all of the $\Phi_i$ by a common phase.  It does not
correspond to any exact symmetry of $M$-theory
even at the critical point.  In  general, when $X$ develops a conical
singularity, $M$-theory on $\R^4\times X$ might develop exact symmetries
that act only on degrees of freedom that are supported near the singularity,
but they cannot be $R$-symmetries.  The reason for this last statement
is that  $R$-symmetries act nontrivially on the gravitino and the gravitino
can propagate to infinity on $X$.

We could remove the $R$-symmetry by adding nonminimal terms to the theory,
for example an additional term $(\Phi_1\Phi_2\Phi_3)^2$ in $W$.
But such terms are irrelevant in the infrared. So a consequence of our
proposal for the dynamics is that the infrared limit of $M$-theory at the
phase transition point has a $U(1)$ $R$-symmetry that is not an exact
$M$-theory symmetry.

\subsec{$\R^4$ Bundle Over $\S^3$}

Now we move on to the last, and, as it turns out, most subtle example.
This is the case of a manifold of $G_2$ holonomy that is asymptotic at
infinity to a cone over $Y=\S^3\times \S^3$.  In this case, the seven-manifold
$X$ is topologically $\R^4\times \S^3$; the ``vanishing cycle'' that
collapses when $X$ develops a conical singularity is the three-manifold
$Q=\S^3$ rather than the four-manifold $M=\S^4$ or $\CP^2$ of the previous
examples.  That is in fact the basic reason for the difference of this
example from the previous ones.

All the manifolds in sight -- $X$, $Y$, and $Q$ -- have vanishing
second cohomology.  So there will be no symmetries coming from the $C$-field.
However, the geometrical symmetries of $Y$ will play an important role.

$Y=\S^3\times \S^3$ admits an obvious Einstein metric that is the product
of the standard round metric on each $\S^3$.  Identifying $\S^3=SU(2)$,
this metric has $SU(2)^4$ symmetry, acting by separate left and right
action of $SU(2)$ on each factor of $Y=SU(2)\times SU(2)$.  However,
the $SU(2)^4$-invariant Einstein metric on $Y$ is not the one that
arises at infinity in the manifold of $G_2$ holonomy.

A second Einstein metric on $\S^3\times\S^3$ can be constructed
as follows. Let $a,b,$ and $c$ be {\it three} elements of
$SU(2)$, constrained to obey \eqn\pilko{abc=1.} The constraint is
obviously compatible with the action of $SU(2)^3$ by $a\to u
av^{-1}$, $ b\to v b w^{-1}$, $c\to wcu^{-1}$, with $u,v,w\in
SU(2)$. Moreover, it is compatible with a cyclic permutation of
$a,b,c$: \eqn\ilko{\beta:(a,b,c)\to (b,c,a)} and with a ``flip''
\eqn\bilko{\alpha:(a,b,c)\to (c^{-1},b^{-1},a^{-1}).} $\alpha$
and $\beta$ obey $\alpha^2=\beta^3=1$,
$\alpha\beta\alpha=\beta^{-1}$. Together they generate the same
``triality'' group $\Sigma_3$ that we met in the previous example.

Another way to see the action of $SU(2)^3\times \Sigma_3$ on $\S^3\times \S^3$
is as follows.  Consider triples $(g_1,g_2,g_3)\in SU(2)^3$
with an equivalence relation $(g_1,g_2,g_3)\cong (g_1h,g_2h,g_3h)$ for
$h\in SU(2)$.  The space of equivalence classes is $\S^3\times\S^3$
(since we can pick $h$ in a unique fashion to map $g_3$ to 1).
On the space $Y$ of equivalence classes, there is an obvious action of
$\Sigma_3$ (permuting the three $g$'s), and of $SU(2)^3$ (acting on the
$g$'s on the left).    The relation between the two descriptions is
to set $a=g_2g_3^{-1}$, $b=g_3g_1^{-1}$, $c=g_1g_2^{-1}$. The description
with the $g_i$ amounts to viewing $\S^3\times\S^3$ as a homogeneous
space $G/H$, with $G=SU(2)^3$, and $H=SU(2)$ the diagonal subgroup of the
product of
three $SU(2)$'s.  There are no nontrivial elements of $G$ (not in $H$)
that centralize $H$, and the $\Sigma_3$ symmetry group comes from outer
automorphisms of $G$.

On $\S^3\times \S^3$ there is, up to scaling, a unique metric with
$SU(2)^3\times\Sigma_3$ symmetry, namely
\eqn\hhurigo{d\Omega^2=-\Tr\left((a^{-1}da)^2+(b^{-1}db)^2+(c^{-1}dc)^2\right),}
where $\Tr$ is the trace in the two-dimensional representation of
$SU(2)$. Just as in our discussion of Einstein metrics on
$SU(3)/U(1)\times U(1)$, the uniqueness implies that it is an
Einstein metric.  We will abbreviate $-\Tr(a^{-1}da)^2$ as
$da^2$, and so write the metric as \eqn\hurigo{d\Omega^2=
da^2+db^2+dc^2.}

The manifold $X=\R^4\times \S^3$ admits a complete metric of
$G_2$ holonomy described in \refs{\bryant, \gibbons}. It can be
written \eqn\burigo{ds^2={dr^2\over 1-(r_0/r)^3}+{r^2\over
72}\left(1-(r_0/r)^3\right) (2 \,da^2 -db^2+2 \,dc^2)+{r^2\over
24} db^2,} where $r$ is a ``radial'' coordinate with $r_0\leq
r<\infty$, and $r_0$ is the ``modulus'' of the solution.\foot{ To
compare to the notation of \gibbons, let $T_i$ be $SU(2)$
generators with $T_iT_j=\delta_{ij}+i\epsilon_{ijk}T_k$. Then
$\sigma_i$ in eqn. (5.1) of \gibbons\ equals $-(i/2)\Tr
\,T_ia^{-1}da$. Also, set $\tilde b=b^{-1}$.  Then
$\Sigma_i=-(i/2)\Tr\, T_i \tilde b^{-1}d\tilde b$.  To match
\burigo\ with the result in \gibbons, one should also solve for
$c$ with $c=\tilde b a^{-1}$.}
  The topology is $\R^4\times \S^3$
since one of the two $\S^3$'s collapses to zero radius for $r\to r_0$.
Near infinity, this metric is
 asymptotic to that of a cone on $Y=\S^3\times \S^3$, with the metric
on $Y$ being the Einstein metric $d\Omega^2$ described above.

The deviation of the metric on $X$ from a conical form is of order $(r_0/r)^3$ for $r\to\infty$,
in contrast to the $(r_0/r)^4$ in the previous examples.  The difference
is essentially because the vanishing cycle is a three-cycle, while in the
previous examples it was a four-cycle.
Because a function that behaves as $(r_0/r)^3$
for large $r$ is not square integrable in seven dimensions (on an
asymptotically conical manifold), the physical interpretation of the modulus
$r_0$ is very different from what it was in the previous examples.
$r_0$ is not free to fluctuate; the kinetic energy in its fluctuation
would be divergent.  It should be interpreted as a boundary condition
that is fixed at infinity.  In the low energy four-dimensional physics,
$r_0$ is a coupling constant.  The fact that a fluctuation in $r_0$
is not square-integrable is related by supersymmetry to the fact that
the $SU(2)^3$-invariant harmonic three-form $\omega$ on this manifold is
not square-integrable \cveticb.

The parameter related to $r_0$ by supersymmetry is
$\theta=\int_QC$, where $Q=\S^3$ is the ``vanishing cycle'' at
the center of $X$.  (In fact, the harmonic three-form $\omega$
obeys $\int_Q\omega\not=0$, so adding to $C$ a multiple of the
zero mode $\omega$ shifts $\theta$.) $r_0$ and $\theta$ combine
into a complex parameter. In the examples based on $\R^3$ bundles
over a four-manifold, there was a massless chiral superfield
whose expectation value parameterized a one complex parameter
family of vacua.  In the present example, there is instead a
complex coupling parameter, and for each value of the coupling,
there is a unique vacuum.   The last statement holds for
sufficiently large $r_0$ because supergravity is valid and gives
a unique vacuum with all interactions vanishing in the infrared.
Then, by holomorphy, uniqueness of the vacuum should hold for all
values of the coupling.

The metric \burigo\ is clearly invariant under $\alpha:(a,b,c)\to
(c^{-1},b^{-1},a^{-1})$, and not under any other nontrivial element of
$\Sigma_3$. So the choice of $X$ has broken $\Sigma_3$ down to the subgroup
$\Sigma_2$ generated by $\alpha$.
We are thus in a situation that is reminiscent of what we found in
section 2.3.
There are three different $X$'s, say $X_1,X_2,X_3$,
permuted by the spontaneously broken ``triality'' symmetry.

To describe the construction of the $X_i$ in topological terms, we
reconsider the description of $Y=\S^3\times \S^3$ in terms of
three $SU(2)$ elements $g_1,g_2,g_3$ with the equivalence relation
\eqn\purple{(g_1,g_2,g_3)=(g_1h,g_2h,g_3h).} To make a
seven-manifold $X'$ bounded by $Y$, we ``fill in'' one of the
three-spheres. To be more precise, we allow one of the $g_i$, say
$g_1$, to take values in ${\B}^4$ -- a four-ball bounded by
$SU(2)$, to which the right action of $SU(2)$ extends in a
natural way -- and we impose the same equivalence relation
\purple.  If we think of $SU(2)$ as the group of unit
quaternions, we can think of ${\B}^4$ as the space of quaternions
of norm no greater than one.  Obviously, we could replace $g_1$
by $g_2$ or $g_3$, so we get in this way three different
seven-manifolds $X'_i$.  The $X'_i$ are compact seven-manifolds
with boundary; if we omit the boundary, we get open
seven-manifolds $X_i$ which are the ones that admit
asymptotically conical metrics of $G_2$ holonomy. (Henceforth, we
will not generally distinguish $X_i$ and $X_i'$.) From this
construction, it is manifest that $X_1$, for example, admits a
$\Z_2$ symmetry that exchanges $g_2$ and $g_3$.

To see the topology of $X_1$, we just set $h=g_3^{-1}$.  So
$X_1=\R^4\times\S^3$, with $\R^4$ parameterized by $g_1$ and
$\S^3$ by $g_2$.  $X_1$ is invariant under an element of
$\Sigma_3$ that exchanges $g_2$ and $g_3$. It maps $(g_1,g_2,1)$
to $(g_1,1,g_2)$ or equivalently $(g_1g_2^{-1},g_2^{-1},1)$.
Looking at the behavior at the tangent space to a fixed point
with $g_2=1$, we see that this reverses the orientation of $X$,
and hence is an $R$-symmetry. So in general, the elements of
$\Sigma_3$ that are of order two are $R$-symmetries, just as in
the model studied in section 2.3.

Using this topological description of the $X_i$, we can compare to the
Type IIA language that was used in \refs{\acharya,\amv} and understand
from that point of view why there are three $X_i$.  To relate to Type IIA,
the idea in \refs{\acharya,\amv} was to divide by a $U(1)$ subgroup
of one of the three $SU(2)$'s, say the first one, and interpret the quotient
space as a Type IIA spacetime.  If we divide $X_1=\R^4\times \S^3$
by a $U(1)$ contained in the first $SU(2)$, it acts only on the $\R^4$ factor,
leaving fixed $\{0\}\times \S^3=\S^3$, where $\{0\}$ is the origin in
$\R^4$.
The quotient $\R^4/U(1)$ is topologically $\R^3$, where
the fixed point at the origin is interpreted in Type IIA as signifying
the presence of a $D$-brane, as in the familiar relation
of a smooth $M$-theory spacetime (the Kaluza-Klein monopole) to
a Type IIA $D6$-brane  \ref\townsend{P. Townsend,
``The Eleven-Dimensional Supermembrane Revisited,''
hep-th/9501068.}.
So $X_1/U(1)$  is an $\R^3$ bundle over $\S^3$ with
a brane wrapped on the zero-section; the $\R^3$ bundle over
$\S^3$ is the deformation of the conifold.
The same $U(1)$, in acting on $X_2$ or $X_3$, acts nontrivially (and freely)
on the $\S^3$ factor of $\R^4\times \S^3$,
giving a quotient that is an $\R^4$ bundle over
$\S^2=\S^3/U(1)$ (with a unit of RR two-form
flux on $\S^2$ because the fibration $\S^3\to\S^2$ has Euler class one).
These quotients of $X_2$ and $X_3$
are the two possible small resolutions
of the conifold.  Thus, the three possibilities -- the deformation
and the two small resolutions -- are familiar in Type IIA.  The surprise
is, perhaps, the triality symmetry between them in the $M$-theory description
in the case with one sixbrane or  unit of flux.

\bigskip\noindent{\it Smooth Continuation?}

\def\M{{\cal M}}
\def\N{{\cal N}}

One of our major conclusions so far is that unlike the models
considered in sections 2.2 and 2.3, which have a family of quantum
vacua depending on one complex parameter, the present example
has a one complex parameter family of possible values of the
``coupling constants.'' For each set of couplings, there is a unique
vacuum.

Let $\N$ be a complex Riemann surface parameterized by the
possible couplings.  Thus, $\N$ might be regarded  from a
four-dimensional point of view as the moduli space of
``theories,'' while in the other examples, the analogous object
would be a moduli space $\M$ of vacua in a fixed theory.

In section 2.3, we argued, in a superficially similar case, that
the moduli space $\M$ of vacua has three distinct components $\M_i$,
one for each classical spacetime.  For the present problem, it has been
proposed \amv\ that the curve $\N$ has only one smooth component,
which interpolates between the possible classical limits.
We can give a quick argument for this based on the relation to
Type IIA that was just explained along with triality symmetry.

In the conformal field theory of the Type IIA conifold, in the
absence of RR flux,  one can interpolate smoothly between the two
small resolutions, without encountering a phase transition
\refs{\agm,\ew}. On the other hand, the transition to the
deformation of the conifold involves a phase transition known as
the conifold transition \refs{\strom,\gms}. Now what happens if
one turns on a unit of RR two-form flux so as to make contact
with $M$-theory on a manifold of $G_2$ holonomy?  The effects of
RR flux are proportional to the string coupling constant $g_s$
and so (assuming that the number of flux quanta is fixed as
$g_s\to 0$) are negligible in the limit of weak string coupling
or conformal field theory.  Existence of a smooth interpolation
between two limits is a stable statement that is not spoiled by a
sufficiently small perturbation, so we can assert that the two
small resolutions are smoothly connected also when the RR flux is
turned on.  What about the deformation?  In the presence of
precisely one unit of RR flux, we can go to $M$-theory and use
the triality symmetry between the three branches; this at once
implies that if two of the branches are smoothly connected to
each other, they must be smoothly connected to the third branch.

We will study the curve $\N$ more comprehensively in sections 4 and 6.
We will obtain
 a quantitative description of $\N$, using arguments that
also apply to the case of more than one unit of RR flux, where
there is no triality symmetry.

\subsec{ Classical Geometry}

We conclude with some more detailed observations on the classical
geometry that will be useful in section 4.

We want to describe the three-dimensional homology and cohomology
of $Y=\S^3\times \S^3$ as well as of the $X_i$.  We regard $Y$ as
the space $SU(2)^3/SU(2)$, obtained by identifying triples
$(g_1,g_2,g_3)$ under right multiplication by $h\in SU(2)$. We
let $\hat D_i\subset SU(2)^3$ be the $i^{th}$ copy of $SU(2)$.
(We will take the index $i$ to be defined mod 3.)  $\hat D_1$, for
example, is the set $(g,1,1)$, $g\in SU(2)$. In
$Y=SU(2)^3/SU(2)$, the $\hat D_i$ project to three-cycles that we
will call $D_i$.  As the third Betti number of $Y$ is two, the
$D_i$ must obey a linear relation.  In view of the triality
symmetry of $Y$, which permutes the $D_i$, this relation is
\eqn\pilgo{D_1+D_2+D_3=0.}

In terms of the description of $Y$ by $SU(2)$ elements $a,b,c$ with
$abc=1$ (where $a=g_2g_3^{-1}$, $b=g_3g_1^{-1}$, $c=g_1g_2^{-1}$),
$D_1$ is $a=1=bc$, and the others are obtained by cyclic permutation of
$a,b,c$.

As before, we can  embed $Y$ in three different manifolds $X_i$
that are each homeomorphic to $\R^4\times \S_3$.
The third Betti number of $X_i$
is one, so in the homology of $X_i$, the $D_i$ obey an additional relation.
Since $X_i$ is obtained by ``filling in'' the $i^{th}$ copy of $SU(2)$,
the relation is just $D_i=0$.  Thus, the homology of $X_i$
is generated by $D_{i-1}$ or $D_{i+1}$ with $D_{i-1}=-D_{i+1}$.  At the ``center'' of
$X_i$, there is a three-sphere $Q_i$ defined by setting $r=r_0$ in \burigo.
In the description of $X_i$ via $(g_1,g_2,g_3)$ (modulo right multiplication
by $h\in SU(2)$) where $g_i$ takes
values in $\R^4$ and the others in $SU(2)$, $Q_i$ corresponds to setting
$g_i=0$.  One can then use right multiplication by $h$ to gauge $g_{i+1}$
or $g_{i-1}$ to 1.  Since $g_i=0$ is homotopic to $g_i=1$, $Q_i$ is
homologous, depending on its orientation, to $\pm D_{i-1}$ and to
$\mp D_{i+1}$.

Next, let us look at the intersection numbers of the $D_i$.  Any
two distinct $D_i$ intersect only at the point $(g_1,g_2,g_3)=(1,1,1)$,
so the intersection numbers are $\pm 1$.  If we orient $Y$ suitably
so that $D_1\cdot D_2=+1$, then the remaining signs are clear from
triality symmetry:
\eqn\jugh{D_i\cdot D_j=\delta_{j,i+1}-\delta_{j,i-1}.}
Note that $D_j\cdot (D_1+D_2+D_3)=0$ for all $j$, consistent with \pilgo.

Finally, let us look at cohomology.  The third  cohomology group of
$SU(2)$ is generated by a three-form $\omega=(1/8\pi^2)\Tr (g^{-1}dg)^3$
that integrates to one.  On $Y$, we consider the
forms $e^1=(1/8\pi^2)\Tr(a^{-1}da)^3$, $e^2=(1/8\pi^2)\Tr (b^{-1}db)^3$,
$e^3=(1/8\pi^2)\Tr(c^{-1}dc)^3$.  (We use the forms $a^{-1}da$, etc.,
rather than $g_i^{-1}dg_i$, as they make sense on the quotient space
$Y=SU(2)^3/SU(2)$.)    We have
$e^1=(1/8\pi^2)\left(\Tr(g_2^{-1}dg_2)^3-\Tr(g_3^{-1}dg_3)^3+3d
\Tr g_2^{-1}dg_2g_3^{-1}dg_3\right),$ and cyclic permutations of
that formula.   Integrating the $e^i$ over $D_j$, we
get
\eqn\huboc{\int_{D_i}e^j=\delta_{j,i+1}-\delta_{j,i-1}.}
Comparing \jugh\ and \huboc, it follows that the map from cohomology
to homology given by Poincar\'e duality is
\eqn\dnin{e^j\to D_{j}.}

In section 6, we will also want some corresponding facts about the classical
geometry of $Y_\Gamma=\S^3/\Gamma\times \S^3$.  Here $\Gamma$ is
a finite subgroup of $SU(2)$, and we understand $Y$ to be, in more
detail, $\Gamma\backslash SU(2)^3/SU(2)$, where as usual $SU(2)$
acts on the right on all three factors of $SU(2)^3$, while $\Gamma$ acts
on the left on only the first $SU(2)$.
Thus, concretely, an element of $Y_\Gamma$ can be represented
by a triple $(g_1,g_2,g_3)$ of $SU(2)$ elements, with $g_1\cong \gamma g_1$
for $\gamma\in\Gamma$ and $(g_1,g_2,g_3)\cong (g_1h,g_2h,g_3h)$ for
$h\in SU(2)$.  Let us rewrite the relation \pilgo\ in terms of $Y_\Gamma$.
The $D_i$ for $i>1$ project to cycles $D'_i\cong \S^3\in Y_\Gamma$,
but $D_1$ projects to an $N$-fold cover of  $D_1'=\S^3/\Gamma$,
which we can regard as the first factor
in $Y_\Gamma=\S^3/\Gamma\times\S^3$.  So we have
\eqn\kilo{ND_1'+D_2'+D_3'=0.}
In the manifold $X_{i,\Gamma}$ obtained by ``filling in'' $g_i$,
there is an additional relation $D_i'=0$.  So the homology of $X_{2,\Gamma}$,
for example, is generated by $D_1'$ with $D_2'=0$ and $D_3'=-ND_1'$.
To compute the intersection numbers of the $D_i'$ in $Y_\Gamma$,
we lift them up to $Y$ and count the intersections there, and then
divide by $N$ (since $N$ points on $Y$ project to one on $Y_\Gamma$).
$D_1'$, $D_2'$, and $D_3'$ lift to $D_1, $ $ND_2$, $ND_3$, so the
intersections are
\eqn\inon{D_1'\cdot D_2'=-D_1'\cdot D_3'=1, ~~D_2'\cdot D_3'=N}
(and $D_i'\cdot D_j'=-D_j'\cdot D_i'$ as the cycles are of odd dimension).
This is consistent with \kilo.

\newsec{Relation To Singularities Of Special Lagrangian Three-Cycles}

\def\I{{\rm I}}
\def\II{\rm II}
\def\III{\rm III}

\subsec{Introduction}

In this section we shall investigate in more detail the geometry
of our three different six-manifolds $Y$, namely \eqn\exy{
(\I)~{\CP}^{3},~~(\II) ~SU(3)/U(1)^{2},~~ (\III) ~\S^{3}\times
\S^{3},} and the associated manifolds $X$ of $G_2$ holonomy.

As we have noted, all of these have Einstein metrics, homogeneous
for the appropriate groups, and giving rise to cones with $G_{2}$
holonomy. Each of these cones admits a deformation to a smooth
seven-manifold $X$ with $G_{2}$ holonomy and the same symmetry
group.

In general, if one is given a free action of $U(1)$ on a
$G_2$-manifold $X$, then $X/U(1)$ is a six-manifold with a
natural symplectic structure. The symplectic form $\omega$ of
$X/U(1)$ is obtained by contracting the covariantly constant
three-form $\Upsilon$  of $X$ with the Killing vector field $K$
that generates the $U(1)$ action on $X$.  (In other words,
$\omega=\pi_*\Upsilon$, where $\pi$ is the projection $\pi:X\to
X/U(1)$.)

In this situation, $M$-theory on $\R^4\times X$ is equivalent to
Type IIA on $\R^4\times X/U(1)$.  $X/U(1)$ is only Calabi-Yau if
the $U(1)$ orbits on $X$ all have the same length.  Otherwise, in
Type IIA language, there is a nonconstant dilaton field, and a
more general form of the condition for unbroken supersymmetry
must be used.

A reduction to Type IIA still exists if the $U(1)$ action on $X$
has fixed points precisely in codimension four.  Because of the
$G_2$ holonomy, the $U(1)$ action in the normal directions to the
fixed point set always looks like \eqn\hutu{ (n_1,n_2)\to
(\lambda n_1,\lambda n_2),~\lambda=e^{i\theta},} with some local
complex coordinates $n_1,n_2$ on the normal bundle $\R^4\cong
\C^2$.\foot{In the tangent space $T_P$ at a fixed point $P$,
$U(1)$ must act as a subgroup of $G_2$, so as to preserve the
$G_2$ structure of $X$. So our statement is that any $U(1)$
subgroup of $G_2$ that leaves fixed a three-dimensional subspace
of $T_P$ acts as in \hutu. Indeed, the Lie algebra of such a
$U(1)$ is orthogonal to a nonzero weight in the seven-dimensional
representation of $G_2$, and this uniquely fixes it, up to
conjugation.} The quotient $\C^2/U(1)$, with this type of $U(1)$
action, is isomorphic to $\R^3$, the natural coordinates on
$\R^3$ being the hyperk\"{a}hler moment map $\vec \mu$, which in
physics notation is written \eqn\nutu{\vec\mu=(n,\vec\sigma n).}
where $n\in \C^2$, $(~,~)$ is a hermitian inner product, and
$\vec\sigma$ are Pauli matrices, a basis of hermitian traceless
$2\times 2$ matrices.

Whenever the  fixed points are in codimension four, it follows by
using this local model at the fixed points that $X/U(1)$ is a
manifold.  Moreover, $X/U(1)$ is symplectic; it can be shown
using the explicit description in \nutu\ and the local form of a
$G_2$ structure that $\omega$ is smooth and nondegenerate even at
points in $X/U(1)$ that descend from fixed points in $X$.  The
fixed point set in $L\subset X$ is three-dimensional (since it is
of codimension four) and maps to a three-manifold, which we will
also call $L$, in $X/U(1)$. $L$ is always Lagrangian, but is not
always special Lagrangian, just as $X$ is not always Calabi-Yau.
Physically, as explained in \townsend, $L$ is the locus of a
$D$-brane (to be precise, a $D6$-brane whose worldvolume is
$\R^4\times L$).

For any $U(1)$ action on $X$ whose fixed points are in
codimension four, this construction gives a way of mapping
$M$-theory on $\R^4\times X$ to an equivalent Type IIA model. In
\amv, this situation was investigated in detail for
$Y=\S^{3}\times \S^{3}$ and a particular choice of $U(1)$. We
shall now investigate a different class of $U(1)$ subgroups that
have fixed points of codimension four, as follows:

(I) For $Y=\CP^3$, the connected global symmetry group is
$Sp(2)$.  We take $U(1)\subset Sp(1)\subset Sp(2)$.

(II) For $Y=SU(3)/U(1)^2$, the connected global symmetry group is
$SU(3)$. We take $U(1)$ to consist of elements ${\rm
diag}(\lambda^{-2},\lambda,\lambda)$, with $\lambda=e^{i\theta}$.

(III) For $Y=\S^3\times \S^3$, the connected global symmetry
group is $SU(2)^3$,  and we take a diagonal $U(1)$ subgroup of
the product of the three $SU(2)$'s.

 In all three
examples, we will show that \eqn\muffin{ \eqalign{Y/U(1) &\cong
\S^{5}  \cr X/U(1) &\cong \R^{6}.\cr}} Moreover, we shall
construct explicit smooth identifications in \muffin\ which
respect the appropriate symmetries.

Since $X/U(1)$ will always be $\R^6$, the interesting dynamics
will depend entirely on the fixed point set $L\subset \R^6$.
Singularities of $X$ will be mapped to singularities of $L$.
Though we will not get the standard metric on $\R^6$ and neither
will $L$ be special Lagrangian, it is reasonable to believe that
near the singularities of $L$, the details of the induced metric
on $\R^6$ and the dilaton field are unimportant. If so, then
since on a Calabi-Yau manifold, supersymmetry requires that $L$
should be special Lagrangian, it should be (approximately)
special Lagrangian near its singularities. Indeed, the
singularities of $L$ that we will find are exactly the simplest
examples of singularities of special Lagrangian submanifolds of
$\C^3$ as investigated in \joycesp, based on earlier work in
\ref\harlaw{R. Harvey and H. B. Lawson, ``Calibrated Geometries,''
Acta Mathematica {\bf 148} (1982) 47.}.

The fixed point sets $F$ of the $U(1)$ action on $Y$ will be
two-manifolds, which descend to two-manifolds in
$Y/U(1)=\S^{5}$.  These will turn out to be \eqn\jury{\eqalign{ Y
=\CP^{3},&~~~~~F=\S^{2}\cup \S^{2}  \cr Y
=SU(3)/U(1)^{2},&~~~~~F=\S^{2}\cup \S^{2}\cup \S^{2}  \cr Y
=\S^{3}\times \S^{3},&~~~~~F=\S^{1}\times \S^{1}  .\cr}}

If we take for $X$ simply a cone on $Y$, then $L$ will be a
(one-sided) cone on $F$.  A cone on $\S^2$ is a copy of $\R^3$.
So in the first two examples, if $X$ is conical, $L$ consists of
two or three copies of $\R^3$, respectively.  The $\R^3$'s
intersect at the origin, because the $\S^2$'s are linked in
$\S^5$.

If we deform to a smooth $X$, $L$ comes out to be \eqn\ujury{
\eqalign{Y =\CP^{3}, &~~~~~L=\S^2\times \R  \cr Y
=SU(3)/U(1)^{2},&~~~~~L=\S^2\times \R\cup \R^3  \cr Y
=\S^{3}\times \S^{3}, &~~~~~L=\S^{1}\times \R^2  .\cr}}

Using these facts, we can compare to the claims in section 2 in
the following way:

(I) Suppose $Y=\CP^3$ and $X$ is a cone on $Y$. As explained
above, the fixed point set in $X$ corresponds to two copies of
$\R^3$, meeting at the origin in $\R^6$. The two $\R^3$'s meet at
special angles such that supersymmetry is preserved
\ref\douglas{M. Berkooz, M. R. Douglas, and R. G. Leigh, ``Branes
Intersecting At Angles,'' Nucl. Phys. {\bf B480} (1996) 265,
hep-th/9606139.}; a massless chiral multiplet $\Phi$ arises at
their intersection point from open strings connecting the two
$\R^3$'s. Generally, a $D$-brane supports a $U(1)$ gauge field,
but in the present case, because of the noncompactness of the
$\R^3$'s, the two $U(1)$'s behave as global symmetries in the
effective four-dimensional physics. $\Phi$ is neutral under the
sum of the two $U(1)$'s, and this sum is irrelevant in the
four-dimensional description.  So the effective four-dimensional
description is  given by a chiral multiplet $\Phi$ with a single
$U(1)$ symmetry.  This agrees with what we found in section 2.2.
Giving an expectation value to $\Phi$ corresponds \ref\cama{C. G.
Callan, Jr. and J. M. Maldacena, ``Brane Dynamics From The
Born-Infeld Action,'' Nucl. Phys. {\bf B513} (1998) 198,
hep-th/9708147.} to deforming the union of the two $\R^3$'s to a
smooth, irreducible special Lagrangian manifold of topology
$\S^2\times \R$.  This deformation has been described in
\refs{\harlaw,\joyce} and from a physical point of view in
\ref\kachru{S. Kachru and J. McGreevy, ``Supersymmetric Three
Cycles And Supersymmetry Breaking,'' Phys. Rev. {\bf D61} (2000)
026001, hep-th/9908135.}.  As stated in \ujury, if we deform the
cone on $Y$ to a smooth $G_2$-manifold $X$, the resulting fixed
point set is indeed $\S^2\times\R$.

(II) For a cone on $Y=SU(3)/U(1)^2$, the fixed point set is three
copies of $\R^3$, meeting at the origin in $\R^6$ in such a way
as to preserve supersymmetry. This corresponds to three
$D$-branes whose worldvolumes we call $D_i$. There are three
massless chiral multiplets, say $\Phi_1,$ $\Phi_2$, and $\Phi_3$,
where $\Phi_i$ arises from open strings connecting $D_{i-1}$ and
$D_{i+1}$.  Each of the $\R^3$'s generates a $U(1)$ global
symmetry, but the sum of the $U(1)$'s decouples, so effectively
the global symmetry group of the $\Phi_i$ is $U(1)\times U(1)$. A
superpotential $\Phi_1\Phi_2\Phi_3$ is generated from worldsheet
instantons with the topology of a disc.  Because of this
superpotential, any one of the $\Phi_i$, but no more, may receive
an expectation value. This agrees with the description in section
2.3. On a branch of the moduli space of vacua on which
$\langle\Phi_i\rangle \not= 0$, for some $i$,  the union of
$D_{i-1}$ and $D_{i+1}$ is deformed just as in case I above to a
smooth $D$-brane with topology $\S^2\times \R$ (and disjoint from
$D_i$), while $D_i$ is undeformed. So $L=\S^2\times \R\cup
\R^3$.  As stated in \ujury, this is indeed the fixed point set
that arises when a cone on $Y$ is deformed to a smooth
$G_2$-manifold.

(III) The cone on $\S^1\times\S^1$ has, as analyzed in \joycesp,
three different special Lagrangian deformations, all  with
topology $\S^1\times \R^2$. (They differ by which one-cycle in
$\S^1\times\S^1$ is a boundary in $\S^1\times \R^2$.) These three
possibilities for $L$ correspond to the three possibilities,
described in section 2.4, for deforming the cone on $Y$ to a
smooth $G_2$-manifold $X$. However, since the cone on $\S^1\times
\S^1$ is singular, and we do not have any previous knowledge of
the behavior of a $D$-brane with this type of singularity, in
this example going from $M$-theory to Type IIA does not lead to
any immediate understanding of the dynamics.  It merely leads to
a restatement of the questions.   Everything we will say in
section 4 could, indeed, be restated in terms of $D$-branes in
$\R^6$ (with worldsheet disc instantons playing the role of
membrane instantons).  This example has also been examined in
\agva.

Our basic approach to establishing that $Y/U(1)=\S^5$ and
$X/U(1)=\R^6$
  will not use a cartesian
coordinate description but what one might call ``generalized polar
coordinates'' better adapted to rotational symmetry. For example,
if we choose an orthogonal decomposition \eqn\orthdec{
\R^{6}=\R^{4}\oplus \R^{2},  } then using polar coordinates
$(r,\theta )$ in $\R^{2}$ we can view $\R^{6}$ as being ``swept
out,'' under the $SO(2)$-action, by the one-parameter family of
half five-spaces \eqn\halfspace{ \R_{+}^{5}\left( \theta \right)
=(x,r), ~~~~~x\in \R^{4}, ~~~~~ r\geq 0,} for fixed $\theta $ in
$\left[ 0,2\pi \right] $. Alternatively, we can restrict $\theta
$ to lie in $\left[ 0,\pi \right] $ and allow negative $r$, thus
sweeping out $\R^{6}$ by copies of $\R^{5}$.

A very similar story applies if we choose a decomposition
\eqn\simsto{ \R^{6}=\R^{3}\oplus \R^{3}  .} This decomposition can
conveniently be viewed in complex notation as the decomposition
of $\C^{3}$ into its real and imaginary parts: $z=x+iy$. Now, for
$x\in \R^3$ with $|x|=1$, we set \eqn\sweepout{
\R^{4}(x)=(tx,y)~{\rm with }~t\in \R,\;y\in \R^{3},} noting that
$\pm x$ give the same four-space. As $x$ varies, the $\R^4(x)$
sweep out $\R^6$.

The key observation about our seven-manifolds $X$ is that they
naturally contain families of six-manifolds or five-manifolds
(depending on the case) so that the quotients by $U(1)$ can be
naturally identified with the relevant polar description of
$\R^{6}$. These manifolds (of dimension six and five) will all be
real vector bundles over the two-sphere $\S^{2}$, with
appropriate $U(1)$-action. As a preliminary we shall need to
understand these actions and their associated quotients. The
nontrivial statements about quotients that we need are all
statements in four dimensions, so we begin there.

\subsec{Some Four-Dimensional Quotients}

We shall be interested in a variety of basic examples of
four-manifolds with $U(1)$ action. The prototype is of course
$\R^{4}=\C^{2}$ with complex scalar action. The quotient is
$\R^{3}$.  If we identify $\C^{2}$ with the quaternions $\H$, as
a flat hyperk\"{a}hler manifold, then $U(1)$ preserves this
structure and the associated hyperk\"{a}hler moment map:
\eqn\hypmom{ \vec\mu :\R^{4}/U(1)\cong\R^{3} } identifies
$\R^{4}/U(1)$ with $\R^{3}$. ($\vec \mu$ was defined in \nutu.)
 Note that $\R^{4}=\H$ gets its
hyperk\"{a}hler structure from right quaternion multiplication, and the $%
U(1) $ action comes from an embedding $\C\rightarrow \H$ using
left quaternion multiplication.

The map \hypmom\ gives a local model for  $U(1)$ actions on more
general four-manifolds $M$.\ If a fixed point $P$ is isolated,
then the linearized action at that point gives a decomposition of
the tangent space \eqn\dectan{ T_P=\R^{2}(a)\oplus \R^{2}(b).}
where the factors denote invariant subspaces $\R^{2}=\C$ on which
$U(1)$ acts by $\lambda \rightarrow \lambda ^{a},\lambda
\rightarrow \lambda ^{b}$, with integers $a,b$. To specify the
signs of $a,b$ we have to pick orientations of the $\R^{2}$'s.
Then the basic result \hypmom\ implies that, near $P$, the
quotient $M/U(1)$ will be a smooth three-manifold provided
$\left| a\right| =\left| b\right| $. If $\left| a\right| =\left|
b\right| =1$ we are, with appropriate orientations, in the case
considered in \hypmom, while if $\left| a\right| =\left| b\right|
=k$ then the $U(1)$ action factors through a cyclic group of
order $k$ and reduces back to \hypmom\ again.

If $b=0$ and $\left| a\right| >0$, then the local model is the action of $%
U(1) $ on $\C^{2}$ given by \eqn\turmo{ \lambda \left(
z_{1},z_{2}\right) =\left( \lambda ^{a}z_{1},z_{2}\right). } The
quotient by $U(1)$ is clearly $\R_{+}\times \R^{2}=\R_{+}^{3}$, a
half-space with boundary $\R^{2}$. This applies locally to a
four-manifold near a fixed surface with such weights $(a,b)$.

With these two model examples, we now want to move on to consider
the case when $M$ is a complex line-bundle over $\CP^{1}=\S^{2}$.
Since \eqn\begeq{ \S^{2}=SU(2)/U(1),} a representation of $U(1)$
given by $\lambda \rightarrow \lambda ^{k}$ defines a complex
line-bundle $H^{k}$ over $\S^{2}$, and moreover this has a
natural action of $SU(2)$ on it.\ With the appropriate
orientation of $\S^{2},$ the line-bundle $H^{k}$ has first Chern
class $c_{1}=k$.

We shall also want to consider an additional $U(1)$-action on the
total space of this line-bundle, given by multiplying by $\lambda
^{n}$ on each fibre. As a bundle with $SU(2)\times U(1)$ action,
we shall denote this by $H^{k}(n)$.

Let $F_1$ be a $U(1)$ subgroup of $SU(2)$, $F_2$ the additional
$U(1)$ that acts only on the fibres, and $F$ the diagonal sum of
$F_1$ and $F_2$. We want to look at the quotient of $H^k(n)$ by
$F$. The fixed points of $F$ are (for generic $n$ and $k$) the
points in the zero-section $\S^2$ that are fixed by $F_1$.  In
fact, $F_1$  has two fixed points, which we will call 0 and
$\infty$. On the tangent planes to $\S^2$ at the two fixed
points, $F_1$
 acts with weights that are respectively 2 and $-2$.  The
$2$ arises because $SU(2)$ is a double cover of $SO(3)$ (so a
spinor on $\S^2$ transforms under $F_1$ with weight $\pm 1$ and a
tangent vector with weight $\pm 2$).  On the fibres of $H^k(n)$
over the two fixed points, $F_1$ acts with respective weights $k$
and $-k$. \foot{The wavefunction of an electron interacting with
a magnetic monopole of charge $k$ is a section of $H^k$.  The
minimum orbital angular momentum of such an electron is $k$,
because a rotation acts on the fibre with weights $k$ and $-k$.}
So its weights are $(2,k)$ at one fixed point and $(-2,-k)$ at
the other.  On the other hand, $F_2$ acts trivially on the
tangent space to $\S^2$, and acts with weight $n$ on the fibre.
So it acts with weights $(0,n)$ at each fixed point.

Adding the weights of $F_1$ and those of $F_2$, we learn that the
diagonal subgroup $F$ acts at the tangent spaces to the fixed
points with weights \eqn\wweights{ \left( 2,n+k\right),~~~~
\left( -2,n-k\right).} The fixed points
 are isolated provided $\left| n\right| \neq \left| k\right| $. The
quotient $H^k(n)/F$ will be a three-manifold provided
\eqn\willman{ \left| n\pm k\right| =2} while if $\left| n\right|
=\left| k\right| >0$ the quotient will be a three-manifold with
an $\R^{2}$-boundary.

We shall be interested, for application to our three
seven-manifolds, in the quotients $H^k(n)/U(1)$ (with $U(1)=F$)
in the following three special cases:

(I) In the first example, we will have $k=2$, $n=0$. Here,
$H^{2}(0)$ is the tangent bundle over $\S^{2}$, with Chern class
$c_{1}=2$ and with standard $U(1)$-action.

(II) In the second example, we will have $k=\pm n = 1$.
$H^{1}(1)$ is the spin bundle $\left( c_{1}=1\right) $, but with
an additional $U(1)$ action by $\lambda $ or $\lambda ^{-1}$ in
the fibres.

(III) In the third example, we will have $k=0$, $n=\pm 2$.
$H^{0}(2)$ is the trivial bundle $\left( c_{1}=0\right) $ with
$\lambda ^{2}$ or $\lambda ^{-2}$ action in the fibres.

{}From our general remarks earlier, we know that $\I$ and $\III$
lead to three-manifold quotients, whereas $\II$ leads to a
three-manifold with boundary. We claim that these quotients are:

\eqn\juggu{ \eqalign{\I &~~~~~~~   H^2(0)/U(1) =\R^{3}\cr
  \II & ~~~~~~~ H^1(1)/U(1)=\R_{+}^{3} \cr
   \III & ~~~~~~~H^0(2)/U(1)=\R^{3}.\cr}}

Note that changing the sign of $n$ in case $\II$ or $\III$ simply
amounts to switching the fixed points $0,\infty $ on the sphere
and does not change the geometry. We shall therefore take $n>0$.

\bigskip
\noindent{Case I: $H^2(0)/U(1)$}

$H^2(0)$ is the resolution  of the $A_{1}$ singularity to a smooth
hyperk\"{a}hler manifold. The resolution of the
$A_{1}$-singularity is the line bundle over $\CP^1$ with $c_1=-2$
and (after changing the orientation to make $c_1=+2$) the map from
$H^2(0)/U(1)$ to $\R^3$
 is given by the hyperk\"{a}hler moment map $\mu_1$ of the $U(1)$ action.
This is a natural generalization of the prototype statement
$\R^4/U(1) =\R^3$, which is really the $N=0$ case of the
$A_{N}$-singularity story.

The proof that $H^2(0)/U(1)=\R^3$ can be made completely explicit
by writing the hyperk\"{a}hler metric on $H^2(0)$ in the form
\ref\gibh{G. W. Gibbons and S. W. Hawking, ``Gravitational
Multi-Instantons,'' Phys. Lett. {\bf B78} (1978) 430.}
\eqn\inin{ds^2=U^{-1}(d\tau+\vec \omega\cdot d\vec x)^2+Ud\vec
x\cdot d\vec x, ~~~\vec x\in \R^3,~~~ U={1\over |\vec x-\vec
x_0|}+{1\over |\vec x-\vec x_1|},~~d\vec \omega = *dU.} $U(1)$
acts by translation of the angular coordinate $\tau$, so dividing
by $U(1)$ is accomplished by forgetting $\tau$; the quotient is
thus $\R^3$, parameterized by $\vec x$.

$H^2(0)$ has another $U(1)$ symmetry that commutes with the group
$F=U(1)$ that we are dividing by, namely the group $F_2=U(1)$
that acts only on the fibres of $H^2(0)$.  $F_2$ corresponds to
the rotation of $\R^3$ that leaves fixed the points $\vec x_0$
and $\vec x_1$.

Since the $A_1$ singularity is conical, its smooth resolution
$H^2(0)$ is
 a particularly simple example of
the process of deforming conical singularities that is the main
theme of this paper, so it is worth examining the geometry here
in more detail.

The map $\mu_{1}$ collapses  $\S^{2}$ (the zero-section) of
$H^{2}(0)$ to an interval in $\R^{3}$ (the interval connecting
$\vec x_0$ and $\vec x_1$), with $0,\infty $ becoming the end
points. In fact the hyperk\"{a}hler metric on $H^{2}(0)=T\S^{2}$
has a parameter $r$, essentially the radius of the sphere, and
this becomes the length of the
interval. Thus as $r\rightarrow 0,$ $T\S^{2}$
degenerates to the cone on $%
\S^{3}/\Z_{2}={\RP}^{3}$ and the interval  shrinks to a point. In
the limit $r=0$,
we still find $\R^{3}$ but this time it is really got from dividing $%
\C^{2}/\Z_{2}$ by $U(1)$, or equivalently by dividing $\C^{2}$ by
the $U(1)$ with weight $(2,2)$. The origin in $\R^{3}$ really has
multiplicity $2$.

Thus, while the deformation of the cone on ${\RP}^{3}$ to a smooth
four-manifold alters the topology, the corresponding deformation
of the quotients by $U(1)$ does not. It simply involves expanding
a point to an interval. Note however that the identification of
$\R^{3}$ before and after the deformation involves a complicated
stretching map in which spheres centred at the origin get
stretched into ellipsoidal shapes surrounding the interval.

Similar remarks apply in fact to all the
$A_{N}$-singularities.\bigskip

\bigskip
\noindent  {Case II: $H^1(1)/U(1)$}

We turn next to case II, the quotient $H^{1}(1)/U(1)$. The
weights at 0 and $\infty$ are respectively $(2,2)$ and $(-2,0)$.
Our aim is to construct a map \eqn\conmap{
f:H^{1}(1)\longrightarrow \R_{+}^{3}=(x,y,z)\in \R^{3},\;z\geq 0}
which identifies the quotient by $U(1)$. The image of the
zero-section will be contained in the half-line $x=y=0$, and
there will be rotational symmetry about this axis.

We shall briefly indicate two different proofs. The first is a
``bare hands'' identification with cutting and gluing, whereas
the second is more elegant but more sophisticated.

The first argument uses, as models, two $U(1)$-actions on $\R^4$
whose quotients we have already discussed. For an action with
weights $(-2,0)$, the quotient is a half-space; this gives a
local model near $\infty$. For an action with weights $(2,2)$,
the quotient is $\R^3$; this gives a local model near 0.

In the figure below, we sketch the ingredients in a direct
analysis of the quotient $H^1(1)/U(1)$. The horizontal direction
in the figure is the $z$ direction; a transverse $\R^2$ is
understood.



{\baselineskip=14pt
\bigskip
{\bf $$\underline{\quad \quad\qquad\qquad \qquad \qquad \;\; {\bf
0} \quad\quad\;\; {\bf 1} \quad\quad\;\; {\bf 2} \qquad \qquad
\qquad \qquad \qquad \qquad\qquad\qquad \,\;\;\;}$$

$$\,\,\,\,\quad \qquad\qquad \!\!\!\;\;\;\;\; \qquad
\qquad\overline{{\bf 0} \quad\quad\;\; {\bf 1} \qquad \qquad
\qquad \qquad \qquad \qquad \qquad \qquad  \qquad
\,\,\,\;\;\;\;\;}$$}}

The first line is the image in $\R^{3}$ of three surfaces in
$H^{2}(0)$. The interval $[0,2]$ with 1 as midpoint represents
the image of the $0$-section of $H^2(0)$ with $\infty \rightarrow
0$ and $0\rightarrow 2$; to the left and right of this interval
are half-lines that are the images of the $\R^{2}$ fibres over
$0,\infty $. Below this is a half-line that represents the
quotient $\R^4/U(1)$ with weights $(-2,0)$. In that  half-line,
the point $1$ has no geometric significance, but is just chosen
to match up with the top line.

We want the quotient $H^1(1)/U(1)$, rather than the quotient
$H^2(0)/U(1)$ which is depicted in the top half of the figure.
To convert $H^2(0)$ into $H^1(1)$, remove the half-line $z<1$ in
the top line and the half-line $z>1$ in the bottom line, and glue
the remaining parts at $z=1$. In the four-manifolds we have, over
this point, copies of $\R^{2}\times U(1)$. We glue these together
with a twist, using $U(1)$ to rotate $\R^{2}$.

It is not hard to see that we have, in this way, constructed
$H^{1}(1)$. The fact that its quotient by $U(1)$ is a half-space
now follows from our construction.

Our second proof uses again the fact that $H^2(0)/U(1)=\R^3$, but
now we make the $H^2(0)=T\S^2$  depend on a parameter $t$, in
such a way that in the limit $t\to 0$, the $\S^2$ splits up (by
pinching at the equator) into two copies of $\S^2$. In this
limit, $H^2(0)$ splits into two $H^1(1)$'s, and its quotient
$\R^3$ splits into two copies of $\R^3_+$.

We start out with the projective plane $\CP^{2}$, with its
standard line-bundle $H$ having $c_{1}=1$, and we take $U(1)$ to
act on the homogeneous coordinates of $\CP^{2}$ by \eqn\makact{
\lambda \left( z_{1},z_{2},z_{3}\right) =\left( z_{1},\lambda
^{2}z_{2},\lambda ^{-2}z_{3}\right)  .} This has three isolated
fixed points \eqn\fixedpts{ A=(1,0,0),\quad B=(0,1,0),\quad
C=(0,0,1).} $U(1)$ acts on the fibres of $H$ over $A,B,C$ with
weights $(0,-2,2)$. Restricting to the projective line $AB$ (the
copy of $\CP^1$ containing $A$ and $B$),
we find the line-bundle $H^{1}(1)$, and over $%
AC $ we find $H^{1}(-1)$. These are the four-manifolds whose quotients by $%
U(1)$ should each be an $\R_{+}^{3}$.

Now we introduce the family of rational curves (i.e. two-spheres)
with equation \eqn\burgo{ z_{2}z_{3}=tz_{1}^{2} ,} with $t$ being
our parameter. Note that each of these is invariant under our
$U(1)$,
and is in fact the closure of an orbit of the complexification $%
\C^{\ast }$. The bundle $H$ over $\CP^{2}$
restricts on each of these rational curves to a copy of $%
H^{2}(0)$ (this follows from the weights at the fixed points $B$
and $C$), so that by our previous analysis, quotienting out
yields an $\R^{3}$ (depending on $t$). As $t\rightarrow 0$, these
$\R^3$'s  tend to the union of the quotients coming from the two
branches of the degenerate curve $z_{2}z_{3}=0$, i.e. the two
lines $AB,\;AC$ (note that the integers add up, so that
$H^{2}(0)$ is the ``sum'' of $H^{1}(1)$ and $H^{1}\left(
-1\right) $). By symmetry this must split the $\R^{3}$ into two
half-spaces, as requested.

In fact, near $A$, the quotient of $H$ (as a bundle over
$\CP^{2}$) is a five-manifold and so there is little difficulty
in checking the details of the limiting process.\bigskip

\bigskip
\noindent {Case III: $H^{0}(2)/U(1)$}

We come finally to case III, the bundle $H^{0}(2)$, i.e. the product $%
\S^{2}\times \C$ with $U(1)$ acting as usual on
$\S^{2}$ (as the standard subgroup of $%
SU(2)$) and acting on $\C$ with weight $2$. We shall show that
$H^{0}(2)$ has the same quotient as $H^{2}(0)$ which, as we have
seen, is $\R^{3}$. To see this, we
observe that both quotients can also be viewed as the quotient by the torus $%
U(1)^{2}$ acting on \eqn\gurfy{ \S^{3}\times \C=\left(
z_{1},z_{2},z_{3}\right) ~{ \rm with }~\left| z_{1}\right|
^{2}+\left| z_{2}^{2}\right| =1} by \eqn\nurfy{\left( \lambda
,\mu \right) \left( z_{1},z_{2},z_{3}\right) =\left( \lambda
z_{1},\mu z_{2},\lambda \mu ^{-1}z_{3}\right) .}
Factoring out by the diagonal $U(1),\;\lambda =\mu $, yields the product $%
H^{0}$, while factoring out by the anti-diagonal $\lambda =\mu
^{-1}$ yields the non-trivial bundle $H^{2}$. The action of the
remaining $U(1)$ identifies these two as $H^{0}(2)$ and
$H^{2}(0)$. This completes the proof that \eqn\jogo{
H^{0}(2)/U(1)=\S^{3}\times \C/U(1)^{2}=H^{2}(0)/U(1)=\R^{3}.}

We could also have used a cutting and gluing argument as in case
II.

Having dealt with these three cases of four-manifolds with $U(1)$
action, we are now ready to apply our results  to the three cases
of asymptotically conical $G_2$-manifolds with $U(1)$ action. All
three are variations on the same theme, with minor differences.
The four-manifold quotients I, II, III that we have just
considered will arise in studying the similarly numbered
$G_2$-manifolds.

\subsec{ Case $\I:\,\,Y=\CP^{3}$ }

We begin with the case when $Y=\CP^{3}$ and $X$ is the
$\R^{3}$-bundle over $\S^{4}$ given by the anti-self-dual
$2$-forms. The action of $U(1)$ on $\CP^{3} $ is given in complex
homogeneous coordinates by \eqn\actgiv{ \lambda \left(
z_{1},z_{2},z_{3},z_{4}\right) =\left( \lambda z_{1},\lambda
z_{2},\lambda ^{-1}z_{3},\lambda ^{-1}z_{4}\right).} It has two
fixed $\CP^{1}$'s, which are of codimension $4$.  The action in
the normal directions to the fixed point set
 has weights $(2,2)$ or $(-2,-2)$. This ensures (as in section 3.1)
that the quotient $\CP^{3}/U(1)$ is a (compact) five-manifold.

The fact that $\CP^3/U(1)=\S^5$ can be shown in a very elementary
fashion.  We can write a point in $\S^5$ in a unique fashion as a
six-vector $(t\vec x,\sqrt{1-t^2}\vec y)$ with $0\leq t\leq 1$
and $\vec x,\vec y\in \S^2$.  We normalize the $z_i$ so
$\sum_i|z_i|^2=1$, and rename them as $(n_1,n_2,m_1,m_2)$. Then
we simply define $t=\sqrt{|n_1|^2+|n_2|^2}$, $\vec x
=(n,\vec\sigma n)/(n,n)$, $\vec y = (m,\vec \sigma m)/(m,m)$, and
this gives an isomorphism
 from $\CP^3/U(1)$ to $\S^5$.

If $X$ is a cone on $Y$, its quotient by $U(1)$ is a cone on
$\S^5$, or in other words $\R^6$.

We want to show that the smooth $G_2$-manifold $X$ that is
asymptotic to a cone on $Y$ also obeys $X/U(1)=\R^6$. We recall
that $X$ is the bundle of anti-self-dual two-forms on $\S^4$. The
$U(1)$ in \actgiv\ is a subgroup of the $Sp(2)$ symmetry group
of  $\CP^3$ (as described in section 2.2). Identifying
$Sp(2)=Spin(5)$ as a symmetry group of $\S^4$, this particular
$U(1)$ acts by rotation on two of the coordinates of $\S^4$. In
other words, if  we regard $\S^{4}$ as the unit sphere in
$\R^{5}$, then there is  a decomposition \eqn\regdec{
\R^{5}=\R^{2}\oplus \R^{3} } such that $U(1)$ is the rotation
group of $\R^2$; we will use coordinates $\vec x=(x_1,x_2)\in
\R^2$, $\vec y=(y_1,y_2,y_3)\in \R^3$. The fixed point set of
$U(1)$ in $\S^4$ is a copy of $\S^2$ (the unit sphere in
$\R^3$).  Over one of the fixed points in $\S^4$, there is up to
a real multiple  just one $U(1)$-invariant anti-self-dual two-form
(it looks like $\epsilon- * \epsilon$, where $\epsilon$ is the
volume form of the fixed $\S^2$ and $*$ is the Hodge dual).  A
fixed point in $X$ is made by choosing a fixed point in $\S^4$
together with an anti-self-dual two-form, so the fixed point set
in $X$ is $\S^2\times \R$.  Thus, for this example, we have
justified the assertions in \jury, \ujury.

\def\od{{\buildrel \lower3pt\hbox{$\scriptscriptstyle\circ$} \over  D}}
\def\ox{{\buildrel \lower3pt\hbox{$\scriptscriptstyle\circ$} \over  X}}
\def\oi{{\buildrel \lower3pt\hbox{$\scriptscriptstyle\circ$} \over  I}}

The quotient $\S^4/U(1)$ is a closed three-disc $D^3$. Indeed, by
a $U(1)$ rotation, we can map any point in $\S^4$ to  $x_2=0$,
$x_1=\sqrt{1-\vec y^2}$, so $\S^4/U(1)$ is parameterized by $\vec
y$ with $|\vec y|\leq 1$. If we omit the fixed $\S^2$, then over
the open three-disc $\od{}^3 $, the bundle of anti-self-dual
two-forms is just the cotangent bundle of the disc. (In fact, if
$\alpha$ is a one-form on the open disc, then on identifying the
open disc with the set $x_1>0$, $x_2=0$, we map $\alpha$ to the
anti-self-dual two-form $dx^1\wedge \alpha -
*(dx^1\wedge\alpha)$.)

We think of $\od{}^3$ as living in $\R^3$.  Its tangent bundle is
trivial and can naturally be embedded in
\eqn\rufog{\C^3=\R^3+i\R^3} with the disc embedded in $\R^3$,
while $i\R^3$ represents the tangent directions.  This is of
course compatible with the natural action of $SO(3)$ on the disc
(which in fact can be extended to $SO(3,1)$ using the conformal
symmetries of $\S^4$).

At this stage one might think we are almost home, in that if
$\ox$ is the part of $X$ that lies over the open disc, we have
identified \eqn\wevid{ \ox/U(1)=T\od{}^{3}=\od{}^3\times\R^3,}
and the right side is diffeomorphic to $\R^{6}.$ However, this
does not allow for the part of $X$ that we have excised, namely
the fibre of $X$ over the fixed $\S^{2}.$ The surprise is that,
including this, we still get $\R^{6},$ with $\R^{3}\times
\od{}^{3}$ as an open dense set.

To deal with this, we shall fix a point $u\in \S^{2}\subset
\R^{3},$ which we shall also identify with its image on the
boundary of the closed disc $ D^{3}. $ Let $I_u$ be the set of
points in the closed disc of the form $ru$,  $-1\leq r\leq 1$,
and let $\oi_u$ be the corresponding open interval with $|u|<1$.
Let $X_u$ be the part of $X$ that lies above $I_u$ (so $\vec y$
is proportional to $u$), and let $\ox_u$ the open part of $X_u$
(with $|\vec y|<1$).

Restricting \wevid\ to $\ox_u$,  we get a ``slice'' of that
fibration: \eqn\getslice{ \ox_u/U(1)=\oi_u\times\R^{3} .} Our aim
is to show that, on passing to the closure $X_{u}$ of $\ox_u$,
we  get \eqn\wegget{ X_{u}/U(1)=\R_u\oplus i\R^{3}  ,} where
$\R_u$ is the line through $u$ in $\R^3$.  We will denote
$\R_u\oplus i\R^3$ as $\R^4_u$. If \wegget\ is true, then simply
upon rotating $u$ by $SO(3),$ the four-spaces $\R_{u}^{4}$ will
sweep out $\R^{6}$ and establish the required identification
\eqn\reqid{ X/U(1)=\R^{3}\oplus i\R^{3}=\C^{3}  .}

We are thus reduced to establishing \getslice,
 and this is where our preliminary
work on $U(1)$-quotients of four-manifolds will pay off. $X_{u}$
is just the restriction of the $\R^{3}$ bundle $X$ from $\S^{4}$
to the two sphere $ \S_{u}^{2}$ which is cut out by the three
space \eqn\thresp{ \R_{u}^{3}=\R^{2}\oplus \R_u } in the original
decomposition \regdec\ of $\R^{5}.$ (In other words, $\S^2_u$ is
characterized by $\vec y$ being a multiple of $u$.)

The bundle $X_{u}$ over $\S_{u}^{2}$ is homogeneous, and we can
identify it by looking at the representation of $U(1)$ at the
point $u.$ An anti-self-dual two-form at a point on
$\S^2_u\subset \S^4$ that has an even number of indices tangent
to $\S^2_u$ is invariant under rotations of $\S^2_u$ around that
point, while those with one index tangent to $\S^2_u$ (and one
normal index) transform as tangent vectors to $\S^2_u$, or
equivalently as vectors in $\R^3$ that are perpendicular to $u$.
So we get a decomposition \eqn\defdec{ \Lambda _{-}^{2}(u)=\R
u\oplus u^{\perp } . }

This decomposition shows that, as a bundle over $\S_{u}^{2},$ the
five-manifold $X_{u}$ is \eqn\itrel{ X_{u}=H^{2}(0)_{u}\times i\R
u.} Hence, using the fact that $H^2(0)/U(1)=\R^3$, its quotient
by $U(1)$ is \eqn\itsquo{ X_{u}/U(1)=\R_{u}^{3}\oplus i\R u , }
and the factor $\R_{u}^{3}$ can naturally be identified as
\eqn\nitsqu{ \R_{u}^{3}=\R u\oplus iu^{\perp }} This implies
\wegget, and hence \reqid.

The fixed point set $L=\S^{2}\times \R$ in $X$ gets mapped to the
three-manifold in $\R^{6}$ given by the union of the lines $u+i\R
u,$ for $u\in \R^{3},\left| u\right| =1.$ But as noted in \harlaw,
 this is the natural embedding
of the normal bundle to $\S^2$ in the tangent bundle \eqn\juffel{
T\R^{3}=\R^{3}\oplus i\R^{3}=\C^{3},} and hence is Lagrangian.
The radius of the $\S^2$ at the ``center'' of $L$ is the modulus
that is related to the radius of the $\S^4$ at the ``center'' of
$X$.

\subsec{Case $\II: ~~Y=U(3)/U(1)^{3} $}

We can now examine the second case, in which our six-manifold is
$Y=SU(3)/U(1)^2=U(3)/U(1)^{3}$, and, as in the previous case, $X$
is an $\R^{3}$ bundle of anti-self-dual two-forms, this time over
$\CP^{2}.$ $Y$ is just the unit sphere bundle in $X$, and the
fibration $Y\rightarrow \CP^{2}$ is given in terms of groups by
\eqn\juvvu{ U(3)/U(1)^{3}\longrightarrow U(3)/U(1)\times U(2).}
The $U(1)$-subgroup which we want to divide by to get to Type IIA
is given  by the left action of the scalars in $U(2)$, or in
other words by $U(3)$ elements of the form \eqn\minno{ \left(
\matrix{ 1 &  &  \cr & \lambda &  \cr &  & \lambda\cr} \right) .}
On $\CP^{2}$, this has an isolated fixed point $A=(1,0,0)$ and
also a fixed $ \CP^{1},$ which we will call $B$,  on which the
first coordinate vanishes. $U(1)$ acts trivially on the fibre
$X_{A}$ over $A$. At each point of $B$, the action of $U(1)$
decomposes the space of anti-self-dual two-forms  into $ \R\oplus
\R^{2}$ with $\R$ fixed and $\R^{2}$ rotated. Thus our fixed
three-manifold $L$ in $X$ this time consists of two components
\eqn\thistime{ L=\R^{3}\cup \S^{2}\times \R  .} This cuts out on
$Y$ the fixed-point set \eqn\tuvu{ F=\S^{2}\cup \S^{2}\cup
\S^{2}  } corresponding to the ends of $L$. Note that the
$\Sigma_3$ symmetry of $F$, permuting the $\S^2$'s, is broken in
$L$, where one of the $\S^{2}$'s has been preferred.  This is in
keeping with the discussion of the $\Sigma_3$ symmetry of $Y$ in
section 2.3.

We now proceed rather as in the previous example. For every $u\in
B$, we denote as $\S^2_u$ the copy of $\CP^1$ that contains $A$
and $u$. We  view $\CP^{2}$ as built up from the two-parameter
family of these $\CP^1$'s.  Of course, the group $SO(3)$ (of
rotations of $B$) acts on this family. Note that each $\S^2_u$ is
acted on by the $U(1)$ in \minno; its fixed points are the points
$A,u$, corresponding to the points called $0,\infty$ in section
3.2.

Given this decomposition of $\CP^{2}$, we consider the
corresponding decomposition of $X$ into a family of
five-manifolds $X_{u}$ which lie above $\S^2_u$ in $X$. Again,
under the action of $SO(3),$ they sweep out $X$ with axis
$X_{A},$ the fibre of $X$ over  $A.$

Each $X_{u}$ is an $\R^{3}$ bundle over the two-sphere
$\S_{u}^{2}$. The representation $\Lambda _{-}^{2}$ of $SO(4)$
(consisting of anti-self-dual two-forms) when restricted to the
subgroup $U(1)\subset U(2)\subset SO(4)$ that is given in $U(2)$
by \eqn\gubbo{ \lambda \longrightarrow \left( \matrix{ \lambda  &
0 \cr 0 & 1 \cr} \right)  } decomposes as \eqn\decas{
\R^{3}=\R\oplus \C} where $U(1)$ acts with weight $1$ on $\C.$ It
follows that $X_{u},$ as a bundle over $\S_{u}^{2},$ splits off a
trivial factor $\R$ and leaves the standard complex line-bundle
$H$ over $S^{2}.$ To emphasize the dependence on $u,$ we shall
write this decomposition as \eqn\toemp{ X_{u}=H_{u}\times \R_{u}
. } Note that the trivial factor $\R_{u}$ enables us to identify
the relevant copies of $\R\subset \R^{3}$ over the two fixed
points $A$ and $u.$ Over $u,$ this $\R_{u}$ is the fibre of $L$
over $u\in \S^{2}_u,$ while over $A$ it lies in the fixed
$\R^{3}=X_{A},$ which is independent of $u.$ If we denote as
$\R_{B}^{3}$ a copy of $\R^3$ which contains $B=\S^{2}$ as unit
sphere, we can think of $L$ as the normal bundle to
$\S^{2}\subset \R_{B}^{3}.$ There is then a natural
identification of $\R_{B}^{3}$ with $\R^{3}=X_{A},$ which matches
up the common factors $\R_u,$ as $u$ varies on $B.$ In other
words there is a natural identification \eqn\huvv{
\R_{B}^{3}=X_{A}  } compatible with the rotation action of
$SU(2)$.

With this analysis of the geometry of $X$ in terms of the family
of $X_{u}$'s, we now move on to consider the quotient by $U(1)$.
First we note that the quotient $\CP^2/U(1)$ is the unit disc $D$
in $\R^3_B$.\foot{This is proved by directly examining the action
of $U(1)$ on the homogeneous coordinates of $\CP^2$.  One
describes $\CP^2$ with coordinates $(z_1,z_2,z_3)$, normalized to
$|z_1|^2+|z_2|^2+|z_3|^2=1$. Dividing by $U(1)$ can be
accomplished by restricting to $z_1>0$.  So the quotient is
parameterized by $z_2,z_3$ with $|z_2|^2+|z_3|^2\leq 1$.} The
center of $D$ is the point $A$, and its boundary is $B=\S^2$.
Each $\S^2_u$ projects to the corresponding radial interval, from
the centre $A$ to the boundary point $u.$ To understand the
quotient of $X_{u}=H_u\times \R_u$ by $U(1)$ we need to know the
``twist'' on $H_{u}$ i.e., in the notation of section 3.2, which
$H^{k}(n)$ we have. We already know that $k=1. $ We could
determine $n$ by carefully examining the group actions but, more
simply, we can observe that, since the fixed point set over $u$ in
$X_{u}$ is $\R_{u},$ and has codimension $4$, the quotient
$X_{u}/U(1)$ is a manifold around this point and this requires
$n=\pm 1.$ But at $A$ we know that we get a manifold with
boundary. This determines the sign of $n,$ giving $n=-1.$ Thus we
can make \toemp\ more precise by writing \eqn\murky{
X_{u}=H_{u}^{1}\left( -1\right) \times \R_{u}  } We are therefore
in case II of section 3.2, and the quotient \eqn\omurky{
H_{u}^{1}(-1)/U(1)=\R_{+}^{3}(u)  } is a half-space, depending on
$u.$  Its boundary lies in $\R^{3}=X_{A}$ and is the orthogonal
complement $\R_{u}^{\perp }$ of $\R_{u}.$ Thus $
X_{u}/U(1)$ is an $\R_{+}^{4},$ depending on $u,$ which has the fixed $%
R^{3}=X_{A}$ as boundary. If we write the right side of \omurky\
as \eqn\tomurky{ \R_{+}^{3}(u)=\R^{2}(u)\oplus \R_{+}(u)  ,} then
the image of the zero-section of the bundle $H_u$ is the unit
interval in $\R_{+}(u)$.  So we have to identify $\R_{+}(u)$ with
the half line in $\R_{B}^{3}$ defined by $u\in \S^{2}.$ The first
factor $\R^{2}(u)$ in \tomurky\
 is
the boundary and we have already seen that this is $\R_{u}^{\perp
}\subset X_{A}.$

To keep track of all this geometry we now introduce \eqn\jomurky{
\R^{6}=\C^{3}=\R^{3}\oplus i\R^{3}=\R_{B}^{3}\oplus X_{A},} using
the identification \huvv. With this notation, \eqn\noddo{
H^{1}(-1)_{u}/U(1)=\R u\oplus i(\R u)^{\perp }} and \eqn\toddo{
X_{u}/U(1)=\R u\oplus i\R^{3}  .} Rotating by $SU(2)$ acting
diagonally  on $\R^{6}=\C^{3},$ \toddo\ implies that \eqn\xoddo{
X/U(1)=\C^{3} , } as required. Moreover this is compatible with
the $SO(3)$ symmetry.

The fixed point set $L$ in $X,$ consists as we have seen of two
components, say $L_{0},L_{1}.$ Under the identification \xoddo,
we see that \eqn\sonon{\eqalign{ L_{0} &=i \R^{3}  \cr L_{1} &=
{\rm{union ~of~ all~ }}u+i\R u,~{\rm { for }}~u\in \S^{2}\subset
\R^{3}. \cr} } The component $L_{1}$ is the same one we met in
case I and, as pointed out there, it is Lagrangian. The component
$L_{0}$ is trivially Lagrangian.

\subsec{Case {\rm III}: $Y=\S^{3}\times \S^{3}$}

For the third case, we take \eqn\burma{
Y=SU(2)^{3}/SU(2)=\S^{3}\times \S^{3} } with $SU(2)^{3}$ and
$\Sigma _{3}$ symmetry. We take our $U(1)$ subgroup of
$SU(2)^{3}$ to be the diagonal subgroup (acting on the left).
Identifying $Y$ with (say) the product of the last two factors of
$SU(2)$, this action becomes conjugation on each factor, with the
fixed point set $F$ being a torus \eqn\turma{ F=\S^{1}\times
\S^{1} .} Since the action in the normal direction is by complex
scalars on $\C^{2},$ the quotient $Y/U(1)$ is again a (compact)
five-manifold.

Our seven-manifold $X$ is, as explained in section 2, an $\R^{4}$
bundle over $\S^{3}$ which is topologically a product. If we
introduce the quaternions $\H$, with standard generators $i,j,k$
and with $SU(2)$ being the unit quaternions then
\eqn\ruma{\eqalign{ Y =&(x,y)~{ \rm with }~x,y\in H, \; \left|
x\right| =\left| y\right| =1 \cr X =&(x,y)~{\rm with }~x,y\in H,
\; \left| x\right| =1 \cr} } and $U(1)\subset SU(2)$ is given by
the embedding $\C\rightarrow \H,$ acting on $x$ and $y$ by
conjugation.

The action of $U(1)$ on $X,Y$ is by conjugation and the
fixed-points are then the pairs $(x,y)$ with $x,y\in \C.$ The
fixed-point set $L$ in $X$ is just \eqn\tuma{ L=\S^{1}\times
\R^{2},} coming from points with $x,y\in \C$ and $|x|=1$.

The quotient $\S^3/U(1)$ is a disc\foot{Describing $\S^3$ by an
equation $x_0^2+x_1^2+x_2^2+x_3^2=1$, $U(1)$ acts by rotations of
the $x_2$-$x_3$ plane; the quotient by $U(1)$ can be taken by
setting $x_3=0$ and $x_2>0$.} \eqn\guma{ \S^{3}/U(1)=D^{2}\subset
\C,  } which we can think of as the unit disc in the complex
$x$-plane with the unit circle $\S^{1}$ coming from the fixed
points. As observed in section 2, this is the main difference
between cases I, II on one hand and case III on the other. Here
we get the two-disc rather than the three-disc.

As before, we pick a point $u\in \S^{1},$ on the boundary of the
disc, and consider the corresponding $U(1)$-invariant two-sphere
$\S_{u}^{2}$ through $ u $.\foot{In the notation of the last
footnote, $u$ is of the form $(x_0,x_1,0,0)$, and $\S^2_u$
consists of points whose first two coordinates are a multiple of
$u$.} This projects onto the unit interval $[-u,u]$ in \guma. Let
$X_{u}$ be restriction of $X$ to $S_{u}^{2}.$ Since $X$ is a
product bundle, this is just \eqn\vuma{ X_{u}=\S_{u}^{2}\times
\R^{4}} and $U(1)$ acts on $\R^{4}$ by conjugation of
quaternions, so that it decomposes into \eqn\quma{
\R^{4}=\C(0)\oplus \C(2)  } where the integer denotes the weight
of the representation of $U(1).$ Thus, in the notation of section
3.2 \eqn\wuma{ X_{u}=H_{u}^{0}(2)\times \R^{2} . } Note, that
unlike cases I, II, the trivial factor $\R^{2}$ here does not
depend on $u. $

Now in section (3.4) we showed that \eqn\oruma{
H_{u}^{0}(2)/U(1)=\R_{u}^{3}.} where here there is a dependence
on $u.$ From \wuma, it follows that \eqn\pruma{
X_{u}/U(1)=\R_{u}^{3}\times \R^{2} } The line in $\R_{u}^{3}$
which contains the image of the zero-section can naturally be
identified with the line $\R u\subset \C$ in $D^2=\S^3/U(1)$, and
the orthogonal $\R^{2}$ can be identified with the $\C(2)$ factor
in \quma. Thus \eqn\numma{ X_{u}/U(1)=\R^{4}\oplus \R u\subset
\R^{4}\oplus \R^{2}=\R^{6} . } Finally, rotating $u$ in the
complex plane leads to the desired identification \eqn\nicid{
X/U(1)=\R^{6}  .} The fixed-point set in $X$ becomes the subspace
\eqn\icid{ L=\C(0)\times \S^{1}  ,} when $\C(0)$ is the first
factor in \quma.

So far we have ignored the extra symmetries of the situation, but
in fact $U(1)^{3}$, modulo the diagonal, acts on $X/U(1)$ and
hence by \nicid\ on $\R^{6}.$ From \quma\ and \numma, we already
have a decomposition of $\R^{6}$ into three copies of $\R^{2}.$
For the right orientations, identifying each $\R^{2}$ with a copy
of $\C,$ a calculation shows that \nicid\ is compatible with the
$U(1)^{3}$ action provided $(\lambda_1 ,\lambda _{2},\lambda
_{3})\in U(1)^{3}$ acts on the three factors by \eqn\humma{
\lambda _{2}\lambda _{3}^{-1},\;\lambda _{3}\lambda
_{2}^{-1},\;\lambda_1 \lambda _{2}^{-1}  .}

\subsec{ The Lebrun Manifolds}

   Cases I and II have much in common.  In each case, the seven-manifold
is the $\R^3$-bundle of anti-self-dual two-forms over a compact
four-manifold $M$. Moreover, there is a $U(1)$-action for which
the quotient is a three-disc,
                    $M/U(1)  =  D^3,$
with the boundary
 $\S^2$ of $D^3$  arising from a component of the
fixed-point-set  of the $U(1)$-action.   The difference between
cases I and II is just that this $\S^2$ is the whole
fixed-point-set in case I, while in case II there is also an
isolated fixed-point which gives a distinguished point interior
to $D^3$. Finally the sphere bundle $Y$ of $X$ is a complex
manifold, being the  twistor  space of a self-dual conformal
structure on $M$.

There is actually a whole sequence of four-manifolds $M(n)$ which
share all these properties, so that
\eqn\nutto{\eqalign{                 M(0) &  =  \S^4  ~~  {\rm
(case~ I)}\cr
                  M(1) & =  \CP^2 ~~{\rm  (case ~II)}\cr
                  M(n) &  =  \CP^2 \# \CP^2 \# \dots \# \CP^2 ~~(n ~{\rm
times}) ,\cr}} where $\# $ denotes the operation of "connected
sum". This means that we excise small balls and attach the
remaining manifolds by small tubes, in the same way as (in
dimension 2) a surface of genus $g$ is a connected sum of tori.

The manifolds $M(n)$ were studied by  Lebrun
\ref\lebrun{C.Lebrun,  ``Explicit Self-Dual Metrics on $\CP^2
\#\dots \#\CP^2$,'' J. Differential Geometry {\bf 34} (1991) 223.}
 and we shall refer to them
as Lebrun  manifolds.  We want to explore the possibility of
deriving from them $M$-theory duals to more general brane
configurations in $\R^6$. We begin by reviewing their
construction and properties.

In  section 3.2 we recalled the Gibbons-Hawking ansatz for
constructing the ALE manifolds of type $A$.  We gave in eqn.
\inin\
the formula for $ A_1$. The  general case is precisely similar,
and it can be generalized further by introducing a ``mass
parameter,'' taking the harmonic function  $U$ on $\R^3$ to be
 \eqn\utto{          U  =  c +\sum_{i=1}^n {1\over |\vec x-\vec x_i|} ,}
 where $c\geq 0$.  If $c = 0$, we get the ALE
manifolds, while if $c\not= 0$, we get the ALF manifolds, their
Taub-NUT counterparts, in which the four-manifold is locally
asymptotic to the product  $\R^3 \times\S^1$.  The parameter $c$
is inverse to the radius of the circle factor so that, as  $c$
tends to 0, the circle becomes a copy of $\R$.

All this can be generalized with $\R^3$ replaced by the hyperbolic
three-space $H^3$, of constant curvature $-1$, to give a complete
Riemannian four-manifold $M_n$.  We simply replace the factor
$1/|\vec x-\vec x_i|$ by the corresponding Green's function for
the hyperbolic metric.  For  small distances, the metric
approximates the flat metric, so the  behaviour near the points
$x_i$ is the same as for $\R^3$.  For $x\to\infty$,
 we get a metric
locally asymptotic to $H^3 \times \S^1$.  Topologically  the
four-manifolds $M_n$ are the same as in the Euclidean case, i.e.
they are the resolution of the $A_{n-1}$ singularity.

If, in particular, we take $c = 1$, then $M_0$ is conformally
flat and by adding a copy of a two-sphere to it at infinity, one
makes $\S^4$.\foot{ To see that $\S^4$ minus a two-sphere is
$H^3\times \S^1$, note that the conformal group $SO(5,1)$ of
$\S^4$ contains a subgroup $ SO(3,1)\times SO(2)$ where $SO(2)$
acts by rotation of two of the coordinates; throwing away the
fixed point set of $SO(2)$, which is a copy of $\S^2$, the rest
is a homogeneous space of $ SO(3,1)\times SO(2)$ which can be
identified as $H^3\times \S^1$.  Of course $\S^4$ is conformally
flat, so $H^3\times \S^1$ is also.} More generally, for $c=1$,
the  four-manifold $M_n$ looks at infinity just like $M_0$ (as
$U\to 1$ at infinity for all $n$), so it can be conformally
compactified at infinity by adding a copy of $\S^2$, to give a
compact manifold $M(n)$.
   Note that the
special value  $c= 1$ is linked to the fact that we took the
curvature $\kappa$
 of $H^3$ to be $-1$: in general we would take  $c^2 \kappa = -1$.

The Lebrun  manifolds $M(n)$ have, from their construction, the
following properties:

 {\it (A)}          $ M(n)$ is conformally
self-dual.

{\it (B)} $M(n)$ has a $U(1)$-action (acting by translations of
$\tau$ just as in eqn. \inin)
with a fixed $\S^2$ and $n$ isolated fixed points
$x_1,\dots,x_n$, so that $M(n)/U(1) = D^3$.

In  {\it (B)}, the fixed points $x_i$ give $n$ distinguished
interior points of $D^3$.

We now want to describe a further property of the Lebrun
manifolds $M(n)$. We pick a point, which will be denoted as
$\infty$, in  the fixed $\S^2$.  We denote $M(n)$ with $\infty$
deleted as $M_\infty(n)$. For $n = 0$ we have the natural
identifications, compatible with $U(1)$, \eqn\rugged{\eqalign{
      M_0 &  = \S^4 - \S^2 = H^3 \times \S^1 = \C \times \C^* \cr
      M_\infty(0)& = \S^4 - \infty = \C^2  \cr}}
These involve the identification  of $H^3$ with the upper-half
3-space $\C \times \R_+$ of pairs $(u,| v|)$ with $u,v$ complex
numbers and $v$ non-zero.

Lebrun shows that $M_\infty(n)$ has the following  further
property

{\it (C)}   There is a map $\pi : M_\infty(n)\rightarrow
M_\infty(0) = \C^2$
 compatible with the
$U(1)$-action, which identifies $M_n(\infty)$ with the blow-up of
$\C^2$ at the $n$ points $  \pi(x_i)$, which are of the form
$(u_i,0)\in \C^2$.

$\pi$ takes the fixed surface in $M_\infty(n)$ into the line $v =
0$
 in
$\C^2$.  The orientation of $M_\infty(n)$ that we will use is the
opposite of the usual orientation of $\C^2$.

For $n= 1$, {\it (C)} gives the well-known fact that
$\overline{\CP}^2$ (that is, $\CP^2$ with the opposite
orientation) is the one-point compactification of the blow-up of
a point in $\C^2$.  The change of orientation reverses the sign
of the self-intersection of the ``exceptional line'' (inverse
image of the blown-up point), turning it from $-1$ into $+1$, and
so agreeing with the self-intersection of a line in  $\CP^2$.
More generally, {\it (C)} implies the assertion made in \nutto\
about the topology of $M(n)$.

It is perhaps worth pointing out that the metric on $M_n$, given
by using the function $U$ of eqn. \utto, depends on the $n$
points $x_1,\dots,x_n$ of $H^3$. But the complex structure of
$M_n$, as an open set of $M_\infty(n)$, given by {\it (C)},
depends only on the points $\pi(x_i)$. (Intrinsically, $\pi(x_i)$
is the other end of the infinite geodesic from $\infty$ to $x_i$
in $H^3$.) Moreover, the complex structure of $M_n$ varies with
the choice of the point $\infty$ in the fixed $\S^2$.   This is
very similar to the story of the complex structures on the
hyperk\"ahler manifolds of type $A_n$. Although the metric on
$M_n$ is not hyperk\"ahler, Lebrun  shows that it is k\"ahler
with zero scalar curvature (this guarantees the conformal
self-duality for the other orientation).

Having summarized the properties of the Lebrun manifolds $M(n)$,
we are now ready to move on to our bundles $X(n)$ and $Y(n)$ over
them.  As before, $X(n)$ is the bundle of anti-self-dual
two-forms  and $Y(n)$ the associated sphere bundle.  Note that
the notion of anti-self-duality makes sense for a conformal
structure, not just for a metric, and so it makes sense for the
conformal compactification $M(n)$ of $M_n$ (actually Lebrun does
give an explicit metric in the conformal class). $Y(n)$ is the
associated sphere bundle and, being the twistor space  of $M(n)$,
it has a natural complex structure (though  we shall not use this
fact).

The action of $U(1)$ on $M(n)$ then  extends naturally to actions
on $X(n)$ and $Y(n)$.  The behaviour at the fixed points in
$M(n)$ follows from that in the model example $n=1$, and gives
fixed manifolds  $F(n)$ inside $Y(n)$ and $L(n)$ inside $X(n)$ as
follows: \eqn\jugop{ F(n)  =  \S^2 \cup \S^2 \cup\dots\cup \S^2  }
\eqn\bugop{ L(n)  =  \R^3 \cup \R^3 \cup\dots\cup \R^3 \cup (\S^2
\times \R). } In \jugop, there are $n+2$ copies of $\S^2$, while
in \bugop\ there are $n$ copies of $\R^3$.  Since the $n$ copies
of $\R^3$ are disjoint, the corresponding $n$ copies of $\S^2$ in
\jugop\ are unlinked.  However, the last 2 copies of $\S^2$ are
linked, and each is linked to the first $n$ copies. This all
follows from the case of $n=1$. This means that the singular
seven-manifold of which $X(n)$ is a deformation is not the cone
on $Y(n)$ for $n\geq 2$.  Instead it has a singularity just like
that for $n=1$, with just one of the first $n$ unlinked $\S^2$
being shrunk to a point, together with the last two.  Thus the
local story does not change when we increase $n$.  If we take a
maximally degenerated Lebrun manifold, with the $\vec x_i$ all
equal, it appears that the $n$ disjoint $\R^3$'s become
coincident, so the branes will consist of three copies of $\R^3$
of multiplicity $(1,1,n)$, rather as in the situation we consider
in section 3.7 below.

We now want to show that \eqn\jurry{\eqalign{
           Y(n)/U(1)  & =  \S^5 \cr
           X(n)/U(1) & =  \R^6    , \cr}}
extending the results for $n=0$ (Case I) and $n=1$ (Case II).
Since for $n
> 1$ we no longer have $SO(3)$-symmetry, we cannot, as before, reduce the problem to one in four dimensions.  Instead we shall use the connected sum decomposition \nutto,
  together with the special cases $n=0,1$ already established.

Before embarking on  the details, we should explain  our
strategy.  As an analogue, recall that the connected sum of two
$n$-spheres is still an $n$-sphere.  This can be seen most
explicitly if we bisect along an equator and then  remove the
southern hemisphere of one and the northern hemisphere of the
other.  Gluing the remaining pieces we clearly get another
$n$-sphere.  There is a similar story if we replace the $n$-sphere
by $\R^n$, bisecting it into two half-spaces.  We shall show that
the connected sum operation on the four-manifolds $M(n)$ leads
essentially to this bisection picture on $\R^6$.

Since we want to keep track of the $U(1)$-action, we want the
more precise description of \nutto\ given by {\it (C)}.  For our
present purposes, since we are not interested in  the complex
structure, we can rephrase this as follows.  Starting with $\S^4$
and its $U(1)$-action with $\S^2$ as fixed-point set, we choose
$n$ distinct points $x_1,\dots,x_n$ on $\S^2$. Blowing up these
we get $M(n)$.   Topologically, the blowup
 means that we excise
small $U(1)$-invariant balls $V_i$ around each $x_i$, whose
boundaries are three-spheres, and then replace them by a
two-plane bundle $W_i$ over $\S^2$  using the standard fibration
$\S^3\rightarrow \S^2$.  Note that such a $W$ is also the
complement of a ball in $\CP^2$,  so that the blow-up operation
is indeed the same as forming the connected sum with $\CP^2$.

If we  blow up just one point, this  gives the manifold $M(1) =
\CP^2$, so that this will provide the local model around each
point.   Each time we modify $\S^4$ near a point $x$ we can
describe the corresponding modification of $X(0)$.  We remove $V
\times \R^3$ (the part of $X(0)$ over $V$) and replace it by the
$\R^3$ bundle over $W$, given by the local model.

Next we need to examine how these modifications behave for the
quotients by $U(1)$.   We have already shown that $X(0)/U(1) =
\R^6$, and our explicit description in section 3.3 shows that the
fibre $\R^3$ over $x_i$
 goes into a
half-plane $H=\R^2_+$.  The neighbourhood $V \times \R^3$
 will therefore go into a
neighbourhood $U$ of $H$ in $\R^6$.  It is not hard to see that
there is a diffeomorphism of the pair $(\R^6,U)$ into
$(\R^6,\R^6_+)$, i.e we have ``bisected'' $\R^6$.  The easiest
way to verify this is to start with a bisection of $\S^4$, i.e.
to choose the original neighbourhood $V$ to be a hemisphere in
$\S^4$.  Then by symmetry $U =\R^6_+$ (see the last part of Case
II in section 3.2).  If we now take a gradient flow along the
meridians shrinking towards the south pole x, then the
hemispheres get shrunk to arbitrarily small size.  This
diffeomorphism of $\S^4$, which is compatible with the
$U(1)$-action, induces the required diffeomorphism on $\R^6$.  We
shall refer to $U$ as a standard neighbourhood of $\R^2_+$ in
$\R^6$.

Consider now the case $n=1$.  As we saw in section 3.4,
$X(1)/U(1) = \R^6$. Moreover the explicit nature of this
identification again shows that the $\R^3$-fibre over a point $x$
of the fixed $\S^2$ in $X(1)$ is  a half-plane $H$. The local
model near here is provided by the case $n=0$ and so a
neighbourhood $V \times \R^3$  in $X(1)$ will go into a standard
neighbourhood $U$ of $H$.  But we have already  seen that such a
$U$ gives a bisection of $\R^6$. It follows that the
$\R^3$-bundle over the complement, i.e. over $W$ in $X(1) =
\CP^2$, goes into the other half of this bisection.  This means
that the decomposition of $\CP^2$ into two unequal parts, one
being a ball and the other being $W$, induces the trivial
bisection of the quotient $\R^6$.

This shows, inductively,  that the process of ``adding''
$\CP^2$'s (by connected sum operations), induces on the quotients
$M(n)/U(1)$, just the trivial operation  on $\R^6$ of bisection
and then gluing two halves together as outlined before.  This
establishes that these quotients are always $\R^6$, as claimed.

The corresponding statement for $Y(n)$ follows at the same time.

Our analysis of the topology of $ X(n)$ and its quotients by
$U(1)$ does show that these manifolds generalize Cases I and II
of the previous sections (for $n=0,1$).  It naturally  raises the
question as to whether, for all $n$, $X(n)$ has a complete metric
with $G_2$ holonomy, with the corresponding fixed-point-set
$L(n)$, given by \bugop, being a special Lagrangian in $\R^6$. It
is in fact easy to find such a special Lagrangian: we just take
the case for $n=1$ given by Joyce and add $n-1$ parallel copies
of the $\R^3$ component.

All of the explicit constructions of explicit $G_2$ holonomy
metrics have used the presence of large symmetry groups, and this
will not work for $n > 1$. What is needed is some kind of gluing
technique, perhaps on the lines of  \ref\kovalev{A. Kovalev,
``Twisted Connected Sums And Special Riemannian Holonomy,'' math.
DG/0012189.}.

\subsec{Another Generalization}

Finally, we want to point out some simple generalizations of our
discussion that may be of physical interest.

Let $X$ be any of the seven-manifolds with $U(1)$ action such
that $X/U(1)=\R^6$.  Let $\Z_n$ be the subgroup of $U(1)$
consisting of the points of order dividing  $n$, and let
$X_n=X/\Z_n$. Obviously, $U(1)$ still acts on $X_n$, and
$X_n/U(1)=X/U(1)=\R^6$.

$X_n$ is an orbifold rather than a manifold.  We want to focus on
the case that the fixed point set of $\Z_n$ is the same as the
fixed point set of $U(1)$.  This will always be so for generic
$n$.  In our examples, it is true for all $n\geq 2$.  This being
so, $X_n$ is an orbifold with a locus of $A_{n-1}$ singularities
that is precisely the fixed point set of the $U(1)$ action on
$X_n$.

The fixed points in the $U(1)$ action on $X_n$ are the same as in
the action on $X$.  So the reduction to a Type IIA model via the
quotient $X_n/U(1)$ leads to branes that occupy the same set in
$\R^6$ that we get from $X/U(1)$.  The difference is that since
the fixed points in $X_n$ are $A_{n-1}$ singularities, the branes
in the $X_n$ model  have multiplicity $n$.

All this holds whether $X$ is a smooth manifold of $G_2$ holonomy
or is conical.  In drawing conclusions, it is helpful to start
with the conical case:

\def\n{{\bf n}}

(I) If $X$ is the cone on $\CP^3$, the brane configuration that
is a Type IIA dual of $X_n$ consists of two copies of $\R^3$
meeting at the origin, each with multiplicity $n$.  The
associated low energy theory is a $U(n)\times U(n)$ theory with
chiral multiplets transforming as $(\n,\overline \n)$.  Deforming
$X$ to a smooth manifold of $G_2$ holonomy and thereby deforming
$\R^3\cup \R^3$ to $\R\times {\S}^2$, the $U(n)\times U(n)$ is
broken to a diagonal $U(n)$.

(II) The case that $X$ is a cone on $SU(3)/U(1)^2$ is similar.
The Type IIA dual of $X_n$ is a brane configuration consisting of
three copies of $\R^3$ meeting at a point, all with multiplicity
$n$. The low energy theory has $U(n)^3$ symmetry and chiral
multiplets transforming as $(\n,\overline \n,{\bf 1})\oplus ({\bf
1},\n,\overline \n) \oplus (\overline \n,{\bf 1},\n)$.  Calling
the three fields $\Phi_1$, $\Phi_2,$ and $\Phi_3$, there is a
superpotential $\Tr\,\Phi_1\Phi_2\Phi_3$, just as in the $n=1$
case, and there are various possibilities of symmetry breaking.

(III) For $X$ a cone on $\S^3\times \S^3$, the dual of $X_n$ is a
brane of multiplicity $n$ that is a cone on $\S^1\times \S^1$. We
can draw no immediate conclusions, as we have no knowledge of the
dynamical behavior of such a brane.

Notice in cases (I) and (II) that if a singularity like that of
$X_n$ would appear in a manifold of $G_2$ holonomy, we would get
a gauge theory with chiral fermions.  This might be of physical
interest. For example, in case (II) with $n=3$, the $U(3)^3 $
gauge group with the indicated representation for the chiral
superfields is very closely related to the standard model of
particle physics with one generation of quarks and leptons.

One might wonder if examples I and II can be further generalized.
For example, could we extend case I so that the branes will be a
pair of $\R^3$'s of respective multiplicities $(m,n)$ for
arbitrary positive integers $m,n$?  Candidate manifolds can be
suggested as follows, though we do not know if they admit metrics
of $G_2$ holonomy.

To build $\CP^3$, we start with $\S^7$, parameterized by four
complex variables $z_1,\dots,z_4$ with $\sum_{i=1}^4|z_i|^2=1$.
Then we divide by a $U(1)$ group that acts by \eqn\nub{z_i\to
e^{i\theta}z_i,\,\,i=1,\dots,4.} The quotient is $Y=\CP^3$.  In
section 3.3, we divided by a second $U(1)$ which acts by
\eqn\tub{(z_1,z_2,z_3,z_4)\to
(e^{i\theta}z_1,e^{i\theta}z_2,z_3,z_4).} The quotient, as we
have seen, is $\S^5$. Now we will use an argument that we have
already used in section 3.2 in case III. If the plan is to divide
$\S^7$ by $U(1)\times U(1)$ to get $\S^5$, we can divide first by
an arbitrary $U(1)$ subgroup of $U(1)\times U(1)$, acting say by
\eqn\gub{(z_1,z_2,z_3,z_4)\to (e^{in\theta}z_1,e^{in\theta}z_2,
e^{im\theta}z_3,e^{im\theta}z_4).} The quotient is a weighted
projective space, $Y(n,m)={\bf WCP}^3_{n,n,m,m}$. Then we divide
by the ``second'' $U(1)$, and we will be left with
$Y(n,m)/U(1)=\S^5$, since we have just looked at the quotient
$Y/U(1)=\S^7/U(1)\times U(1)$ in a different way.

By the same argument, if $X(n,m)$ is a cone on $Y(n,m)$, then
$X(n,m)/U(1)=X/U(1)=\R^6$.  What branes appear in the Type IIA
model derived in this way from $X(n,m)$?  We may as well assume
that $n$ and $m$ are relatively prime (as a common factor can be
removed by rescaling $\theta$ in \gub).  The fixed points of the
``second'' $U(1)$ are the points with $z_1=z_2=0$ or $z_3=z_4=0.$
These consist of two copies of $\S^2$, so on passing to the cone
the branes fill two copies of $\R^3\subset \R^6$. The two
components of the fixed point set are $A_{m-1}$ and $A_{n-1}$
singularities, respectively, so the branes have multiplicities
$m$ and $n$.

More generally, suppose we want multiplicities $(m,n)=r(a,b)$
where $a$ and $b$ are relatively prime.  We do this by combining
the two constructions explained above.  We start with $X(a,b)$,
which gives
 a model with multiplicities $(a,b)$, and then we divide by the $\Z_r$
subgroup of the ``second'' $U(1)$.  The last step multiplies the
multiplicities by $r$, so the quotient $X_r(a,b)=X(a,b)/\Z_r$
leads to a model in which the brane multiplicities are $(m,n)$.

An analogous construction can be carried out in case II.

\newsec{Quantum Parameter Space Of Cone On $\S^3\times \S^3$}

\subsec{Nature Of The Problem}

In this section, we return to  the problem
of $M$-theory on a $G_2$ manifold that is asymptotic to a cone on
$Y=\S^3\times \S^3$.  We have seen in section 2.5 that
there are three $G_2$ manifolds
$X_i$ all asymptotic to the same cone.  It has been proposed
\amv\ that there is a smooth curve $\N$ of theories that interpolates
between different classical limits corresponding to the $X_i$.  We will
offer further support for this, but first let us explain
why it might appear problematical.

First we recall how we showed, in section 2.3, in a superficially similar
problem, involving a cone on $SU(3)/U(1)^2$,
that there were three distinct branches $\M_i$ of the quantum
moduli space corresponding to three classical spacetimes.  We showed
that the quantum problem had a symmetry group $K=U(1)\times U(1)$,
determined by the symmetries of the $C$-field at infinity,
and that on the three different spacetimes, there are three different
unbroken $U(1)$ subgroups.  The classification of $U(1)$ subgroups of
$U(1)\times U(1)$ is discrete, and an observer at infinity can
determine which branch the vacuum  is in by determining which $U(1)$
is unbroken.

In the case of a cone on $\S^3\times\S^3$, there are no global symmetries
associated with the $C$-field, but we can try to make a somewhat
similar (though ultimately fallacious)
argument using the periods of the $C$-field.
An observer at infinity measures a flat $C$-field, as otherwise
the energy would be infinite.  A flat $C$-field at infinity takes
values in $E=H^3(Y;U(1))=U(1)\times U(1)$.  But not all possibilities
are realized.  For unbroken supersymmetry, the $C$-field on $X_i$ must be flat
at the classical level,
so it must take values in $H^3(X_i;U(1))$.  There is a natural
map from $H^3(X_i;U(1))$ to $H^3(Y;U(1))$ which amounts to restricting
to $Y$ a flat $C$-field on $X_i$.  However, this map is not an isomorphism;
not all flat $C$-fields on $Y$ extend over $X_i$ as flat $C$-fields.
In fact, $H^3(X_i;U(1))=U(1)$ is mapped to a rank one subgroup $E_i$ of
$H^3(Y;U(1))$.

For different $i$, the $E_i$ are different.
In fact, they are permuted by triality, just like the generators
$D_i$ of $H_3(Y;\Z)$
that were found in section 2.5.  (In fact, the $D_i$ are Poincar\'e
dual to the $E_i$.)

Thus, an observer at infinity can measure the $C$-field as an element
of $H^3(Y;U(1))$ and -- classically -- will find it to belong to
one of the three distinguished $U(1)$ subgroups $E_i$. By finding which
subgroup the $C$-field at infinity belongs to, the classical observer can
thereby generically determine which spacetime  $X_1$, $X_2$, or $X_3$
is present in the interior.  If this procedure is valid quantum
mechanically, then
the moduli space of theories has distinct branches $\N_1$, $\N_2$,
and $\N_3$.  Of course, even classically there is an exceptional possibility
that the $C$-field belongs to more than one of the $U(1)$'s (in which
case it must vanish); in this
case, the measurement of the $C$-field does not determine the manifold
in the interior.  This might seem to be a hint that the $\N_i$ intersect
at some exceptional points, rather as we found in the superficially
similar model of section 2.3.

In this discussion, we have put the emphasis on measurements at infinity,
where semiclassical concepts apply, since there is no useful way to describe
measurements in the interior, where the quantum gravity effects may be big.

Let us contrast this with a more familiar situation in Type II
superstring theory:  the
``flop'' between the two small resolutions $Z_1$ and $Z_2$
of the conifold singularity
of a Calabi-Yau threefold in Type II superstring theory.  In this case, the Neveu-Schwarz two-form
field $B$ plays the role analogous to $C$ in $M$-theory.
The $B$-field
 periods takes values in $H^2(Z_1;U(1))$ or $H^2(Z_2;U(1))$.  Both
of these groups are isomorphic to $U(1)$, and in this case
the two $U(1)$'s are canonically the same.  That is because
both $Z_1$ and $Z_2$ are asymptotic to a cone on a five-manifold $B$
that is topologically $\S^2\times \S^3$.  The second Betti number
of $B$ is one, equal to the second Betti number of the $Z_i$, and
$H^2(B;U(1))=U(1)$. The restriction maps from $H^2(Z_i;U(1))$ to
$H^2(B;U(1))$ are isomorphisms, so in fact the three groups
$H^2(B;U(1))$, $H^2(Z_1;U(1))$, and $H^2(Z_2;U(1))$ are all naturally
isomorphic.  So by a measurement of the $B$-field period at infinity,
one cannot distinguish the manifolds $Z_1$ and $Z_2$.

Since supersymmetry
relates the Kahler moduli to the $B$-field periods, this leads to
 the fact that the Kahler moduli of $Z_1$ and $Z_2$ fit together in
a natural way at the classical level.  (Moreover, all this
carries over to other small resolutions of Calabi-Yau threefolds,
even when the second Betti number is greater than one.)  In fact,
each $Z_i$ contains an exceptional curve that is a two-sphere
$\S^2_i$; one can naturally think of $\S^2_2$ as $\S^2_1$
continued to negative area. (The last assertion is related to the
signs in the isomorphisms mentioned in the last paragraph.)  By
contrast, in $M$-theory on a manifold of $G_2$ holonomy, the
metric moduli are related by supersymmetry to the $C$-field
periods.  The fact that the classical $C$-field periods on the
$X_i$ take values in different groups $E_i$ also means that there
is no way, classically, to match up the metric moduli of the
three manifolds $X_i$.

We can be more explicit about this.
The metric modulus of $X_i$ is the volume $V_i$ of the
three-sphere
$Q_i\cong\S^3_i$ at the ``center'' of $X_i$.  Each $V_i$, classically, takes
values in the set $[0,\infty\}$ and so runs over a ray, or half-line.
In a copy of $\R^2$ that contains the lattice $\Lambda$, these rays
(being permuted by triality)
are at $2\pi/3$ angles to one another.  They
 do not join smoothly.

How can we hope nevertheless to find a single smooth curve $\N$
that interpolates between the $X_i$?  We must  find a quantum
correction to the claim that the $C$-field period measured at infinity
on $X_i$ takes values in the subgroup $E_i$ of $H^3(Y;U(1))$.  It must
be that the $C$-field period takes values that are very close to $E_i$
if the volume $V_i$ is large, but not close when $V_i$ becomes small.
Then, one might continuously interpolate from $X_i$ to $X_j$, with
the period taking values in $E_i$ in one limit, and in $E_j$ in the other.

Let us see a little more concretely what is involved in getting such a
quantum correction.  When $Y=\S^3\times \S^3$ is realized as the boundary
of $X_i$, the three-sphere $D_i$ defined in
section 2.5 is ``filled in'' -- it lies at infinity in the $\R^4$ factor
of $X_i=\R^4\times \S^3$.  So, for a flat $C$-field,
with $G=dC$ vanishing, we have
\eqn\iiko{\int_{D_i}C =\int_{\R^4}G=0.}
Classically, we impose $G=0$ to achieve supersymmetry.  Quantum mechanically,
we consider fluctuations around the classical supersymmetric state.
If quantum corrections modify the statement $\int_{D_i}C=0$,
this would correct the statement that the $C$-field periods lie in $E_i$,
and perhaps
enable a smooth interpolation between the different classical manifolds
$X_i$.  As we will show in section 4.4, perturbative quantum corrections
do not modify the statement that $\int_{D_i}C=0$, but membrane instanton
corrections do modify this statement.

In section 4.2, we will interpret the $C$-field periods as the arguments
of holomorphic functions on $\N$.
By exploiting the existence of those functions, we will, in section 4.3, making
a reasonable assumption about a sense in which the $X_i$ are the
only classical limits of this problem, argue  that
$\N$ {\it must} have just one branch connnecting the $X_i$
and give a precise description of $\N$.
In section 4.4, we analyze the membrane instanton effects and show
that they give the requisite corrections to $\N$.
Details of the solution found in section 4.3 will be compared
in section 5 to topological subtleties concerning the $C$-field.

\subsec{Holomorphic Observables}

The $C$-field periods on $Y$ must be related by supersymmetry to some
other observables that can be measured by an observer at infinity.
Supersymmetry relates fluctuations in the $C$-field to fluctuations
in the metric, so these other observables must involve the metric.

We are here considering $C$-fields that are flat
near infinity, so that if $D_i$ is any of the cycles at infinity
discussed at the end of section 2.5, the periods
\eqn\nindo{\int_{D_i}C}
are independent of the radial coordinate $r$.  Since the radius
of $D_i$ is proportional to $r$, this means that the components of
$C$ are of order $1/r^3$.  So supersymmetry relates the $C$-field
to a metric perturbation that is of relative order $1/r^3$, compared
to the conical metric.

Moreover, a flat  $C$-field preserves supersymmetry, so a
perturbation of $C$ that preserves the flatness is related to a
perturbation of the metric that preserves the condition for $G_2$
holonomy. To find the metric perturbations that have this
property, let us examine more closely the behavior of the
 metric \burigo\ near $r=\infty$.  We introduce a new radial
coordinate $y$ such that $dy^2=dr^2/(1-(r_0/r)^3)$.  To the
accuracy that we will need, it suffices to take
\eqn\onon{y=r-{r_0^3\over 4r^2}+O(1/r^5).} The metric is then
\eqn\turigo{ds^2=  dy^2+{y^2\over 36}\Bigl(da^2 +db^2+dc^2 -
{r_0^3\over 2y^3} \left(f_1\,da^2+f_2\,db^2+f_3\,dc^2\right)
+O(r_0^6/y^6)\Bigr)} with $(f_1,f_2,f_3)=(1,-2,1)$. If we set
$r_0=0$, we get the conical metric with the full $\Sigma_3$
symmetry.  This is valid near $y=\infty$ even if $r_0\not= 0$.
Expanding in powers of $r_0/y$, the first correction to the
conical metric is proportional to $(r_0/y)^3$ and is given
explicitly in \turigo.

Obviously, if we make a cyclic permutation of $a,b,c$, this will cyclically
permute the $f_i$.  So a metric of the form in \turigo\ with
$(f_1,f_2,f_3)$ equal to
$(1,1,-2)$ or $(-2,1,1)$ also has $G_2$ holonomy, to the given order
in $r_0/y$.  Moreover, since the term
of lowest order in $r_0/y$ obeys a {\it linear equation}, namely
the linearization of the Einstein equation around the cone metric
(nonlinearities determine the terms of higher order in $r_0/y$),
we can take linear combinations of these solutions if we are only
interested in the part of the metric of order $(r_0/y)^3$.  Thus, the metric
has $G_2$ holonomy to this order in $r_0/y$  if $(f_1,f_2,f_3)$
are taken to be any linear combination of $(1,-2,1)$ and its cyclic
permutations.   Hence, $G_2$ holonomy is respected to this order if the
coefficients $f_i$ obey the one relation
\eqn\hoboc{f_1+f_2+f_3=0.}

A flat $C$-field at infinity has periods
\eqn\noboc{\alpha_i=\int_{D_i}C~~{\rm mod}~2\pi.}
Actually, there is a subtlety in the definition of the $\alpha_i$,
because of a global anomaly in the membrane effective action;
we postpone a discussion of this to section 5.
Since $D_1+D_2+D_3=0$ in homology, as we explained in section 2.5,
one would guess that $\alpha_1+\alpha_2+
\alpha_3=0~{\rm mod}~2\pi$, but for reasons we explain in section 5,
the correct relation is
\eqn\toboc{\alpha_1+\alpha_2+\alpha_3=\pi ~{\rm mod}~2\pi.}

Our discussion in section 4.1 started with the fact that an observer
at infinity can measure the $\alpha_i$.  Now we can extend this to
a supersymmetric set of measurements: the observer at infinity can
also measure the $f_i$.

A classical physicist at infinity would
expect the $    f_i$ to be a positive multiple of
 $(1,1,-2)$ or a cyclic permutation
thereof, and would expect one of the $\alpha_i$ to vanish.  The
reason for this is that classically, while one can obey the
Einstein equations near infinity with any set of $f_i$ that sum
to zero, to obey the nonlinear Einstein equations in the interior
and get a smooth manifold $X_i$ of $G_2$ holonomy, the $f_i$ must
be of the form found in \onon, or a cyclic permutation thereof.
(See \cvglp\ for an analysis of the equations.)  Likewise, on any
$X_i$, the corresponding cycle $D_i$ is contractible, and so
$\alpha_i$ must vanish.

Let us consider the action of the permutation group $\Sigma_3$.
The $f_i$  and $\alpha_i$ are cyclically permuted under a cyclic
permutation of $a,b,c$. Under the flip $(a,b,c)\to
(c^{-1},b^{-1},a^{-1})$, we have $(f_1,f_2,f_3)\to
(f_3,f_2,f_1)$, but $(\alpha_1,\alpha_2,\alpha_3) \to
(-\alpha_3,-\alpha_2,-\alpha_1)$.  The reason for the sign is that
the flip reverses the orientation of $Y$, and this gives an extra
minus sign in the transformation of $C$.  So the holomorphic
combinations of the $f_i$ and $\alpha_j$ must be, for some
constant $k$ (we indicate later how $k$ could be computed),
$kf_1+i(\alpha_2-\alpha_3)$, $kf_2+i(\alpha_3-\alpha_1)$, and
$kf_3+i(\alpha_1-\alpha_2)$. These combinations are mapped to
themselves by $\Sigma_3$. Since the $\alpha_i$ are only defined
modulo $2\pi$, it is more convenient to work with combinations
such as \eqn\nonf{y_i=\exp(kf_i+i(\alpha_{i+1}-\alpha_{i-1})).}
The $y_i$, however, do not quite generate the ring of holomorphic
observables.  We can do better to define
\eqn\bonfly{\eta_i=\exp((2k/3)f_{i-1}+(k/3)f_{i}+i\alpha_i).} (So
$\eta_i=(y_{i-1}^2y_i)^{1/3}$.) Any holomorphic function of the
$f$'s and $\alpha$'s that is invariant under $2\pi$ shifts of the
$\alpha$'s can be expressed in terms of the $\eta$'s. The
$\eta_i$ are not independent; they obey
\eqn\rufgo{\eta_1\eta_2\eta_3=-1.} Each $\eta_i$ can be neither 0
nor $\infty$ without some of the $f_i$ diverging to $\pm
\infty$.  So at finite points in the moduli space, the $\eta_i$
 take values in $\C^*$ (the complex plane with the origin omitted),
and because of the constraint \rufgo, the $\eta_i$ taken together
define a point in
$W=\C^*\times \C^*$.

Let us verify that in the classical approximation, $\N$ is a holomorphic
curve in $W$.  On the branch of $\N$ corresponding to the manifold
$X_2$, the $f_i$ are in the ratio $(1,-2,1)$.  Moreover,
on this manifold, $\alpha_2=0$.  Altogether, $\eta_2=1$, and hence,
given \rufgo, $\eta_1\eta_3=-1$.  These conditions
define a holomorphic curve in $W$, as expected.  More generally, on the branch corresponding to the classical
manifold $X_i$ one has, in the classical approximation,
\eqn\minin{\eta_i=1, ~~\eta_{i-1}\eta_{i+1}=-1.}
In the classical
description, $\N$ consists of those three distinct branches.

In what limit is classical geometry valid?
To see the manifold $X_i$ semiclassically, its length scale $r_0$ must
be large.
In the limit $r_0\to\infty$, we have
 $f_{i\pm 1}\to +\infty$ and $f_i\sim -2f_{i\pm 1}$,
so
\eqn\jaguar{\eta_{i-1}\to \infty,~~ \eta_{i+1}\to 0.}
We will
argue later that, in fact, $\eta_{i-1}$ has a simple pole and $\eta_{i+1}$ a
simple zero as $r_0\to\infty$.

Let us record how the symmetries of the
problem act on the $\eta$'s.
A cyclic permutation in $\Sigma_3$ permutes the $\eta$'s in the obvious
way, while the flip $(a,b,c)\to (c^{-1},b^{-1},a^{-1})$ acts by
\eqn\tufgo{(\eta_1,\eta_2,\eta_3)\to (\eta_3^{-1},\eta_2^{-1},\eta_1^{-1}).}
There is one more symmetry to consider; as explained
at the beginning of section 2.1, a reflection in the first factor
of the spacetime $\R^4\times X$ exchanges chiral and antichiral fields;
it reverses the sign of the $C$-field while leaving fixed the metric
parameters $f_i$, so it acts antiholomorphically, by
\eqn\jufgo{\eta_i\to\bar\eta_i.}

The $f_i$ have an intuitive meaning as ``volume defects.'' Let us
recall that in section 2.5, we defined three-cycles $D_i\cong \S^3$
in $Y$.  $D_1$ was defined by the conditions \eqn\jukob{a=1=bc.}
The others
are obtained by cyclic permutation.  We can embed $D_1 $ in any of the
$X_i$ by setting, in addition, the radial coordinate $y$ to an arbitrary
constant.  If we do so, $D_1$ has a $y$-dependent volume that
behaves for $y\to\infty $  as
\eqn\hutty{{2\pi^2  y^3\over 27} + {\pi^2r_0^3f_1\over 36}+O(r_0^6/y^3).}  Thus, subtracting
the divergent multiple of $y^3$, there is a finite volume defect
$\pi^2r_0^3f_1/36$ at infinity.  Likewise,
all the $D_i$ have volume defects $\pi^2r_0^3f_i/36$.
The fact that for $r_0\to\infty$, up to a cyclic permutation,
the $f_i$ are a positive multiple of $(1,1,-2)$ means
that the volume defects are also a positive multiple of  $(1,1,-2)$,
and in particular precisely one of them is negative.  This fact
 has an intuitive meaning.
In the ``interior'' of $X_i$, precisely one of the $D$'s, namely $D_i$,
is ``filled in'' and has its volume go to zero.  This is the one
whose volume defect at infinity is negative.  There is no smooth manifold of
$G_2$ holonomy in which the volume defects are a {\it negative} multiple
of $(1,1,-2)$, roughly since there is no way to make a smooth manifold
by filling in two of the $D$'s.

\subsec{ Quantum Curve}

To understand supersymmetric dynamics via holomorphy, one must
understand the singularities.  In the present case, a singularity arises
when some of the $f_i$ diverge to $\pm \infty$, and hence some $\eta_i$
have zeroes or poles.  In our definition of $\N$, we will include points
where there are such zeroes or poles,
so our $\N$ is really a compactification of the moduli space of coupling
parameters.

For $f_i$ to diverge to $\pm\infty$ means that the volume defects
are diverging.  It is reasonable to expect that such behavior
can be understood classically.  In the present problem, we will
assume that the only way to get a zero or pole of the $\eta_i$ is
to take $r_0\to\infty$ on one of the classical manifolds $X_i$. Our assumption
could be wrong, for example, if there are additional smooth manifolds
of $G_2$ holonomy that are likewise asymptotic to a cone on $Y$.
In section 6, we will meet cases in which the enumeration of the possible
singularities contains some surprises.

Now we can explain why there must be corrections to the classical limit
described in \minin.  The curve described in \minin\ has an end with
$\eta_{i-1}\to\infty$, $\eta_{i+1}\to 0$, and a second end with
$\eta_{i-1}\to 0$, $\eta_{i+1}\to \infty$.  The first end has the $f_i$ diverging
as a positive multiple of $(1,1,-2)$ or a cyclic permutation thereof,
but at the second end, the $f_i$ diverge as a {\it negative}
multiple of $(1,1,-2)$.  This does not correspond to any known
classical limit of the theory, and according to our hypothesis,
there is nowhere in the moduli space that the $f_i$ diverge in this
way.  So \minin\ cannot be the exact answer.

On the other hand, a holomorphic function that has a pole also has
a zero, so a component of $\N$ that contains a point with
$\eta_{i-1}\to\infty$ must also contain a point with $\eta_{i-1}\to 0$.
By our hypothesis, this must come from a classical limit associated
with one of the $X_i$.  By the $\Sigma_3$
symmetry, if two ends are in the same component, the third must be also,
so given our assumptions, we have proved that $\N$ has a single component
that contains all three ends.  Therefore, it is possible to interpolate
between $X_1, $ $X_2, $ $X_3$ without a phase transition.

On any additional branches of $\N$, the $\eta_i$ have neither zeroes
nor poles and hence are simply constant.  This would correspond to a
hypothetical quantum $M$-theory vacuum that is asymptotic to a cone
on $Y$ but whose ``interior'' has no classical limit, perhaps because
it has a frozen singularity (analogous to frozen singularities that
will appear in sections 6.3 and 6.4).  If such a component exists,
new tools are needed to understand it.
We have no way to probe for the existence
of such vacua in $M$-theory, and will focus our attention on the known
branch of $\N$ that interpolates between the three classical manifolds $X_i$.
In fact, we will henceforth use the name $\N$ to refer just to this branch.

We know of three such points $P_i$, $i=1,2,3$ corresponding to the
points at which one observes the classical manifolds $X_i$ with large
$r_0$.  Near $P_i$, a local holomorphic parameter is expected to be
the expansion
parameter
for membrane instantons on $X_i$.  In fact, the three-sphere $Q_i$ at
the ``center''
of $X_i$ (defined by
setting $r=r_0$ in \burigo) is a supersymmetric cycle.  The amplitude
for a membrane instanton wrapped on this cycle is
\eqn\hindon{u=\exp\left(-TV(Q_i)+i\int_{Q_i}C\right).}
Here $T$ is the membrane tension, and $V(Q_i)$ is the volume of $Q_i$.
For an antimembrane instanton, the phase $\int_{Q_i}C$ in \hindon\ has the opposite
sign.  In any event, to define the sign of $\int_{Q_i}C$, one must be careful
with the orientation of $Q_i$.  According to \jaguar, we know already that
at $P_i$,  $\eta_{i-1}$ has a pole and $\eta_{i+1}$ has
 a zero.  It must then be that $\eta_{i-1}\sim u^{-s}$
and $\eta_{i+1}\sim u^t$ near $P_i$, with some $s,t>0$.  To
determine $s$ and $t$, we need only compare the phase of
$\eta_{i\pm 1}$, which is $\int_{D_{i\pm 1}}C$, to the phase
$\int_{Q_i}C$ of $u$.  We saw in section 2.5 that $Q_i$ is
homologous (depending on its orientation) to $\pm D_{i-1}$ and to
$\mp D_{i+1}$, so $s=t=1$.  Thus, $\eta_{i-1}$ has a simple pole,
and $\eta_{i+1}$ has a simple zero. It should also be clear that
the constant $k$ in the definition of the $\eta$'s could be
determined (in terms of $T$) by comparing the modulus of
$\eta_{i\pm 1}$ to that of $u$.

Now we have enough information to describe $\N$ precisely.  Each $\eta_i$
has a simple pole at $P_{i+1}$, a simple zero at $P_{i-1}$, and no other
zeroes or poles.  The existence of a holomorphic function with just one
zero and one pole implies that $\N$ is of genus zero.  We could pick
any $i$ and identify $\N$ as the complex $\eta_i$ plane (including the
point at infinity), but proceeding in this way would obscure the $\Sigma_3$
symmetry.  Instead, we pick an auxiliary parameter $t$ such that the
points $P_i$ are at $t^3=1$, with the goal of expressing everything
in terms of $t$.  The action of $\Sigma_3$ on $t$ can be determined
from the fact that it must permute the cube roots of 1.  Thus,
$\Sigma_3$ is generated by an element of order three
\eqn\ininp{t\to \omega t,    ~~ \omega=\exp(2\pi i/3),}
and an element of order two,
\eqn\hininp{t\to 1/t.}
The antiholomorphic symmetry \jufgo\ will
be
\eqn\jininp{t\to 1/\bar t.}

We identify $P_i$ with the points $t=\omega^{i+1}$.
$\eta_i$ should equal 1 at $P_i$ and should have a simple pole at
$P_{i+1}$ and a simple zero at $P_{i-1}$.  This gives
\eqn\jininko{\eta_i=-\omega{t-\omega^{i}\over t-\omega^{i-1}}.}
This is obviously invariant under the cyclic permutation of the $\eta_i$
together with $t\to\omega^{-1} t$.
It is invariant under elements of $\Sigma_3$ of order two since
\eqn\guyy{\eta_1(1/t)=\eta_3(t)^{-1},~\eta_2(1/t)=\eta_2(t)^{-1}.}
It is invariant under the antiholomorphic symmetry since
\eqn\nuyy{\eta_i(1/\bar t)=\bar\eta_i(t).}
Finally, \jininko\ implies the expected
relation
\eqn\pininko{ \eta_1\eta_2\eta_3=-1.}
Thus, we have a unique candidate for $\N$, and it has all the expected
properties.

\bigskip\noindent{\it Superpotential?}

Another holomorphic quantity of interest is the superpotential $W$
that arises from the sum over membrane instantons that are wrapped,
or multiply wrapped, on the supersymmetric cycle $Q_i\subset X_i$.
(For each $i=1,2,3$, this method of computing $W$ is valid near the
point $P_i\in\N$ that describes the manifold $X_i$ with large volume.)
If the conical singularity that we have been studying is embedded
in a compact manifold $\hat X$ of $G_2$ holonomy, then the moduli
of $\hat X$, including the volume of $Q_i$, are dynamical, and the
superpotential has a straightforward physical interpretation:
it determines which points in the parameter space actually do correspond
to supersymmetric vacua.

The physical interpretation of the superpotential is less compelling
in the case  considered in this paper of
an asymptotically conical $X$, since the variables on which the superpotential depends are nondynamical,
because of the infinite kinetic energy in their fluctuations, and
behave as coupling constants in an effective four-dimensional
theory rather than as dynamical fields.
  If there were more than one quantum vacuum for
each point in $\N$, then the differences between the values of
$W$ for the different vacua would give tensions of BPS domain walls;
but there is actually only one vacuum for each point in $\N$.

At any rate, let us see how far we can get toward determining $W$.
$W$ must have a simple zero at each of the $P_i$, since it
vanishes in the absence of membrane instantons, and in an
expansion in powers of instantons, it receives a one-instanton
contribution proportional to the instanton coupling parameter
$u$.  (The analysis in \harvey\ makes it clear that, since $Q$ is
an isolated and nondegenerate supersymmetric cycle, an instanton
wrapped once on $Q$ makes a nonvanishing contribution to the
superpotential.)  Since $W$ has at least three zeroes, it has at
least three poles.  If we assume that the number of poles is
precisely three, we can determine $W$ uniquely.  The positions of
the three poles must form an orbit of the group $\Sigma_3$, and
so these points must be permuted both by $t\to\omega t$ and by
$t\to 1/t$.  The only possibility (apart from $t=\omega^i$ where
we have placed the zeroes of $W$) is that the poles are at
$t=-\omega^i$, $i=1,2,3$. The superpotential is then
\eqn\hininp{W=ic{t^3-1\over t^3+1},} where the constant $c$ could
be determined from a one-instanton computation using the analysis
in \harvey.

Note that $W(\omega t)=W(t)$, but
\eqn\jininp{W(1/t)=-W(t),}
so that the transformations of order two in $\Sigma_3$ are $R$-symmetries
that reverse the sign of $W$.  This is expected for geometrical
reasons explained in
section 2.4.  Also, $W(1/\bar t)=\bar W(t)$ if $c$ is real,
so this candidate for $W$ is compatible with the real structure of the
problem.

Thus, we have a minimal candidate for $W$ that is fairly natural,
but we do not have enough information to be sure it is right;
one could consider another function with more zeroes and poles,
at the cost of introducing some unknown parameters.

One important fact is clearly that $W$ must have some poles.
What is their physical significance?  This is not at all clear.
A rough analogy showing the possible importance of the question is
with the ``flop'' transition of Type II conformal field theory.
In that problem, there is a complex moduli space $\tilde \N$, analogous
to $\N$ in the problem studied in the present paper, that interpolates
between the two possible small resolutions of the conifold singularity.
On $\tilde\N$ there is a natural holomorphic function $F$, the ``Yukawa coupling,''
that has a pole at a certain point of $\tilde \N$.  (In fact, $F$ only
has a straightforward interpretation as a four-dimensional Yukawa coupling
if the conifold singularity is embedded in a compact Calabi-Yau manifold;
in the noncompact case, the relevant modes are not square-integrable and
are nondynamical.
This is analogous to the status of the superpotential in our
present problem.)  At the pole,
the Type II conformal field theory becomes singular.
The singularity was mysterious for some time, but it was ultimately
understood \refs{\strom,\gms} that
at this point one can make a phase transition to a different
branch of vacua, corresponding to the deformation (rather than small
resolution) of the conifold.  The poles in $W$ might similarly be related
to novel phenomena.\foot{For example, a familiar mechanism \ref\ads{I. Affleck, M. Dine, and N. Seiberg,
``Dynamical Supersymmetry Breaking In Four Dimensions And
Its Phenomenological Implications,'' Nucl. Phys. {\bf B256} (1985) 557.}
for generating a pole in a superpotential
in four dimensions involves an $SU(2)$ gauge theory with a pair of doublets,
so perhaps the theory near the poles has a description in such terms.}

\subsec{Membrane Instantons}

On the classical manifold $X_i$, the three-cycle $D_i$ is a boundary
and hence, in a classical supersymmetric configuration,
$\alpha_i=\int_{D_i}C$ vanishes.
As we have seen in section 4.1, to get a smooth curve $\N$ interpolating
between the different classical limits, we need to find a quantum
correction to this statement.

$D_i$ is defined by setting the radial coordinate $r$ in \burigo\
equal to a large constant $t$, which should be taken to infinity,
and also imposing a certain relation on the $SU(2)$ elements
$g_i$. For $r\to\infty$, $X_i$ becomes flat, with the curvature
at $r=t$ vanishing as $1/t^2$.  The volume of $D_i$ grows as
$t^3$, so to get a nonzero value of $\alpha_i$, $C$ must vanish
as $1/t^3$. Perturbative corrections to the classical limit
vanish faster than this. Consider Feynman diagram contributions
to the expectation value of  $C$ at a point $P\in X_i$.  We assume
$P$ is at $r=t$  and ask what happens as  $t\to\infty$. If all
vertices in the diagram are separated from $P$ by a distance much
less than $t$, we may get a contribution to $\langle C(P)\rangle$
that is proportional to some three-form built locally from the
Riemann tensor $R$ and its covariant derivatives.  Any such
three-form vanishes faster than $1/t^3$ for $t\to\infty$. ($R$
itself is of order $1/t^2$, so its covariant derivative $DR$ is
of order $1/t^3$.  But a three-form proportional to $DR $ vanishes
using the properties of the Riemann tensor. Other expressions
such as $RDR$ are of higher dimension and vanish faster than
$1/t^3$.)  Things are only worse if we consider Feynman diagrams
in which some of the vertices are separated from $P$ by a distance
comparable to $t$.  Such diagrams can give nonlocal contributions;
those vanish at least as fast as $t^{-9}$, which is the order of
vanishing at big distances of the massless propagator in eleven
dimensions.

More fundamentally, the reason that perturbative corrections
on $X_i$ do not modify $\alpha_i$ is holomorphy.  As we have seen,
$\alpha_i$ is
the argument of the holomorphic observable $\eta_i$, while
a  local holomorphic parameter at $X_i$ is the membrane amplitude
\eqn\imon{u=\exp\left(-T\int_{Q_i}d^3x\sqrt g +i\int_{Q_i}C\right).}
 $\eta_i$ must be a holomorphic function of $u$, and
perturbative corrections to $\alpha_i$ or $\eta_i$ must vanish as they
are functions only of $|u|$, being independent of the argument
$\int_{Q_i}C$ of $u$.

\nref\defspace{N. Seiberg, ``Exact Results On The Space Of Vacua
Of Four-Dimensional SUSY Gauge Theories,'' Phys. Rev. {\bf D49} (1994) 6857,
hep-th/9402044.}%
\nref\kaplouis{V. Kaplunovsky and J. Louis, ``Field Dependent Gauge
Couplings In Locally Supersymmetric Effective Quantum Field Theories,''
Nucl. Phys. {\bf B422} (1994) 57, hep-th/9402005.}%
This makes it clear where, in $M$-theory on $\R^4\times X_i$,
we must look to find a correction to
the statement that $\alpha_i=0$.
The correction must come from membrane instantons, that is from
membranes whose world-volume is $y\times Q_i$, with $y$ a point in $\R^4$.
\foot{Using instantons to deform a moduli space is familiar in
four-dimensional supersymmetric gauge theories \refs{\seiberg,
\defspace,\kaplouis}.}

The quantity $u$ is really a superspace interaction or
superpotential. To convert it to an ordinary interaction, one
must integrate over the collective coordinates of the membrane
instanton. This integration is $\int d^4y \,d^2\theta$, where
$d^2\theta$ is a chiral superspace integral over the fermionic
collective coordinates of the membrane, and the $y$ integral is
the integral over the membrane position in $\R^4$.

Presently we will show that $\int d^4y d^2\theta\, u$ can be replaced
by a ($u$-dependent) constant times $\int_{\R^4\times Q_i}*G$.
Here, $*$ is the Hodge duality operator, so, in eleven dimensions,
$*G$ is a seven-form that is integrated over the seven-manifold
$\R^4\times Q_i$.  We postpone the evaluation of $\int d^4y \,d^2\theta\,u$
 momentarily, and first
concentrate on showing that, if the result is as claimed, this will solve our problem.

We will show  that adding to the effective action a multiple
of $\int_{\R^4\times Q_i}*G$ will induce a nonzero value for
$\int_{D_i}
C$.  For this, we must analyze the correlation function
\eqn\kilo{\left\langle\int_{\R^4\times Q_i}*G \cdot \int_{D_i}C\right\rangle,}
and show that it is nonzero.

First of all, let us check the scaling.  For propagation at large
distance $t$, the two point function $\langle G\cdot C \rangle$
is proportional to $1/t^{10}$.  But the integration in \kilo\ is
carried out over the seven-manifold $\R^4\times Q_i$ times the
three-manifold $D_i$, and so altogether over ten dimensions.
Hence the powers of $t$ cancel out, and also  $C$ can be treated
as a free field, since corrections to the free propagator would
vanish faster than $1/t^{10}$.

In the free field approximation,
the action for $C$ is a multiple of $\half\int d^{11}x \sqrt g |G|^2$.
The free field equations of motion, in the absence of sources,
are $dG=d*G=0$.

A simple way to evaluate the correlation function
 is to think of $\int_{D_i}C$ as a source
that creates a classical $G$-field, after which $*G$ is then integrated over $\R^4\times
Q_i$.   Thus, we look for the classical solution of the action
\eqn\olop{{1\over 2}\int_{\R^4\times X_{1,\Gamma}}|G|^2+\int_{D_i}C.}
The classical field created by the source is determined by the equations
\eqn\hinu{\eqalign{dG & = 0 \cr
                   d*G & = \delta_{D_i}.\cr}}
The first equation is just the Bianchi identity.  The second
contains as a source $\delta_{D_i}$, a delta function form that
is Poincar\'e dual to $D_i$.

We do not need to solve for $G$ in detail.  In order to evaluate
$\int_{\R^4\times Q_i}*G$, it suffices to know $*G$ modulo an
exact form.  Any solution of the second equation in \hinu\ (with
$G$ vanishing fast enough at infinity) will do.  A convenient
solution can be found as follows. Let $B$ be a ball in $X_i$
whose boundary is $D_i$. (Existence of $B$ is  the reason for the
classical relation $\int_{D_i}C=0$!) We can obey
$d*G=\delta_{D_i}$ by $*G=\delta_B$.  Hence
\eqn\kkilo{\int_{\R^4\times Q_i}*G=\int_{\R^4\times Q_i}\delta_B.}
The latter integral just counts the intersection number of the
manifolds $\R^4\times Q_i$ and $B$; it is the number of their
intersection points,
 weighted by orientation.  If $B$ is obtained
by ``filling in'' $D_i$ in the obvious way, then there is precisely one
intersection point, so the integral is 1.

The intersection number of $B$ with $\R^4\times Q_i$ is a version, adapted
to this noncompact situation, of the ``linking number'' of the
submanifolds $\R^4\times Q_i$ and $D_i$.  In essence, we have deduced
the desired result about the deformation of the moduli space from this
linking number.

\bigskip
\noindent{\it Evaluation Of Superspace Integral}

It remains to show that the integral $\int d^4y\,d^2\theta\,u$
has the right properties for the above computation.  We write the
two components of $\theta$ as $\theta^1$ and $\theta^2$, so $d^2\theta
=d\theta^1\,d\theta^2$.

Let us write $u=e^w$, with $w=-TV(Q_i)+i\int_{Q_i}C$. Since a
fermion integral has the properties of a derivative or a
derivation, we have $\int d^2\theta\,u=u\left(\int d^2\theta
w+\int d\theta^1 w \int d\theta^2 w\right).$  Here the second
term $\int d\theta^1 w \int d\theta^2 w$ contributes a fermion
bilinear (analogous to a Yukawa coupling in four-dimensional
supersymmetric field theory).  It can be omitted for our present
purposes, since it can contribute to $\int_{D_i}C$ only via
Feynman diagrams containing one boson and two fermion
propagators, and such diagrams
  vanish for large $t$ much faster than $1/t^{10}$.

We are left with computing $\int d^2\theta w = \int d^2\theta
\left(-TV(Q_i)+i\int_{Q_i}C\right).$ Because $w$ is a chiral or
holomorphic field, we would have $\int d^2\theta \,\bar w=0$, and
hence \eqn\noggo{\int d^2\theta\, w = -2T\int d^2\theta V(Q_i) =
2i\int d^2\theta \int_{Q_i}C.} Because it is slightly shorter, we
will compute $\int d^2\theta V(Q_i)$. But obviously, the result
also determines $\int d^2\theta\int_{Q_i}C$. This fact will be
useful in section 6.1, where we will need the latter integral.

Let $\epsilon$ be a covariantly constant spinor on $\R^4\times X_i$.
Under a supersymmetry generated by $\epsilon$, the variation
of the volume $V(Q_i)=\int_{Q_i}d^3x \sqrt g$ is, using the
supersymmetry transformation laws of eleven-dimensional supergravity
\ref\crem{E. Cremmer, B. Julia, and J. Scherk, ``Supergravity in
11 Dimensions,'' Phys. Lett. {\bf 76B} (1978) 409.},
\eqn\rugo{\delta_\epsilon V(Q_i)=-i\kappa \int_{Q_i}d^3x \sqrt g
g^{ab}\bar\epsilon \Gamma_a\psi_b.}
Here indices $a,b,c$ run over tangent directions to $Q_i$, while
indices $A,B,C$ will run over tangent directions to $\R^4\times X_i$.
Also, $\kappa$ is the gravitational coupling, $\psi$ the gravitino,
 $\Gamma_A$ are gamma matrices, and likewise $\Gamma^{A_1A_2\dots A_k}$
will denote an antisymmetrized product of gamma matrices.

To compute $\int d^2\theta\, V$, we let $\epsilon_1$ and $\epsilon_2$
be covariantly constant spinors of positive chirality on $\R^4$,
and compute the second variation $\delta_{\epsilon_2}
\delta_{\epsilon_1}V$.  ($\epsilon_1$ and $\epsilon_2$ are the tensor
products of the same covariantly constant spinor on the $G_2$ manifold
$X_i$ times a constant positive chirality spinor on $\R^4$.)
If $\epsilon_1$ and $\epsilon_2$ are properly normalized, this equals
$\int d^2\theta \,V$.
Ignoring terms proportional to $\psi^2$ (as their contributions
vanish too fast for large $t$), we get
\eqn\urugo{\delta_{\epsilon_2}\delta_{\epsilon_1}V(Q_i)
={\kappa\over 144}\int_{Q_i}d^3x\sqrt g\bar\epsilon_1\Gamma_a
\left(\Gamma^{ABCDa}-8\Gamma^{BCD}\delta^A_a\right)\epsilon_2\,G_{ABCD}.}

The field $G$ created by a delta function source on $y\times D_i$
has all indices tangent to $X_i$ (or more precisely $y\times
X_i\subset \R^4\times X_i$) since the source has that property.
Moreover, it follows from the symmetries of $X_i$ that when
restricted to $Q_i$, $G$ can be written \eqn\hbci{G=G'+G'',}
where $G'$ has all four indices normal to $Q_i$ and $G''$ has
precisely two indices in the normal directions.  To make the
notation simple in justifying this claim, take $i=1$ so we are on
$X_1$.  Then $D_1$ is the set $(g_1,1,1)$, and is mapped to
itself by $g_1\to ug_1$, $u\in SU(2)$, so $G$ has that symmetry.
This symmetry acts trivially on $Q_1$, which (if we gauge away
$g_3$ and then set $g_1$ to zero to get $Q_1$) is the set
$(0,g_2,1)$; the symmetry transforms the normal bundle to $Q_1$
in the fundamental representation of $SU(2)$ (which is of complex
dimension two or real dimension four). As there are no odd order
invariants in this representation, invariance of $G$ under this
$SU(2)$ action implies that all terms in $G$, when restricted to
$Q_1$, have an even number of indices in the directions tangent
to $Q_1$; the number can only be zero or two as $Q_1$ is
three-dimensional.

Using this decomposition, we can simplify \urugo, getting
\eqn\turugo{\delta_{\epsilon_2}\delta_{\epsilon_1}V(Q_i)
= {\kappa\over 24}\int_{Q_i}d^3x\sqrt g\bar\epsilon_1\Gamma^{ABCD}\epsilon_2
G'_{ABCD}-{\kappa\over 48}\int_{Q_i}d^3x\sqrt g\bar\epsilon_1\Gamma^{ABCD}\epsilon_2
G_{ABCD}.}

In fact, with $\epsilon_1$ and $\epsilon_2$ being covariantly
constant spinors,
\eqn\rutip{\bar\epsilon_1\Gamma^{ABCD}\epsilon_2 G_{ABCD}=0.}
This follows from the following facts.  The space $\Omega^4$ of
four-forms on a $G_2$-manifold has a decomposition as $\Omega^4_1
\oplus \Omega^4_{7}\oplus \Omega^4_{27}$, where the subscript
refers to the transformation under the group $G_2$ acting in the
tangent space at a point.  This decomposition is described in
\joyce, sections 3.5 and 10.4, where it is  proved (Theorem 3.5.3)
that it commutes with the Laplacian $\Delta = d^*d+dd^*$.  The
statement \rutip\ means that $G$ has no component in
$\Omega^4_1$, since with the $\epsilon$'s being covariantly
constant spinors,
$\Psi^{ABCD}=\bar\epsilon_1\Gamma^{ABCD}\epsilon_2$ is a
covariantly constant antisymmetric tensor, and contracting it with
$G$ is the projection onto $\Omega^4_1$. Since (upon solving
\hinu) the $G$-field produced by a source $\delta_{D_i}$ is
\eqn\nonon{G={1\over \Delta} d*\delta_{D_i},} to show that $G$
has no component in $\Omega^4_1$, it suffices to prove that
$d*\delta_{D_i}$ has no component in $\Omega^4_1$. It is
equivalent to show that, if $\Upsilon$ is the covariantly
constant three-form of the $G_2$-manifold, then
\eqn\icoc{\Upsilon\wedge d*\delta_{D_i}=0.} In fact, if we
suitably normalize the $\epsilon$'s, then $\Upsilon = *\Psi$
(where we have ``lowered indices'' to interpret $\Psi$ as a
four-form) so contracting $d*\delta_{D_i}$  with $\Psi$ is
equivalent to taking a wedge product with $\Upsilon$.

To verify \icoc, we again simplify the notation by choosing $i=1$
and work on $X_1$. Using the description of $Y$ by group elements
$a,b,c$ with $abc=1$, $D_1$ is given by the equations $r=t$ and
$a=1=bc$.  $\delta_{D_1}$ is then a multiple of
$\delta(r-t)\delta^3(a-1) \,dr\,\Tr(a^{-1}da)^3$, and
$*\delta_{D_1}$ is a multiple of
$\delta(r-t)\delta^3(a-1)\,\Tr\,(b^{-1}db)^3$. Finally,
$d*\delta_{D_1}$ is a multiple of
\eqn\polly{d(\delta(r-t)\delta^3(a-1))\,\Tr\,(b^{-1}db)^3.} Now we
use the explicit description of $\Upsilon$ in \cvetica, eqn.
(6.10), where it is called $Q_{(3)}$:
\eqn\lolly{Q_{(3)}=e^0\wedge e^i\wedge e^{\bar i}+{1\over 2}
\epsilon_{ijk} e^i\wedge e^{\bar j}\wedge e^{\bar k}-e^1\wedge
e^2\wedge e^3.} Here $e^i=\gamma\Sigma^i$ is a multiple of what
in  our notation is $db$. (For more details on the relation of
our notation to that in \cvetica, see the footnote after eqn.
\burigo.)  Since every term in $\Upsilon=Q_{(3)}$ is proportional
to at least one factor of $db$, and $*\delta_{D_1}$ has three of
them, which is the maximum possible, $\Upsilon\wedge
\delta_{D_1}=0$.

So we can reduce \turugo\ to
\eqn\turugot{\delta_{\epsilon_2}\delta_{\epsilon_1}V(Q_i) =
{\kappa\over 24}\int_{Q_i}d^3x\sqrt g \Psi^{ABCD}G'_{ABCD}.} Now
we have to use the fact that $Q_i$ is a supersymmetric or
calibrated cycle.  This means that $\Upsilon$ is the volume form
of $Q_i$.   It also means with $G'$ being a four-form in the
normal directions, the map $G'\to *G'$, when restricted to $Q_i$,
is equivalent to $G'\to \Upsilon\cdot \Psi^{ABCD}G_{ABCD}$. Here
$*$ is understood as the Hodge duality operator in the
seven-dimensional sense.  So finally, \turugot\ is a constant
multiple of $\int_{Q_i}*G$.

Finally, when we incorporate the collective coordinate describing the membrane
position $y\in \R^4$ and integrate over $y$, we get
$\int_{\R^4\times Q_i}*G$, where now $*$ is understood in the
eleven-dimensional sense.

\newsec{Role Of A Fermion Anomaly}

In any careful study of the $C$-field in $M$-theory, one
encounters a fermion anomaly.  A brief explanation of the reason
is as follows. Let us call spacetime $M$. Let $Q$ be a
three-dimensional submanifold\foot{It suffices for $Q$ to be an
immersion.  We will actually need this later for the examples.}
 of
$M$, and consider a membrane whose worldvolume is $Q$. In the
worldvolume path integral for such a membrane, we meet a
classical phase factor $\exp(i\int_QC)$.  But we also meet a
fermion path integral.  The classical phase factor must really be
combined with a sign coming from the fermion path integral.  It
turns out that only their product is well-defined.

\def\D{{\cal D}}
\def\Pf{{\rm Pf}}
To describe the worldvolume fermions, let $N_Q$ denote the normal
bundle to $Q$ in $M$. For simplicity, we will here assume that
$M$ and $Q$ are spin, but in any event, for $M$-theory membranes,
$N_Q$ is always spin. This being so,  we let $S(N_Q)$ be the
spinor bundle of $N_Q$ and decompose it in pieces of definite
chirality as $S(N_Q)=S_+(N_Q)\oplus S_-(N_Q)$. Since $N_Q$ has
rank $8$, $S_+(N_Q)$ is real and has rank eight. The worldvolume
fermions are spinors on $Q$ with values in $S_+(N_Q)$. We let
$\D$ denote the Dirac operator on $Q$ with values in $S_+(N_Q)$.
The fermion path integral is the square root of the determinant
of $\D$, or as we will write  it, the Pfaffian of $\D$, or
$\Pf(\D)$. Because spinors on a three-manifold are pseudoreal,
and $S_+(N_Q)$ is real, $\Pf(\D)$ is naturally real. Its absolute
value  can be naturally defined using zeta function
regularization.  But there is no natural way to define the sign
of $\Pf(\D)$.
  One cannot remove this indeterminacy by arbitrarily declaring
$\Pf(\D)$ to be, say, positive, because in general as $Q$ moves
in $M$, eigenvalue pairs  of $\D$ can pass through zero and one
wants $\Pf(\D)$ to change sign.  When $Q$ is followed around a
one-parameter family, $\Pf(\D)$ may in general come back with the
opposite sign. In that case, the fermion path integral has an
anomaly which one cancels by modifying the quantization law for
the periods of the curvature $G=dC$. The modified quantization law
\ref\cper{E. Witten, ``On Flux Quantization In $M$-Theory And The
Effective Action,''  J. Geom. Phys. {\bf 22} (1997) 1.} says that
for any four-cycle $B$ in $M$, \eqn\plok{\int_B{G\over
2\pi}={1\over 2}\int_B{p_1(M)\over 2}~{\rm mod}~\Z,} where here
for a spin manifold $M$, $p_1(M)/2$ is integral but may not be
even.

Mathematically, one can define a real line bundle, the ``Pfaffian
line bundle,'' in which $\Pf(\D)$ takes values.  Here we will
focus on the fact that $\Pf(\D)$ appears in the worldvolume path
integral together with the classical phase factor coming from the
$C$-field. It is really the product
\eqn\inumu{\Pf(\D)\,\exp\left(i\int_QC\right)} that must be
well-defined.  This means that $\exp\left(i\int_QC\right)$ is not
well-defined as a number; it must take values in the
(complexified) Pfaffian line bundle.  If we define $\mu(Q)$ to be
0 or 1 depending on whether $\Pf(\D)$ is positive or negative,
then the phase of the path integral is really
\eqn\tinumu{\phi(Q)=\int_QC+\pi\mu(Q)~{\rm mod}~2\pi.} In
general, only the sum of these two terms is well-defined.

This implies that the geometrical nature of the $C$-field is
somewhat more subtle than one might have assumed; it is not the
three-form analog of a $U(1)$ gauge field but of a ${\rm Spin}^c$
structure.  One can make this analogy  rather precise. For a spin
$1/2$ particle propagating around a loop $S\subset M$ and
interacting with a ``$U(1)$ gauge field'' $A$, the phase of the
path integral comes from a product
\eqn\polko{\Pf(D/Dt)\,\exp\left(i\int_SA\right),} which is the
analog of \inumu. Here $t$ is an angular parameter on $S$, and
$D/Dt$ is the Dirac operator on $S$ acting on sections of the
tangent bundle  to $M$. A spin structure on $M$ gives a way of
defining the sign of $\Pf(D/Dt)$.  On a spin manifold, $A$ is an
ordinary $U(1)$ gauge field and the two factors in \polko\ are
separately well-defined.  In the ${\rm Spin}^c$ case, there is no
definition of the sign of $\Pf(D/Dt)$ as a number, the
geometrical meaning of $A$ is modified, and only the product in
\polko\ is well-defined.

Because only the total phase $\phi(Q)$ is well-defined, the
definition of the periods $\alpha_i=\int_{D_i}C$ in section 4
should be modified to \eqn\limunu{\alpha_i=\int_{D_i}C+\pi
\mu(D_i).} The goal of the present discussion is to determine how
the correction to the definition of $\alpha_i$ should enter the
formulas in section 4.

For our present purposes, we do not need to know much about
Pfaffian line bundles, because everything we need can be deduced
from a situation in which the separate contributions to the phase
 actually {\it are} well-defined.
This is the case in which we are given a four-manifold $B$ in
spacetime with boundary $Q$.  For simplicity we will assume $B$
to be spin.
  Then we would like to write
\eqn\bimmu{\exp\left(i\int_QC\right)=\exp\left(i\int_BG\right).}
The right hand side is well-defined, as $G$ is gauge-invariant.
However, the right hand side may depend on the choice of $B$;
according to \plok, this will occur when $p_1(M)/2$ is not even.
At any rate, since a choice of $B$ enables us to make a natural
definition of $\int_QC$ mod $2\pi$, and since the total phase
$\phi(Q)$ is always well-defined, it must be that once $B$ is
chosen, one also has a natural definition of $\mu$.  The
appropriate definition (which was originally pointed out by D.
Freed) is as follows. Let $N_B$ be the normal bundle to $B$ in
$M$.  Let $S(N_B)$ be the spin bundle of $N_B$; it is a real
bundle of rank eight that on $Q$ reduces to $S_+(N_Q)$.  Let
$\D_B$ be the Dirac operator on $B$ with values in $S(N_B)$, and
with Atiyah-Patodi-Singer (APS) boundary conditions\foot{The APS
boundary conditions are usually formulated for metrics which are
a product near the boundary, but they extend by continuity to
other metrics and the index remains unaltered, provided the
metric on the boundary remains fixed.} along $Q$. Its index is
even, since in general, the Dirac index in four dimensions with
values in a real bundle, such as $S(N_B)$, is even.  Let $i(B)$
be the index of $\D_B$ and let $\nu(B)=i(B)/2$. Then in this
situation, we define \eqn\cimmu{\mu(Q)={\nu(B)}~{\rm mod}~2.} The
justification for this definition is that if \bimmu\ and \cimmu\
are used, one gets a definition of the total phase $\phi(Q)$ that
is independent of the choice of $B$.

This is proved as follows. If $B_1$ and $B_2$ are two spin
manifolds in $M$ with boundary $Q$, one forms the closed
four-manifold $B=B_1-B_2$, where the minus sign refers to a
reversal of orientation of $B_2$ so that $B_1$ and $B_2$ join
smoothly on their common boundary.  The gluing theorem for the
APS index gives $\nu(B_1)-\nu(B_2)=\nu(B)$.  The index theorem
for the Dirac operator on a closed four-manifold gives
$\nu(B)=\half \int_B(p_1(M)/2)$ mod 2, and then using \plok, this
implies that when $B_1$ is replaced by $B_2$, the change in the
period $\int_QC$ just cancels the change in $\mu(Q)$. We shall
apply a variant of this argument later, in section 5.2, to give
an explicit topological formula for $\nu(B)$.

Before presenting some relevant examples in which there is an
anomaly, let us describe some simple cases in which an anomaly
involving the index $\nu(B)$ does {\it not} appear. The most
basic case is that $Q$ is a copy of $\S^3$, embedded in
$M=\R^{11}$ in the standard way. Then we can take $B$ to be a
four-ball, with a standard embedding in $\R^4$.  In this case,
$N_B$ is a rank seven bundle with a trivial flat connection, and
$S(N_B)$ is a trivial flat bundle of rank 8. So $i(B)$ is
divisible by 8, and hence $\nu(B)$ is zero mod 2. (In fact, it
can be shown that in this example, $i(B)$ vanishes.)

This example has the following generalization.  Let $M=\R^4\times
X$, with any seven-manifold $X$.  Suppose that $Q$ and $B$ are
submanifolds of $X$.  Then $N_B$ is a direct sum $\R^4\oplus N'$,
where $N'$ is the rank three normal bundle to $B$ in $X$, and
$\R^4$ is a trivial flat bundle of rank four.  In this situation,
$S(N_B)$ is (when complexified) the sum of four copies of $S(N')$
(the spinors of $N'$), so $i(B)$ is divisible by four and hence
$\nu(B)$ is even and $\mu(Q)=0$.

Let now $X_i$ be one of the three familiar seven-manifolds of
$G_2$ holonomy that is asymptotic to a cone on $\S^3\times
\S^3$.  Let $D_i$ be a three-sphere in $\S^3\times \S^3$ that
bounds a ball $B$ in $X_i$. Classically, the curvature $G=dC$
vanishes for supersymmetry, and hence $\int_{D_i}C=\int_BG=0$.
Moreover, from what has just been observed, $\nu(B)$ is zero mod
2 in this example, and  $\mu(D_i)=0$. So finally we learn that
the ``period'' $\alpha_i$, correctly defined as in \limunu,
vanishes.  This completes the justification of the assertion made
in section 4 that $\alpha_i=0$ on $X_i$ in the classical limit.

Now let us consider a somewhat analogous question that arose in
section 4.  In $\S^3\times \S^3$, we defined three-spheres
$D_1,D_2,D_3$, with $D_1+D_2+D_3=0$ in $H_3(\S^3\times \S^3;\Z)$.
It follows that \eqn\agnu{\sum_i\int_{D_i}C=0,} since the left
hand side can be written as $\int_BG$, where $B$ is a
four-dimensional chain in $\S^3\times \S^3$ with boundary
$D_1+D_2+D_3$.\foot{If one could pick $B$ to be a smooth manifold
in $\S^3\times\S^3$, then $\sum_i\mu(D_i)$ would vanish by the
argument given above.  However,  the $D_i$ intersect, so $B$
cannot be a manifold. Later, we will see that after perturbing
the $D_i$ slightly, we can take $B$ to be a smooth manifold in
$\R^2\times\S^3\times\S^3$, but this does not lead to vanishing
of $\sum_i\mu(D_i)$.} We have therefore
\eqn\pagnu{\alpha_1+\alpha_2+\alpha_3= \pi\sum_i\mu(D_i).} We
claim that $\sum_i\mu(D_i)=1$, and hence that
\eqn\dagnu{\alpha_1+\alpha_2+\alpha_3=\pi.} To demonstrate this,
we first need to describe some properties of the index function
$\nu(B)$ and some methods for calculating it.

\subsec{Stiefel-Whitney Classes }

The mod 2 invariants, such as $\nu (B),$ that we shall be dealing
with are best described in terms of the Stiefel-Whitney classes
$w_{i}.$ These are, in a sense, the real counterparts of the more
familiar Chern classes $c_{i}.$  They may be less familiar, and
so we shall briefly review them at this stage.

For an $O(n)$ bundle (or equivalently for a real vector bundle)
over a space $Y$, the $w_i$ are characteristic classes
\eqn\charclass{ w_{i}\in H^{2}(Y;\Z_{2})} such that $w_{0}=1,$
$\;w_{i}=0$ for $i>n$.  They have the following properties:

{\it (1)} $w_1$ measures the obstruction to orientability of a
vector bundle; to a circle $C\subset Y$ it assigns the value $1$
or $-1$ depending on whether the restriction of the bundle to $C$
is orientable. In particular, for the two-sheeted cover $\S^n$ of
$\RP^n$, $w_1$ assigns the nontrivial element of
$H^1(\RP^n;\Z_2)$.
\medskip

{\it (2)} Let $w(E)=w_0(E)+w_1(E)+\dots$ be the total
Stiefel-Whitney class. For a direct sum of real vector bundles
$E,F$ we have the product formula \eqn\dirsum{ w(E\oplus
F)=w(E)\cdot w(F).  }
 Taking $F$ to be a trivial bundle,
$%
w(F)=1,$ and so the product formula implies that the $w_{i}$ are
{\it stable}, i.e. unchanged by $E\to E\oplus F$.\medskip

{\it (3)} For an $SO(2n)$-bundle, $w_{2n}$ is the mod 2 reduction
of the Euler class $e\in H^{2n}(Y;\Z)$: \eqn\eulcla{ e\equiv
w_{2n}~{\rm mod}~2 .}

Recall that, for $Y$ a manifold, $e$ may be defined by the locus
of zeros of a generic section $s$ of a rank $2n$ vector bundle.
The fact that $w_{2n}$ is the mod 2 reduction of $e$ actually
follows from the corresponding statement mod 2 for a generic
section of any real vector bundle (oriented or not, even or
odd-dimensional). More generally we have the following:
\noindent  {} For a real vector bundle $E $ of rank $n$, let
$s_{1},...,s_{n-i+1} $ be generic sections. Then $w_{i}(E)$  is
represented by the mod 2  cycle of points where the sections
become linearly dependent. \medskip

Note that the Chern classes may be defined by a similar property
for sections of a complex vector bundle, except that dimensions
are doubled and we work with integer cohomology.

The first two classes $w_{1},\;w_{2}$ characterize orientability
and spin, i.e. they vanish for $SO(n)$ and Spin($ n $) bundles,
respectively. For a Spin($n)$-bundle we also have
\eqn\alshav{w_{3}=0,\;w_{4}=p_{1}/2.} Here $p_{1}$ is the first
Pontrjagin class (which is naturally divisible by two for a
spin-bundle). By the stability of the $w_{i}$ (and of $p_{1}$),
we can check \alshav\ by looking at Spin($3$) and Spin($4$)
bundles. But \eqn\buthav{ {\rm Spin}(3)=SU(2),~~ {\rm
Spin}(4)=SU(2)\times SU(2).} Since $\pi_i(SU(2))=0$ for $i\leq
2$, the first statement implies that a Spin($3$)-bundle over $Y$
is always trivial on the three-skeleton of $Y$, showing that
$w_{3}=0.$ The second statement implies that a Spin($4)$-bundle
has, in dimension 4, two integral characteristic classes (say
$a,b$) coming from the Chern classes of the two factors. They are
related to the Pontrjagin class $p_{1}$ and the Euler class $e$ by
\eqn\relby{\eqalign{ a & ={p_1\over 4}+{e\over 2} \cr b &
={p_{1}\over 4}-{e\over 2}.\cr}}
\medskip

These formulae show that, as asserted in \alshav, $p_{1}/2$ is
naturally an integral class, namely $2a-e.$ Moreover, reducing
modulo $2$, and using the fact that $w_4$ is the mod 2 reduction
of $e$, we deduce \eqn\weded{ p_{1}/2=w_{4}\quad {\rm mod }~2 .}
Although we have only verified this for Spin($4)$-bundles, it
follows for all Spin($n$)-bundles ($n>4)$. This is an aspect of
stability: since
Spin($%
n+1)/$Spin($n)=\S^{n}$, no new relations can be introduced on
$q$-dimensional characteristic classes for $q\leq n,$ when
passing from $n$ to $n+1$.

For our applications, \weded\ is the key formula, and the reason
for our interest in Stiefel-Whitney classes.

For a Spin($3)$-bundle, we have $w_{4}=0$.  (This is a special
case of the vanishing of $w_k$ for a bundle of rank less than
$k$.)
 This implies that, if a Spin($n)$-bundle over $Y$ can be
reduced over a subspace $Y_{0}$ to a Spin($3$)-bundle, then
$w_{4}$ can be lifted
back from $H^{4}(Y;\Z_{2})$ to a {\it relative class} in $%
H^{4}(Y,Y_{0};\Z_{2}).$ More precisely, a choice of reduction over
$Y_{0}$ gives a definite choice of relative class. This is
because such a reduction is given by ($n-3)$ sections
$s_{1},...,s_{n-3}$ which are independent
over $%
Y_{0}$: their locus of dependence then gives a representative cycle in $%
Y-Y_{0} $ for the relative $w_{4}$.

Actually we can define a relative $w_{4}$ in the more general
situation of a Spin$(n)$\ bundle over $Y$ with a reduction to an
$H$-bundle over $Y_{0},$ where \eqn\whereh{ H=\prod_{i}{\rm
Spin}(n_{i})\quad \sum_i n_{i}=n,\;n_{i}\leq 3.} (Such an $H$ is
not always a subgroup of Spin$(n)$.  It may be a finite covering
of such, so a reduction means a reduction and a lifting.) Indeed,
each Spin$(n_{i})$ bundle, with $n_{i}\leq 3,$ has trivial \
Stiefel-Whitney classes ($w=1),$ and so by the product formula
\dirsum, $w=1$ for an $H$-bundle. Thus again $w_{4}$ lifts back
to a relative class. The uniqueness can be seen from the
universal case when $Y,Y_{0}$ are the appropriate Grassmannians,
using the fact that the Grassmannian for $H$ (i.e. the product of
the Grassmannians for the factors) has no cohomology in dimension
3.

Similar reasoning enables us to use the product formula in the
relative case to show that if, $E,F$ are spin-bundles over $Y$
with reductions to groups of type $H$ over $Y_{0},$ then the
relative $w_{4}$ is additive: \eqn\lolly{ w_{4}(E\oplus
F)=w_{4}(E)+w_{4}(F) .} Here one must use the fact that, for
spin-bundles, $w_{1}=w_{2}=w_{3}=0$.

\subsec{Topological formula for $\nu (B)$}

After this digression about Stiefel-Whitney classes, we return to
our problem of computing the index function $\nu (B)$ and the
corresponding
invariant $%
\mu (Q)$ introduced in eqn. \cimmu.

We shall give a topological way of computing these mod 2
invariants under the assumption that at least near $Q$, our
space-time manifold is $M=\R^{5}\times Y$ where $Y$ is a spin
six-manifold and  $Q\subset Y.$  However we shall {\it not}
assume that $Q$ is the boundary of some  $B\subset Y$ since, as
we argued earlier, this would make our invariant automatically
zero.

In fact, for our applications, we will allow $Q$ to be not quite
a smooth submanifold of $Y$, but  the union of a number of smooth
submanifolds \eqn\tuvuu{ Q=Q_{1}\cup Q_{2}\cup ...\cup Q_{k}. }
It may be that  the $Q_i$ intersect in $Y$, but if this happens,
 we can separate these components, so that they do not
intersect, by using some of the additional $\R^{5}$-variables.
Thus we
 take $k$ distinct vectors $u_{1},...,u_{k}$ in $\R^{5}$ and shift
the component $Q_{j}$ to lie over the point $u_{j}.$ \ With this
understanding, $ Q $ becomes a genuine submanifold of $M,$ and
$\mu(Q)$ is unchanged by such a shift since the metrics induced
by $Q$ remain the same and $\mu$ is defined for immersions.

We now assume that $B$ is a compact spin four-manifold, embedded
in $M,$ with boundary $Q.$ This implies that, in homology,
\eqn\imhom{ Q=\sum_{j}Q_{j}=0{\quad \quad {\rm in}
}~H_{3}(M)=H_{3}(Y).} We shall also assume that, near each
$Q_{j}$, $B$ is the product of $Q_{j}$ with the half-line
$ru_{j},\;r\geq 0,$ so that $u_{j}$ is normal to $Q_{j}$ in $B.$

Consider now  the normal bundle $N_{B}$ to $B$ in $M.$ \ This is a
Spin$%
(7)$-bundle. \ Over each $Q_{j}$\ it splits off a trivial
$\R^{4}$ factor (orthogonal to $u_{j}$ in $\R^{5})$ and hence
reduces to a Spin$(3)$-bundle.
 In this situation, as explained above, we have a relative class
\eqn\nurho{ w_{4}(N_{B})\in H^{4}(B,Q;\Z_{2}).}

We claim  that \eqn\urgho{ \nu (B)=w_{4}(N_{B})  } where on the
right side we evaluate the relative class $w_4$ on the top cycle
of $B$, to get an element of $\Z_2$.

To argue this, note first that over
 each component $Q_{j}$ we have a natural decomposition
\eqn\natdec{ N_{B}|_{Q_{j}}=N_{j}\oplus R^{4},  } where $N_{j}$
is the normal Spin$(3)$-bundle to $Q_{j}$ in $Y.$  Since $\dim
Q_{j}=3,$ $N_{j}$ is actually trivial, so that $N_{B}$ also gets
trivialized over $Q=\partial B.$ \ Now take two copies of $B,$
and put a rank $7$ vector bundle over each copy. \ Over the first
copy $B_{1}$ we take $N_{1}=N_{B}$ while over the second copy
$B_{2}$ we take the trivial
bundle $%
N_{2}.$ \ Gluing $B_{1}$ and $-B_{2}$ together to form a closed
spin four-manifold $\hat{B},$ we can also glue together the two
vector bundles, using the trivialization coming from \natdec, to
get a vector bundle $\hat{N}. $ \ We now compute the index of the
Dirac operator, with coefficients in the spinors of this rank $7$
bundle, for the two $B_{i}$ and for $\hat{B}.$ We denote the
spinors of $N$ by $S(N)$, and for the manifolds with boundary
$B_i$, we take APS boundary conditions. The additivity of the APS
index gives \eqn\adindex{ {\rm index }D_{B_{1}}(S(N_{1}))-{\rm
index }D_{B_{2}}(S(N_{2}))={\rm index }D_{\hat{B}}S(\hat{N}). }

Now over $B_{2},$ $N_{2}$ is trivial, so the index is divisible
by $8.$
 On
the closed manifold $\hat{B}$, the index theorem gives
\eqn\closind{ {1\over 2}{{\rm index
}}D_{\hat{B}}(S(\hat{N}))={p_{1}(\hat{N})\over 2} ~{\rm mod} ~2,}
where the right-side is viewed as an integer by evaluation on the
top cycle of $\hat{B}.$

Hence \adindex\ gives \eqn\purho{ \nu (B)={p_{1}(\hat{N})\over
2}\quad {\rm mod}~2. } But $p_{1}(\hat{N})$ is just the relative
Pontrjagin class of $N_{B}$ given by the trivialization \natdec\
on $\partial B.$  As explained in section 5.1, this can be
re-written in terms of the relative $w_{4}$ to give \urgho.

At this stage we could just as well have stuck to (half) the
relative Pontrjagin class. The advantages of using $w_{4}$ will
appear later.

\subsec{Topological Formula For $\mu (Q)$}

If we can explicitly find a convenient four-manifold $B\subset M$
with boundary $Q,$ then \urgho\ gives an effective way to
calculate $\nu (B)$ and hence our
anomaly $%
\mu (Q).$ \ We shall exhibit a concrete example of this later for
the case when $Y=\S^{3}\times \S^{3}$ and each $Q_{j}$ is also
$\S^{3}.$  However, as we shall now explain, it is possible to
give a useful formula for $\mu (Q)$ for some cases
 even when we do not know how to
construct $B$.

The first step is to observe that \eqn\jukk{ N_{B}\oplus
T_{B}=T_{M}|_B  } where $T_{B},\;T_{M}$ are the respective
tangent bundles.  Moreover, we are in the general situation,
explained in the discussion on Stiefel-Whitney classes, where we
can apply the  formula \lolly\ for relative $w _{4}$ (here
$\,H=\;$Spin$(3)\times $ Spin$(3)$ $\rightarrow $ Spin$(11)),$ so
that \eqn\eqfor{ w _{4}(N_{B})+w _{4}(T_{B})=\langle w
_{4}(T_{M}),B\rangle .} Thus, in \urgho, we can replace $w
_{4}(N_{B})$ by the other two terms in
 \eqfor. \ We examine each in turn.

The easy one is $w_{4}(T_{B}).$ By \eulcla, we have
\eqn\relcal{\eqalign{ w_{4}(T_B)=&e(B,\partial B) ~ {\rm mod}~2
\cr =&\chi (B) ~ {\rm mod}~2,}} where $\chi $ is the usual
Euler-characteristic.  For a four-dimensional spin-manifold $B$
with boundary, $\chi (B)~{\rm mod}~2$ depends only on $\partial
B.$
 This follows from the fact that $\chi $ is additive under gluing,
while for a closed spin manifold $\chi$ is congruent to the
signature mod 2, and hence even.\foot{ The signature of a
four-dimensional spin-manifold is even (and in fact divisible by
16) by the index theorem.} Notice that, in this elementary
argument, given $Q$ we can choose {\it any }spin manifold $B$
with $\partial B=Q.$  We do not need $B$ to be embedded in $M.$
In particular, if $Q$ is a union of components $Q_j$, we can take
$B$ to be a disjoint union of $ B_{j}$ with $\partial
B_{j}=Q_{j}.$ Thus, if (as in our example) $Q$ is a union of
three three-spheres $Q_j$, we can take each $B_{j}$ to be a
four-ball so that $\chi (B_{j})=1$ and hence  $\chi (B)\equiv 1$
mod 2.

It remains for us to dispose of the term in \eqfor\ coming from
the relative $w_{4}$ of $T_{M}.$  We would like to find
conditions that make this zero. One can actually show that the
absolute $w_{4}$ is always zero for $M=Y\times \R^{5},$ but we
need the relative version. \ From the definition and properties of
the relative $w_{4}$, we see that to show that this object
vanishes, it is sufficient to find a reduction of $T_{M}$ to
Spin$(3)\times $ Spin$(3)$ which agrees with the natural
decomposition over each $Q_{j}$ (with the first factor being
tangent to $Q_{j}$ and the second normal to $Q_{j}$ in $Y)$. \
Essentially, all we need is a (spin) $3$-dimensional sub-bundle
of $T_{Y}$ which is transversal to each $Q_{j}.$ \ This is easy
to do if we make the following further assumptions (which hold in
the examples we need):

{\it (A)} $Y=Y_{1}\times Y_{2}$, where the $Y_i$ are
three-dimensional spin-manifolds.

{\it (B)} Each $Q_{j}$  is either a cross-section or a fibre of
the projection $\pi _{2}:Y\rightarrow Y_{2}$.

 To get our sub-bundle of $T_{Y}$ we start with the bundle
$\pi_{1}^{\ast }T_{Y_{1}}.$ \ This is certainly transversal to all
cross-sections of $\pi _{2},$ but of course it is not transversal
to the fibres of $\pi _{2}.$ \ However spin three-manifolds are
always parallelizable so that $\pi _{1}^{\ast }T_{Y_{1}}\cong \pi
_{2}^{\ast }T_{Y_{2}}.$ \ Choosing such an isomorphism we can
rotate $\pi _{1}^{\ast }T_{Y_{1}}$ slightly in the
$Y_{2}$-direction. \ If this rotation is small enough it does not
destroy transversality to a finite set of cross-sections. \ But
any non-zero rotation (with all ``rotation angles'' nonzero)
gives us transversality to all fibres. \ Thus conditions {\it
(A)} and {\it (B)} are sufficient to ensure that \eqn\huglo{
\langle w_{4}(T_{M}),B\rangle =0} and hence \urgho,\eqfor, and
\relcal\ give us our final formula \eqn\juglo{ \mu (Q)=\chi
(B)=\sum_{j}\chi (B_{j})~{\rm mod}~2,  } where $B_{j}$ are any
spin four-manifolds (not necessarily in $M$) with $\partial
B_{j}=Q_{j}.$

\subsec{The Examples}

Our first example is the familiar case \eqn\famcase{
Y=SU(2)^{3}/SU(2).} There are three projections $\pi
_{j}:Y\rightarrow \S^{3}$ (given by omitting the $j^{{th}}$
coordinate), and the $\S^{3}$-fibres of the $\pi _{j}$ we have
denoted by $D_{j}.$ \ We noted in section 2.5 that, in
$H_{3}(Y;\Z),$ we have \eqn\telfor{ D_{1}+D_{2}+D_{3}=0. } We
want to compute $\mu(Q)$ with $Q=D_1\cup D_2\cup D_3$.

 $D_{2}$ and D$_{3}$ are both cross-sections of
the projection $\pi _{1}$: in fact, if we identify $Y$ with the
product of the first two factors, then $D_{2}$ is the graph of a
constant map $ \S^{3}\rightarrow \S^{3}$, while $D_{3}$ is the
graph of the identity map.

Thus, the conditions {\it (A)} and {\it (B)} are satisfied, so
that formula \juglo\ applies.  Taking $B$ to be a union of balls
bounded by the $D_i$, we get \eqn\pugulo{ \mu (Q)=1~{\rm mod}~2 .}
 This establishes \dagnu\ as promised.

The second example is the generalization discussed in Section 2.5
and further explored in section 6.  It involves the six-manifold
\eqn\furex{ Y_{\Gamma }=\S^{3}/\Gamma \times \S^{3},} where
$\Gamma $ is a finite subgroup of $SU(2).$ \ This can also be
viewed as the quotient of \eqn\nurex{ Y=SU(2)^{3}/SU(2)} by the
action of $\Gamma $ on the left on the first factor.  The three
projections of $Y$ to $\S^{3}$ now give rise to projections
\eqn\turex{\eqalign{ \pi_{1}^{\prime }: &Y_{\Gamma }\rightarrow
\S^{3} \cr \pi _{2}^\prime: &Y_{\Gamma }\rightarrow \S^{3}/\Gamma
\cr \pi _{3}^{{\prime }}: &Y_{\Gamma }\rightarrow
\S^{3}/\Gamma.\cr}} The fibres of these projections are denoted by
$D_{j}^{{\prime }}:D_{1}^{{\prime }}=\S^{3}/\Gamma
,\;D_{2}^{{\prime }}=D_{3}^{{\prime }}=\S^{3}.$  Again
$D_{2}^{{\prime }}$ and $D_{3}^{\prime }$ are cross-sections of
the projection $\pi _{1}^{{\prime }}.$  The homology relation in
$H_{3}(Y_{\Gamma };\Z)$ is given as in section 2.5 by
\eqn\retherf{ ND_{1}^{{\prime }}+D_{2}^{{\prime }}+D_{3}^{{\prime
}}=0} where $N$ is the order of $\Gamma .$

We therefore take our three-manifold $Q$ to have $N+2$ components
\eqn\belfor{ Q=Q_{1}\cup Q_{2}\cup \dots\cup Q_{N+2},} where the
first $N$ are parallel copies of $D_{1}^{{\prime }}$ (the fibre
of $\pi _{1}^{{\prime }})$ and the last two are $D_{2}^{{\prime
}}$
and $%
D_{3}^{{\prime }}.$ \ The conditions {\it (A)} and {\it (B)} are
again satisfied, so that formula \juglo\ gives \eqn\fihar{ \mu
(Q)=\sum_{j=1}^{N+2}\chi (B_{j}) ,} where $B_{j}$ is any
spin-manifold with boundary $Q_{j}.$ \ For $j=N+1$ or $N+2$, we
can take $B_{j}$ to be a four-ball, while for $j\leq N$ we can
take $B_{j} $ to be the resolution of the singular complex
surface $\C^{2}/\Gamma .$ This has non-zero Betti numbers
\eqn\elfal{\eqalign{ b_{0} =&1 \cr b_{2} =&r,\cr}} where $r$ is
the rank of the corresponding Lie group (of type $A,$ $D$ or $E$).
 Thus, mod 2,
\eqn\haffot{ \mu (Q)=N(1+r).} But $Nr$ is always even (in fact,
$N$ is even except for a group of type $A$ with even $r$), so
finally \eqn\gaffot{ \mu (Q)=N~{\rm mod}~2 , } generalizing
\pugulo. \ This result will be used in section 6.2 (formula
(6.7)).

\def\hpo{{\bf HP}^2_0}

Finally, we return to the first example of $Y=\S^{3}\times \S^{3}$
and exhibit an explicit four-manifold $B$ with boundary $Q,$ with
the aim of giving a more direct proof of \pugulo.  For this
purpose we shall introduce the quaternionic projective plane
$\HP^{2}.$ \ By definition, this is the eight-manifold
parameterized by triples of quaternions $(u_1,u_2,u_3)$, not all
zero, modulo right multiplication by an element of the group
$\H^{\ast }$ of non-zero quaternions. This group is \eqn\nonzh{
\H^{\ast }=SU(2)\times \R^{+},} the product of the unit
quaternions and a radial coordinate. The subspace $\hpo$ of
$\HP^2$ in which the homogeneous quaternionic coordinates
$(u_1,u_2,u_3)$ are each nonzero is a copy of $\H^\ast\times
\H^\ast=SU(2)\times SU(2)\times \R^2$. So our spacetime
$M=\R^5\times Y$ can be identified as $M=\R^3\times \hpo$.
 $Y$ is naturally embedded in $\hpo$ as the subspace whose homogeneous
quaternionic coordinates $(u_{1},u_{2},u_{3})$ satisfy \eqn\nobob{
\left| u_{1}\right| =\left| u_{2}\right| =\left| u_{3}\right| .}
Similarly the three seven-manifolds $X_{i}$ are also embedded in
$\HP^{2}.$ \ For instance,  $X_{1}$ is given by \eqn\obob{ \left|
u_{1}\right| <\left| u_{2}\right| =\left| u_{3}\right| .}

The  homology class $D_{j}$ is represented in $Y$ by the
three-sphere with $u_k=1$ for $k\not= j$.  In $\hpo,$ $D_j$ can
be deformed to a three-sphere that links around the line
$u_{j}=0$ in $\HP^2$.

We get a simple choice for our required four-manifold $B$ as
follows. Let $\overline B$ be a quaternionic line in $\HP^2$
given  by, say,  the equation \eqn\nueff{ u_{1}+u_{2}+u_{3}=0. }
Now define $B$ by removing small neighbourhoods of the three
points where $\overline B$ meets the coordinate lines.  Thus $B$
is a four-sphere with three open balls removed.

It is now easy to compute the topological invariant
$w_{4}(N_{B}),$ which by \urgho\ determines our invariant $\nu
(B).$  First we can replace the rank $7$ bundle $N_{B}$ by the
rank $4$ bundle $N_{B}^{\prime }$ which is just the normal to $B$
in $\HP^{2}$ (since the remaining $\R^{3}$ of $M=Y\times
\R^{2}\times \R^{3}$ gives a trivial factor).  The reduction of
$N_{B}^\prime$ to a Spin(3)-bundle over $Q=\partial B$ amounts to
fixing a normal direction to the compactification $\overline B$
at the three ends. \ The relative $w_{4}(N_{B}^\prime)$ is thus
just $ w_{4}(N_{\bar{B}}^\prime)$, which  is the reduction mod 2
of the Euler class of the normal bundle to $\HP^{1}$ in
$\HP^{2}.$ But this is just $1,$ since two quaternionic lines in
the quaternionic plane meet in just one point.  This shows that
\eqn\melifo{ \nu (B)=1,} in agreement with \pugulo.

\subsec{  More on the Quaternionic Projective Plane}

  As we have just seen, the quaternionic projective plane $\HP^2$ provides a
convenient compact eight-manifold which  naturally contains, as in
(5.47), the three seven-manifolds $X_i$, and it enabled us to
give a direct computation of the fermionic anomaly in one of the
main examples.   We shall now point out some  further geometric
properties of $\HP^2$, which are closely related to the
discussion in section 3.  (The rest of the paper does not depend
on the following discussion.)

  Since $\HP^2$ is the quotient of $\H^3 - \{0\}$ by the right
action of $\H^*$, it admits the  left action of $Sp(3)$ and, in
particular, of its subgroup $U(3)$. The complex scalars $U(1)
\subset U(3)$ therefore act on $\HP^2$ and commute with the
action  of $SU(3)$.   We shall show that \eqn\algo{ \HP^2 /
U(1)   =    \S^7   .} Note that the fixed-point set of $U(1)$ is
$\CP^2$ (the subset of $\HP^2$ in which the homogeneous
coordinates are all complex numbers), which has codimension 4.
$U(1)$ acts on the normal bundle in the usual way,
 so that the quotient is indeed a
manifold.

   In fact we will indicate two proofs of \algo.  The first is in the
spirit of the proofs in section 3 and actually builds on the
results of section 3.5.  The second is more elegant, and exploits
the symmetry to show that  the diffeomorphism \algo\ is actually
compatible with the action of $SU(3)$, where $\S^7$ is the unit
sphere in the adjoint representation.

Since writing the first version of this paper, we have learned
that the result \algo\ is actually a special case of a result by
Arnold \arnold.  However, it seems worth keeping our
presentation, partly because of the way it fits into the context
of this paper, and partly because our approach is different from
that of Arnold, and may generalize in different directions.

   For our first proof we take $(x,y,z)$ as homogeneous quaternionic
coordinates for $\HP^2$.  Then $\lambda = |x|^2$, $\mu = |y|^2$,
and $\nu = |z|^2$ are naturally  homogeneous coordinates for
$\RP^2$, and fill out the interior of a triangle in $\RP^2$
(since $\lambda,\mu, \nu$ are all positive or zero).  The quotient
of $\HP^2$ by $SU(2)^3$ (acting by conjugation on each of
$x,y,z$) is this triangle.  The point $A = (1,1,1)$ is a
distinguished  point in the triangle and corresponds to (that is,
its inverse image in $\HP^2$ is) our six-manifold $Y = \S^3 \times
\S^3$.  Any line in $\RP^2$ is given by an equation
\eqn\balgo{          a \lambda  + b \mu + c \nu = 0,} with
$a,b,c$ real (and not all zero).  It passes through $A$ provided
       $a + b + c = 0$.

\def\bar{\overline}
The region of $\HP^2$ corresponding to such a line is, in
general, a smooth compact seven-manifold $\bar X(a,b,c)$
containing $Y$. If any one of $a,b,c$ is zero, say $c = 0$, then
the line joins a vertex of the triangle to the mid-point of the
opposite side.   We then recognize $\bar X(a,b,c)$ as the
one-point  compactification of  one of our seven-manifolds $X_i$,
and so it has just one singular point. For a general line, as
indicated by the diagram,
we have points $B,C$ on two sides of the triangle.  The segment
$AB$ corresponds to a manifold diffeomorphic to the compact form
of $X_2$ ( with $Y$ as boundary), while the segment $AC$ gives
$X_3$ (with $Y$ as boundary).   Thus the whole segment expresses
$\bar X(a,b,c)$ as the double
              $ \bar X(a,b,c)  =  X_2\cup_Y X_3$, where the symbol $\cup_Y$
              refers to an operation of gluing two manifolds along $Y$.

\centerline{\psfig{figure=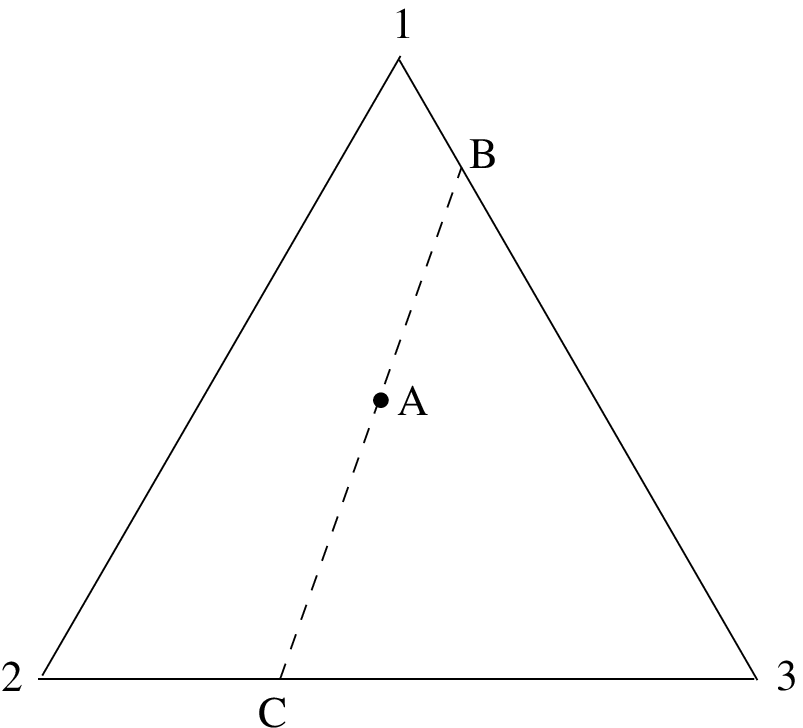,width=2in}}

\vskip 1cm

 Clearly, as the lines  through $A$ sweep out the
(closed) triangle  the manifolds $\bar X(a,b,c)$ sweep out
$\HP^2$, with axis $Y$.   Now divide  by $U(1)$.  From the
results of section 3.5, it follows that \eqn\calgo{    \bar
X(a,b,c) / U(1)  = \S^6.} In fact, for general $(a,b,c)$ we get
$\S^6$ as the  union of two closed six-balls (glued along $Y/U(1)
= \S^5$), while if one of $a,b,c$ is zero we get $\S^6$ as the
one-point compactification of $\R^6$.

Thus we have exhibited the quotient $\HP^2 / U(1)$ as being swept
out by a one-parameter family of 6-spheres, having $\S^5$ as
axis.   It only remains to check the local behaviour near the
three special six-spheres (where $a,b,$ or $c$ is zero), and this
can be done explicitly by using local coordinates.  This
completes the first proof of \algo.

   We turn now to the second proof, which will be based on studying the
orbit structure  of $\HP^2$ under the action of $SU(3)$.  The
generic orbit has codimension one.  Whenever this happens on a
compact manifold, there must be two special orbits of lower (and
possibly unequal) dimension, and the local behaviour near such
special orbits is determined by the normal representation of the
isotropy group.   Thus, if the generic orbit is $W=G/K$ and the
special orbits are \eqn\dalgo{W_1 = G/K_1,~~~~~   W_2 = G/K_2,}
 then the projections
\eqn\ealgo{ G/K \rightarrow G/K_1,~~~~~   G/K \rightarrow G/K_2}
give the normal sphere-bundles of the special orbits.  In
particular $K_1/K$ and $K_2/K$ must both be spheres.   Thus $G/K$
is a sphere-bundle in two different ways.  ``Filling in'' these
sphere-bundles, joined along their common boundary, gives us back
the original $G$-manifold, which
 is thus completely determined by the subgroups $K_1$, $K_2$, and $K$  of $G$,
with $K\subset K_1 \cap  K_2$.

   A standard example with such codimension one orbits is the action of
$SO(p) \times SO(q)$ on $\S^{p+q-1}$.  The generic orbit has
isotropy group $SO(p-1) \times SO(q-1)$ while the two special
orbits are
  $\S^{p-1}$ with $K_1 = SO(p-1) \times  SO(q)$ and
   $\S^{q-1}$ with $K_2 = SO(p) \times SO(q-1)$.
The particular case $p=2,q=3$ already occurred in section 3 and
will be used again shortly.

We now return  to the  case of interest, where the manifold is
$\HP^2$ and $G = U(3)$.   We claim that the isotropy groups are
\eqn\galgo{
   K = U(1)^2, ~~          K_1 = U(1) \times U(2),~~    K_2 = SU(2)\times U(1)}
and the orbits are hence \eqn\malgo{   W = U(3)/U(1)^2,~~     W_1
= \CP^2,~~ W_2 = \S^5.}

More precisely $K$,   $K_1$,  and $K_2$ are represented by $U(3)$
matrices of the form \eqn\halgo{ K:~~\left(\matrix{
 \lambda & 0  &0\cr
    0    & \lambda^{-1} & 0 \cr
    0    &  0  & \mu \cr}\right),~~K_1:~~\left(\matrix{
       \lambda & 0 & 0 \cr
         0     & a & b \cr
         0     & c & d \cr}\right),~~
         K_2:~~\left(\matrix{ a & b & 0\cr
                              c & d & 0\cr
                              0 & 0 & \mu \cr}\right)}
                              (with $ad-bc=1$ in $K_2$).
This checks with the requirement that $K \subset K_1 \cap K_2$.

Deferring, for the moment, the proof of \galgo, we move on to
consider the quotient by the action of the central $U(1) \subset
U(3)$. We will get an induced orbit structure on the quotient of
$\HP^2 / U(1)$  for the action of $U(3)$ and, since $U(1)$ now
acts trivially, this is essentially an $ SU(3)$-orbit structure.
{}From \galgo, we see that, using $K'$, $W'$,  etc. for the
isotropy groups and orbits in the action of $U(3)$ on
$\HP^2/U(1)$,   we have
\eqn\nalgo{    K' = U(1)^3 ,  ~~          K'_1 = U(1) \times U(2),
~~K'_2 = U(2) \times U(1)} and hence \eqn\palgo{
    W' = U(3)/U(1)^3,~~     W'_1 = \CP^2,~~          W'_2 = \CP^2.}

   Since our aim is to identify $\HP^2/U(1)$ with $\S^7 $, as a manifold with
$SU(3)$-action, all we have to show is that  the orbit structure
described by \nalgo\ coincides with that of $\S^7$, regarded as
the unit sphere in the adjoint representation of $SU(3)$. But the
adjoint orbits are characterized by three imaginary eigenvalues
$i \lambda_1$, $i\lambda_2$, $i\lambda_3$ (with $\sum_i \lambda_i
= 0$), and we can arrange these so that $\lambda_1 \leq
\lambda_2\leq  \lambda_3.$   To be on the unit sphere we have in
addition the equation \eqn\oalgo{     \sum_i \lambda_i^2 = 1.}

For generic $\lambda_i$, the isotropy group in $U(3)$ is the
maximal torus $U(1)^3$, and there are just two special cases
$(-1,-1,2)/\sqrt 6$  and $(-2,1,1)/\sqrt 6$ where the isotropy
group is greater. These are both $\CP^2$-orbits with isotropy
$U(2)\times U(1)$ .  The  orbits can be viewed as parameterized
by the middle eigenvalue $i\lambda_2$ , with $1/\sqrt 6 \geq
\lambda_2\geq -1/\sqrt 6$.  The generic orbit has stabilizer
$U(1)^3$ and at the endpoint, the stabilizer is $U(2)\times U(1)$.
But this, in view of \nalgo, is identical with the $U(3)$-orbit
structure of $\HP^2/U(1)$, and hence establishes that
$\HP^2/U(1)=\S^7$.  (The embedding of $\CP^2$ in $\S^7$ that we
have used here was described in \ref\massey{W. S. Massey,
``Imbeddings Of The Projective Plane And Related Manifolds In
Spheres,'' Indiana Univ. Math. Journal {\bf 23} (1974).}.)

   It remains to verify \galgo.  In fact this is just the special case $n
= 3$ for the $U(n)$-action on $\HP^{n-1}$, and the  general case
is no more difficult.  Start first with $n = 2$, i.e. the action
of $U(2)$ on $\HP^1 = \S^4$.  This acts through its quotient by
$\{\pm 1\}$.  The quotient is $SO(2)\times SO(3)$,  and acts by
the natural action of $SO(2) \times  SO(3)\subset SO(5)$  on
$\S^4$. As noted above (and also used in section 3), the generic
orbit  is $\S^1 \times \S^2$ with $\S^1$ and $\S^2$ being the
special orbits. $\S^2  = {\CP}^1$ is fixed by $U(1)$ and $\S^1$ is
fixed by $SU(2)$.

Now we move to the case of general $n$.  Consider the following
involution of $\H$ (given by conjugation by ${\bf i}$):
\eqn\gurmo{ \sigma(u+{\bf j}v)=u-{\bf j}v.} Here $u,v\in\C$ and
the quaternion units are ${\bf i},{\bf j},{\bf k}$. We also denote
as $\sigma$ the corresponding involution of $\HP^{n-1}$ given by
the left action of the complex scalar $i\in U(1)\subset U(n)$.
Every $\CP^1$ in $\CP^{n-1}$ generates an $\HP^1$ in $\HP^{n-1}$
and every point $\xi \in \HP^{n-1} - \CP^{n-1}$ lies in a unique
such $\HP^1$: it is the quaternionic projective line joining
$\xi$ and $\sigma(\xi)$. Indeed, for $\xi$  not in $\CP^{n-1}$,
at least one component $v_i$ is non-zero, and so $\sigma(\xi)$ is
not equal to $\xi$; they are joined by a unique $\HP^1$.   This
means that $\HP^{n-1} - \CP^{n-1}$ is fibred over the complex
Grassmannian $Gr(2,n)$ of lines in $\CP^{n-1}$, with fibre $\HP^1
-\CP^1 = \S^4 - \S^2 = \S^1 \times D^3$ .  The origin of $D^3$
yields the special orbit, the $\S^1$-bundle over $Gr(2,n)$ which
is $U(n)/SU(2) \times U(n-2)$. The generic orbit, given by the
non-zero points of $D^3$, is of codimension one and is $U(n)/U(1)
\times U(n-2)$, where $U(1)\subset SU(2)$.  The other special
orbit is of course $\CP^{n-1} = U(n)/U(1) \times U(n-1)$.  This
describes the $U(n)$-orbit structure of $\HP^{n-1}$.

We want to apply this to the case $n=3$, when the Grassmannian
$Gr(2,n)$ is again $\CP^2$ (the dual of the first $\CP^2$) and
the generic orbit is $U(3)/U(1)^2$. More precisely, we get just
the description in \nalgo.

In comparing the second proof with the methods of section 3.5, we
see that the key advantage of moving from the seven-manifold $X$
to the eight-manifold $\HP^2$ is that the $U(1)$-action we want to
divide out by has a large commutant $SU(3)$ in $U(3)$.  This gives
rise to the codimension one orbits in the quotient $\S^7$, whereas
the $U(1)$-action on $X$ only had a $U(1)^2$ commutant in
$SU(2)^2$, with correspondingly smaller orbits.

Our first proof of \algo\  shows that $\overline X/U(1) = \S^6$ is
diffeomorphic to an equatorial $\S^6$ in $\HP^2/U(1) = \S^7$. This
is not so transparent from the second point of view, because this
equatorial division of $\S^7$ is not compatible with the
$SU(3)$-action.

\bigskip\noindent{\it Further Comparison To Section Three}

   So far in this section we have studied $\HP^2$ and its quotient by
   $U(1)$
in relation to Case III of section 3.5 . But it is intriguing to
observe that the $SU(3)$-orbit structure of $\S^7$ which we have
encountered (in the adjoint action) is also related to Case II,
as  studied in section 3.4.   In fact we have shown that $\S^7$
contains two dual copies of $\CP^2$ and  that the open set
obtained by deleting one is the $\R^3$ bundle over the other
(with sphere-bundle the flag manifold $U(3)/U(1)^3$). But these
are just the seven-manifolds X of Case II.   Thus $\S^7$ can be
described as the double
             \eqn\descoub{ \S^7  =  X_1\cup_Y X_2  ,}
where $Y$ is the flag manifold, and the $X_i$ are two copies of
the compact form of $X$, glued along their common  boundary. There
is an involution on $\S^7$ which  interchanges the $X_i$ and comes
from an involution of the Weyl group $\Sigma_3$  on Y.

   The results of section 3.4 applied to the decomposition \descoub\ then
show that, on dividing by the same $U(1)$-subgroup of $SU(3)$ as
in section 3.4,
 \eqn\newdo{      \S^7 / U(1)  =  B^6_{(1)}\cup_{\S^5} B^6_{(2)}  =
 \S^6}
where $B^6$ is a six-ball. Moreover this identification is
compatible with the natural $SU(3)$-action on both sides.

\bigskip\noindent
{\it A Real Analogue}

   In conclusion, it is perhaps  worth pointing out that there is a
``real'' analogue of the  geometry of section 3, in which $O(1) =
\Z_2$ replaces U(1).   To get a manifold as a $\Z_2$-quotient, we
need this time a fixed-point set of codimension 2, the familiar
branched  locus of complex variable theory.  The prototype is of
course the quotient
       \eqn\tudo{ \C/\Z_2  =\C}
given by the map $ w = z^2 $.  This can be embedded in the
four-dimensional prototype
  \eqn\cudo{    \C^2 / U(1)  =  \R^3.}
To see this note that $\R^3$ in \cudo\ is naturally
  \eqn\kudo{    \R^3  = {\rm Im}\, \H =  \R {\bf i} + \R {\bf j} + \R {\bf
  k}.}
and that reflection in $\R^2 = \R{\bf i} + \R{\bf k}$  corresponds
to complex conjugation for the complex structure of $\H$ defined
by ${\bf i}$.  The fixed-point set of this is \eqn\hombo{ \R^2  =
\R + \R{\bf j} =  \C({\bf j}).} Thus it is the real part of $\H$
for the complex structure ${\bf i}$, but it has another complex
structure defined by ${\bf j}$. Taking the real part (for ${\bf
i}$ ) of the relation in \cudo\ gives the assertion \tudo, in
which the complex structure of $\C$ is given by ${\bf j}$.

   More generally the same story applies with $\H$ replaced by the ALE
spaces $M(n)$ of type $A_{n-1}$. They have a quotient map
\eqn\quomap{ M(n)/ U(1)  =  \R^3,}
with distinguished points $a_1,\dots,a_n$ . If these lie in a
plane $\R^2$ in $\R^3$ then, as for the case $n=1$, one can take
a real form of the assertion in \quomap,  giving \eqn\romap{
M^{\R}(n) / \Z_2  =  \R^2 .}
Both sides have a complex structure given by ${\bf j}$, which
exhibits $M^{ \R}(n)$ as the Riemann surface given by \eqn\fibby{
y^2 = \prod_{i=1}^n (x - a_i)}

   A very similar situation occurs with projective spaces.  Decomposing
$\H^n$ as a left $\C({\bf i})$ vector space \eqn\decas{
      \H^n
= \C^n + {\bf j} \C^n} and  taking the real part gives \eqn\tecas{
      \C^n({\bf j}) = \R^n  +  {\bf j} \R^n,}
which is a $\C({\bf j})$ vector space.   Passing to the right
projective spaces, we  get \eqn\ecas{      \CP^{n-1}({\bf j})
\subset \HP^{n-1}} as the ``real part.''   Dividing by the left
$U(1)$ in $\C({\bf i})$ gives an embedding \eqn\mecas{
\CP^{n-1}/\Z_2\subset \HP^{n-1}/U(1),} where $\Z_2$ acts as
complex conjugation.  Taking $n=3$ and using \algo\ we get an
embedding
      $\CP^2 / \Z_2  \hookrightarrow  \S^7$.

    The orbit structure of $\HP^{n-1}$ under $U(n)$ has an exact
counterpart in the orbit structure of $\CP^{n-1}$ under $O(n)$.
For $n=3$,  the $U(3)$-orbit structure of $\S^7$ in the adjoint
representation corresponds to a similar $O(3)$-orbit structure of
$\S^4$, but this time  it is in the representation of real
symmetric  $3\times 3$ matrices of trace zero.  Again the
eigenvalues $\lambda_1 \leq  \lambda_2 \leq \lambda_3$
parameterize the orbits.   The counterpart of \algo\ is the
well-known identification
       $\CP^2/\Z_2  =  \S^4,$
where $\Z_2$ acts by complex conjugation and the two special
$\RP^2$ orbits in $\S^4$ arise from $\RP^2$ in $\CP^2$ and from
$\S^2$ in $\CP^2$, the latter being the totally imaginary conic
$\sum_i z_i^2 = 0$.   It is worth noting that, removing one
$\RP^2$ from $\S^4$, we are left with the moduli space of centred
$SU(2)$-monopoles of charge 2, studied in \ref\ah{M. F. Atiyah and
N. Hitchin, {\it The Geometry and Dynamics Of Magnetic Monopoles}
(Princeton University Press, 1988).}, in which the generic orbit
is \eqn\mobo{      O(3)/O(1)^3  =  SO(3)/\Z_2\times \Z_2  =
SU(2)/\Gamma,} with $\Gamma$ of order 8, and where the remaining
$\RP^2$ represents the ``coincident'' monopoles.

    Just as $\HP^2$ is a compact eight-manifold containing the three
seven-manifolds$ X_i$ of section 3.5, so $\CP^2$ is the compact
four-manifold which contains three three-manifolds $X'_i$.  Thus
$X'_i$ is given, relative to complex homogeneous coordinates
$(z_1,z_2,z_3)$ of $\CP^2$, by the equation
 \eqn\toggo{     |z_1|  = |z_2|  \not= 0.}
$X'_i $, which can be viewed as the ``real part'' of $X_i$, is
clearly $\S^1 \times  \R^2$.   The analogue  of the statement
$X_i/U(1) = \R^6$ is now
 \eqn\oggo{     (\S^1 \times  \R^2)/\Z_2  =  \R^3,}
where $\Z_2$ acts by complex conjugation  on both factors with
$\S^0 \times  \R$ as fixed points.  Similarly the analogue of $Y
= SU(2)^3/SU(2)$ is now $Y' = U(1)^3/U(1)$, and the statement that
$Y/U(1) = \S^5$  corresponds  to \eqn\joggop{    (\S^1 \times
\S^1)/ \Z_2 = \S^2,}
exhibiting an elliptic curve as the double covering of $\S^2$ with
four branch points.  The action of the symmetric group $\Sigma_3$
on the homology group $H_3(Y)$ has its counterpart in the action
of $\Sigma_3$ on $H_1(Y')$.

   There is a parallel story for Case I, with $Y'$ being the real flag
manifold of \mobo\ and $X'$ the corresponding $\R^2$-bundle over
the Atiyah-Hitchin manifold \ah.  The involution on $\RP^2$ given
by changing the sign of one of the three coordinates induces  an
action on $X'$ with quotient $\R^4$ and $Y'/\Z_2 = \S^3$.  The
branch locus $F'$ in $\S^3$ is the union of three linked circles.

  This suggests that the analogue of Case I is given by the $\R^2$-bundle
over $\S^2$ with Euler number 4.  Taking the involution given by
reflection in the equator of $\S^2$ we find  that $X'/\Z_2 = \R^4$
and $Y'/Z_2 = \S^3$ with branch locus $F$  the union of two linked
circles.

   Note that, for the analogues of Cases I and II,  the manifolds
   $X'$
have dimension 4, while for Case III,  $X'$ is of dimension 3.

   There is even a real analogue of the Lebrun manifolds of
section 3.6.  We take the connected sum  $M'(n)$ of $n$ copies of
$\RP^2$ (giving a non-orientable manifold whose oriented cover has
genus $n-1$) and then take the appropriate $\R^2$-bundle over it
(so that its lift to the oriented cover has Euler number 4).

\newsec{$M$-Theory Curve And Four-Dimensional Gauge Theory}

In this section, we apply the methods of section 4 to the case,
considered in \refs{\acharya,\amv}, of
a manifold of $G_2$ holonomy that is asymptotic  to a cone on
$Y_\Gamma=(\S^3\times\S^3)/\Gamma$, where $\Gamma$ is a finite
subgroup of $SU(2)$.  To be more specific, if $\S^3\times\S^3$
is understood as the space of triples $(g_1,g_2,g_3)\in SU(2)$ modulo
the right action of $h\in SU(2)$, then in the definition of $Y_\Gamma$,
we will take $\Gamma$ to act on the left on $g_1$.

Three different manifolds $X_{i,\Gamma}$ of $G_2$ holonomy can be
made by ``filling in'' one of the copies of $SU(2)$.  If we fill
in the first copy of $SU(2)$ -- and so allow $g_1$ to vanish --
we get a singular manifold $X_{1,\Gamma}=\S^3\times \R^4/\Gamma$,
where the singularity arises because $\Gamma$ acts trivially at
$g_1=0$. (Here we can ``gauge away'' $g_3$ by $(g_1,g_2,g_3)\to
(g_1h,g_2h,g_3h)$ with $h=g_3^{-1}$.  Then $\S^3$ is
parameterized by $g_2$ and $\R^4/\Gamma$ by $g_1$.) If we fill in
$g_2$ or $g_3$, we get smooth manifolds $X_{2,\Gamma}$ and
$X_{3,\Gamma}$ both diffeomorphic to $\S^3/\Gamma\times\R^4$.
(For example, if one ``fills in'' $g_2$, one gauges away $g_3$,
after which $g_1$, modulo the action of $\Gamma$, parameterizes
$\S^3/\Gamma,$ and $g_2$ parameterizes $\R^4$.) Obviously, by
choosing $\Gamma$ to act in the first $SU(2)$, we have broken the
symmetry $\Sigma_3$ of permutations of the $g_i$ to
$\Sigma_2=\Z_2$, the group that exchanges $g_2$ and $g_3$.  So
$X_{2,\Gamma}$ and $X_{3,\Gamma}$ are equivalent, but
$X_{1,\Gamma}$ is completely different.

\def\N{{\cal N}}
Now consider $M$-theory on $\R^4\times X_{i,\Gamma}$,
where $\R^4$ is understood as four-dimensional Minkowski space.
What does the low energy physics look like?

In the case of $X_{1,\Gamma}$, the $A$, $D$, or $E$ singularity
$\R^4/\Gamma$ produces $A$, $D$, or $E$ gauge symmetry on the fixed
point set $\R^4\times \S^3$.  At low energies, as $\S^3$ is compact
and simply-connected, this gives $A$, $D$, or $E$ gauge theory
on the four-dimensional spacetime
$\R^4$.  To be more precise, because of the $G_2$ holonomy,
one gets in four dimensions an $\N=1$ supersymmetric gauge theory;
it is the minimal $\N=1$ theory, with only the $A$, $D$, or $E$ vector
multiplet (and no chiral multiplets).  About this theory,
there  are some standard and rather deep conjectures; it
is believed to generate
at low energies a mass gap, and to exhibit confinement, magnetic screening,
and spontaneous breaking of chiral symmetry.

$X_{2,\Gamma}$ and $X_{3,\Gamma}$ are smooth, and admit no normalizable
zero modes for supergravity fields (as the covering spaces $X_2$ and
$X_3$ have none).  So $M$-theory on $\R^4\times X_{i,\Gamma}$, $i>1$,
has no massless fields that are localized in four dimensions.
Therefore, if it is possible to smoothly interpolate from $X_{1,\Gamma}$
to $X_{i,\Gamma}$, $i>1$, then $M$-theory on $\R^4\times X_{1,\Gamma}$
likewise has no massless four-dimensional fields.
That in turn means that the minimal, ${\cal N}=1$ supersymmetric
four-dimensional gauge theory with $A$, $D$, or $E$ gauge group has
a mass gap.  This framework for explaining the mass gap was proposed
in \refs{\acharya,\amv}.  Related explanations have also been proposed
for  chiral symmetry
breaking \amv\ and confinement \achtwo.

Just as we explained in section 2.4 in the simply-connected case,
the moduli of $X_{i,\Gamma}$ behave as coupling constants (rather
than expectation values of massless fields) from a
four-dimensional point of view.   Our goal is to describe a
Riemann surface $\N_\Gamma$ that parameterizes the possible
couplings.  We will see that $\N_\Gamma$ does indeed interpolate
between the classical limits based on the various $X_{i,\Gamma}$.

\subsec{Imprint At Infinity Of Chiral Symmetry Breaking}

At first sight, there is an immediate contradiction in the claim
of a smooth interpolation from $X_{1,\Gamma}$ to the other classical
limits.

Let $h$ be the dual Coxeter number of the $A$, $D$, or $E$ group
associated with $\Gamma$.  (Thus, $h$ is $n+1$ for $A_n=SU(n+1)$,
$2n-2$ for $D_n=SO(2n)$, and $12$, $18$, or $30$ for $E_6,$ $E_7$,
or $E_8$.)  The minimal $\N=1$ supersymmetric gauge theory
has a discrete chiral symmetry group $\Z_{2h}$.
It is believed that this is spontaneously broken to $\Z_2$,
producing $h$ vacua.
The $\Z_{2h}$ is not an exact symmetry of $M$-theory on $\R^4\times
X_{1,\Gamma}$, but the explicit
breaking of $\Z_{2h}$ is unimportant in the infrared
(if the length scale of $X_{1,\Gamma}$ is large).  The $h$ supersymmetric
vacua that in the field theory limit arise from chiral symmetry
breaking are protected by supersymmetry and so should
be present in some form in $M$-theory on $\R^4\times X_{1,\Gamma}$.
\foot{A rough analogy is to a recent framework
\ref\polstrass{J. Polchinski and M. Strassler, ``The String Dual
Of A Confining Four-Dimensional Gauge Theory,'' hep-th/0003136.} for
studying the same model (for the $A_n$ case)
in the AdS/CFT correspondence.  Here
there is no exact $\Z_{2h}$ symmetry, but the $h$ supersymmetric vacua
do appear.}

Thus, if we are near $X_{1,\Gamma}$ (or more precisely in a limit
in which $X_{1,\Gamma}$ appears with large volume), it seems that
there are $h$ vacua for each point in $\N_\Gamma$. So the branch
of $\N_\Gamma$ that contains $X_{1,\Gamma}$ appears to have an
$h$-fold cover $\N_\Gamma'$ that parameterizes quantum vacua,
while $\N$ itself presumably parameterizes quantum theories or
possible values of coupling constants.

On the other hand, near
$X_{i,\Gamma}$, $i>1$, the infrared dynamics is trivial and there is
precisely one vacuum for each point in $\N_\Gamma$.  So on the branch
of $\N_\Gamma$   containing these classical limits, the moduli space of couplings
and the moduli space of quantum vacua coincide.

This qualitative difference seems to show
that $X_{1,\Gamma}$ and the other $X_{i,\Gamma}$ are contained
on different branches of $\N_\Gamma$.
However, we will claim otherwise.
We will argue that the $h$ different vacua of the gauge theory
can all be distinguished at infinity -- by the same sort of measurements
of periods and volume defects that were considered in section 4 --
and hence that on each branch of the moduli space, once the measurements
at infinity have all been made, there is only a unique quantum vacuum.

First we give a heuristic argument in this direction.  In \oldvafa, the $A_{n-1}$ case
of these models was considered, in the Type IIA language.  The vacuum
degeneracy associated with chiral
symmetry breaking was exhibited.  The model, from the Type IIA
standpoint, involves a resolved conifold singularity, of topology
 $W=\R^4\times \S^2$.  (The metric on $W$ is asymptotic to
a cone on $\S^3\times \S^2$.)   To get a dual description of $A_{n-1}$
gauge theory, $n$ units of Ramond-Ramond (RR) two-form flux were placed
on $\S^2$.  In addition, RR  four-form flux was placed on the $\R^4$
fibers of the conifold.  This last step makes sense if the conifold
singularity is embedded in a compact Calabi-Yau threefold $\hat W$;
the RR four-form flux is then Poincar\'e dual to the $\S^2$.
But  if $\hat W$ is decompactified to
the conifold $W=\R^4\times \S^2$,
 then the four-form flux spreads to infinity
(there being no normalizable harmonic four-form on $W$).
This suggests
that, when one decompactifies, some of the information about chiral symmetry
breaking is stored at infinity.

Next, we will give a simple model of what we think is happening.
We consider a supersymmetric theory on $\R^7=\R^4\times \R^3$,
with a threebrane whose worldvolume is $\R^4\times \{0\}$, with
$\{0\}$ being the origin in the second factor $\R^3$. We suppose
that on the threebrane, there are $\N=1$ supersymmetric $A$, $D$,
or $E$ vector multiplets. The fermionic components of these
multiplets will be denoted as $\lambda$. Thus, the effective
theory is four-dimensional supersymmetric gauge theory. We
suppose further that in bulk there is a complex massless scalar
field $\phi$, with a coupling on the brane of the form
\eqn\turbob{\int_{\R^4}d^4x\,\bar \phi\,\,\Tr\lambda\lambda
+{c.c.}} Then, when $\Tr\lambda\lambda$ gets a vacuum expectation
value, it will act as a source for $\phi$.  The $\phi$ field on
$\R^3$ will obey the Laplace equation with a delta function
source at the origin, proportional to
$\langle\Tr\lambda\lambda\rangle$, and so will be a multiple of
\eqn\duffo{{\Tr\,\lambda\lambda\over |\vec x|},} where $|\vec x|$
is the distance from the origin in $\R^3$. Since the function
$1/|\vec x|$ is not square-integrable in three dimensions, the
coefficient of $1/|\vec x|$ is part of what we would regard as a
``coupling constant,'' and parameterize by a point in $\N_\Gamma$.
As this coefficient is $\Tr\,\lambda\lambda$,
 in this simple model, the fields at infinity
are different for the different chiral vacua.

Finally, we will try to show in a precise way that something just
like this happens in $M$-theory on $\R^4\times X_{1,\Gamma}$,
but with the $M$-theory three-form field $C$
playing the role of $\phi$.  The analysis is very similar
to the membrane instanton computation in section 4.4, and we will
be schematic.

In this spacetime, the gauge fields, whose curvature we will call $F$,
 are supported on the locus of the singularity, which is $\R^4\times\S^3
\subset \R^4\times X_{1,\Gamma}$.
There is a coupling
\eqn\inop{\int_{\R^4\times \S^3}C\wedge \Tr\,  F\wedge F,}
which is familiar principally because it causes instantons to carry
a membrane charge.  The contribution to \inop\ that we are really interested
in for the present is the term in which $C$ is integrated on $\S^3$ and
$\Tr\,F\wedge F$ is integrated on $\R^4$.
In four-dimensional supersymmetric gauge theory,
$\Tr\,F\wedge F$ arises as the imaginary part of
 a superspace interaction $\int d^2\theta
\,\Tr\,W_\alpha W^\alpha$, where $d^2\theta$ is the chiral measure
in superspace and $W_\alpha$ is the superfield that contains $\lambda$
and $F$.  The supersymmetric extension of \inop\ is thus
the imaginary part of
 $\int_{\R^4\times\S^3}\int d^2\theta \,\,C\cdot \Tr W_\alpha
W^\alpha \cdot \epsilon$ (with $\epsilon$ the volume form of
$\R^4$). \foot{Here $C$ really should be completed to a chiral
object, rather as we did in section 4.4.} In section 4.4, we
learned that $\int d^2\theta \,C$ has a contribution proportional
to $*G$. So in the present context, evaluating the $d^2\theta$
integral generates a coupling \eqn\tinop{\int_{\R^4\times
\S^3}*G\cdot \Tr\,\lambda\lambda,} which is part of the
supersymmetric completion of \inop. (The detailed analysis of
what components of $G$ enter is just as in the membrane instanton
calculation.)

Once we assume that $\Tr\,\lambda\lambda$ has a vacuum expectation value
and replace it by a $c$-number, \tinop\ is the same interaction that
we arrived at in section 4.4.
The derivation was a little different in that case, of course.
At any rate, since the effective interaction is the same,
the rest of the analysis is the same.  The
addition of the term \tinop\ to the Lagrangian shifts the value of
$\int_{D_1}C$ as measured at infinity, just as in section 4.4.

The upshot is that the different vacua resulting from chiral symmetry
breaking can be distinguished at infinity because of the way that
the eleven-dimensional massless fields respond to four-dimensional
chiral symmetry breaking.  This information will enable us to proceed,
even though, at some level, it is  not entirely attractive.  It prevents a
clean separation between ``problems'' or ``couplings'' specified or measured
at infinity, and ``answers'' resulting from the quantum dynamics in the
interior.   Instead, what is measured at infinity
is a mixture of what a four-dimensional physicist would usually
regard as a ``coupling constant'' and what such a physicist would
 regard as a dynamically generated answer.

\subsec{First Approach To The Curve}

In section 2.5, we defined three-cycles $D_i'$, $i=1,2,3$, that
generate the homology of $Y_\Gamma$.  They obey
\eqn\ploop{ND_1'+D_2'+D_3'=0}
in $H_3(Y_\Gamma;\Z)$.  Here $N$ is the order of the finite group $\Gamma$.
The periods of the $C$-field at infinity are defined naively as
\eqn\oloop{\alpha'_i=\int_{D_i'}C,}
though some more care is needed because of a fermion anomaly that
was explained in section 5.
They obey
\eqn\kloop{N\alpha'_1+\alpha'_2+\alpha'_3=N\pi,}
where the contribution $N\pi$ comes from the anomaly and
was explained in section 5.
The $\alpha_i'$ can be measured by an observer at infinity.

As in section 4, we complete the $\alpha_i'$
to a set of holomorphic observables by considering
also the metric parameters $f_i$ that appear in \turigo.
(This metric is $\Gamma$-invariant, so it makes sense in the present
context where we are dividing by $\Gamma$.)
To find the holomorphic combinations of the $f_i$ and $\alpha'_j$,
it is convenient to work on the $N$-fold cover $Y$ of $Y_\Gamma$.
Pulling back $C$ to $Y$, its periods $\alpha_i=\int_{D_i}C$ (with
$D_i$ as defined in section 2.5)
obey
\eqn\ununu{\alpha_1=N\alpha_1',~~\alpha_i=\alpha'_i ~{\rm for}~i>1.}
This is because $D_1$ projects to $ND_1'$, and $D_i,$ $i>1$,
projects to $D_i'$.  The holomorphic combinations are found by
making the substitutions \ununu\ in the holomorphic combinations
found in section 4.2, and hence are
 $kf_1+i(\alpha'_2-\alpha'_3)$,
$kf_2+i(\alpha'_3-N\alpha'_1)$, and $kf_3+i(N\alpha'_1-\alpha'_2)$.
The single-valued holomorphic functions are
\eqn\murko{\eqalign{\eta_1 & = \exp\left({2k\over 3N}f_3+{k\over 3N}f_1
+i\alpha'_1\right)   \cr
\eta_i & = \exp\left({2k\over 3}f_{i-1}+{k\over 3}f_i+i\alpha_i'\right),~
{\rm for}~i>1.\cr}}
They obey
\eqn\huvigo{\eta_1^N\eta_2\eta_3=(-1)^N.}

The $\Z_2$ symmetry that exchanges $g_2$ and $g_3$ should act
on these variables by
\eqn\olpo{(\eta_1,\eta_2,\eta_3)\to (\eta_1^{-1},\eta_3^{-1},\eta_2^{-1}).}
The antiholomorphic symmetry that comes from a parity reflection in
space acts by
\eqn\tolpo{\eta_i\to \bar\eta_i.}

We assume that $\N_\Gamma$ has distinguished points $P_i$, $i=1,2,3$
corresponding to classical limits with the spacetimes $X_{i,\Gamma}$.
On $X_{i,\Gamma}$, $\alpha_i'$ vanishes (since $D_i'$ is ``filled in'').
The modulus of $\eta_i$ is 1 at $P_i$ because of the reasoning in section
4 about the volume defects $f_i$ in the metric \turigo.
(Dividing by the finite group $\Gamma$ does not change the ratios
of the $f_i$, which remain $(-2,1,1)$ up to permutation.)
So just as in section 4, we have $\eta_i=1$ at $P_i$.

Near $P_i$, the $f$'s diverge to $\pm\infty$ exactly
as in section 4.  This causes $\eta_{i-1}$ to have a pole and
$\eta_{i+1}$ to have a zero at $P_i$ for the same reasons as there.  However,
the orders of the zeroes and poles may be different, because of factors
of $N$ in the above formulas and because of chiral symmetry breaking.
For $i>1$, we can proceed precisely as in section 4.3.  At the
``center'' of $X_{i,\Gamma}=\S^3/\Gamma\times \R^4$ is a three-cycle
$Q_i'\cong \S^3/\Gamma$.  The membrane instanton amplitude is
\eqn\memamp{u=\exp\left(-TV(Q_i')+i\int_{Q_i'}C\right).}  It is a local
parameter at $P_i$.   The cycle $D_1'$ is contractible in $X_{i,\Gamma}$,
$i>1$, to $\pm Q_i'$, where the sign depends on orientations.
Hence $\alpha_1'=\pm \int_{Q_i'}C$ near $P_i$.  Since its argument
is $\pm 1$ times that of the local parameter $u$,  $\eta_1$
has a simple zero or pole at $P_2$ and $P_3$.  Since we know that
$\eta_1$ has a pole at $P_2$ and a zero at $P_3$, these are in fact
a simple pole and a simple zero.

If the $P_i$ are the only points at which the $f_i$ diverge to $\pm \infty$,
then $\eta_1$ has only the one simple zero and simple pole that we have
just found.  Existence of a holomorphic function $\eta_1$ with only
one zero and one pole implies that $\N_\Gamma$ is of genus zero,
and we can in fact identify $\N_\Gamma$ as the complex $\eta_1$ plane,
including the point at infinity.

 In section 4.3, we introduced an auxiliary parameter
$t$ to make the triality symmetry manifest, but there is no
triality symmetry in the present problem and we may as well
parameterize the moduli space via $\eta_1$.  The global
symmetries \olpo\ and \tolpo\ act by $\eta_1\to\eta_1^{-1}$ and
$\eta_1\to\bar\eta_1$; they can be perfectly clear in a
description with $\eta_1$ as the parameter. Taking $\eta_1$ as
the parameter and using our knowledge of how it behaves at the
$P_i$, we simply identify $P_1$, $P_2$, and $P_3$ as the points
$\eta_1=1,$ $\infty$, and $0$, respectively.

Since $\eta_1^N\eta_3=(-1)^N$ near $P_2$, and $\eta_1^N\eta_2=(-1)^N$
near $P_3$, it follows that $\eta_3$ has a zero of order $N$ at $P_2$,
and $\eta_2$ has a pole of order $N$ at $P_3$.  What happens at $P_1$?
We know already that $\eta_2$ has a zero
and $\eta_3$ has a pole at $P_1$, but to determine the orders of these
zeroes and poles,
 we must be careful.  With $Q_1'$ understood as the $\S^3$ at the
``center'' of $X_{1,\Gamma}=\S^3\times \R^4/\Gamma$, the quantity
$u$ defined in \memamp\ is the membrane instanton amplitude, but
it is not a good local parameter for the curve $\N_\Gamma$ at
$P_1$. In fact, the membrane instanton in this singular geometry
is equivalent to a (pointlike) Yang-Mills instanton in the
four-dimensional supersymmetric $A$, $D$, or $E$ gauge theory.
Chiral symmetry breaking means that the gluino condensate is
proportional to $u^{1/h}$, and hence that a local parameter in a
curve that parameterizes quantum vacua must be not $u$ but
$u^{1/h}$.  We have argued that $\N_\Gamma$ is such a curve, so
$u^{1/h}$ is a good local parameter for $\N_\Gamma$ near $P_1$.
$\eta_2$ and $\eta_3$
 are proportional to $u^{\pm 1}$ near $P_1$, by the same reasoning
as in section 4.3 (the
three-spheres $D_2'$ and $D_3'$ are contractible to $Q_1'$ in $X_{1,\Gamma}$).
They are thus proportional to the $\pm h$ power of the local parameter
$u^{1/h}$,  so finally
(since we know $\eta_2$ vanishes at $P_1$ and $\eta_3$ diverges)
$\eta_2$ has a zero of order $h$ at $P_1$ and $\eta_3$ has a pole of
order $h$.  The results are summarized in the table.

\input tables
\centerline{\bf Table 1}
\bigskip
\begintable
|~~~~$P_1$~~~~|~~~~$P_2$~~~~|~~~~$P_3$~~~~\crthick
~~~$\eta_1$~~~|$1$|$\infty$|$0$\cr
~~~$\eta_2$~~~|$0^h$|$1$|$\infty^N$\cr
~~~$\eta_3$~~~|$\infty^h$|$0^N$|$1$\endtable
\centerline{\ninerm
This table shows the behavior of the $\eta_i$ at the special points
$P_j$. }\centerline{\ninerm  $\infty^N$ denotes a pole of order $N$, etc.}

\bigskip

Now let us see if we can find $\eta_2$ and $\eta_3$ as functions of
$\eta_1$.  From what we have seen so far, $\eta_2$, for example,
has a zero of order $h$ at $P_1$,  equals 1 at $P_2$, and has a pole
of order $N$ at $P_3$.  If $h=N$, we can proceed on the assumption
that these are the only zeroes and poles; for $h\not=N$ there
must be additional singularities.

Among the $A$, $D$, and $E$ groups, $h=N$ for and only for the  $A$ series.
Indeed, for $SU(N)$, $h$ and the order of $\Gamma$ both equal $N$.
So for the moment, we limit ourselves to $SU(N)$.  $\eta_2$ and $\eta_3$
are uniquely determined by the facts stated in the last paragraph to
be
\eqn\inono{\eqalign{\eta_2 & = \eta_1^{-N}{(\eta_1-1)^N} \cr
                    \eta_3 & =  { (1-\eta_1)^{-N}}.\cr}}
Here, $\eta_2$ was determined by requiring that it have a zero
of order $N$ at $\eta_1=1$ (which we have identified with $P_1$),
equals 1 at $\eta_1=\infty$ (which is $P_2$), and has a pole of
order $N$ at $\eta_1=0$ (which is $P_3$).  Similarly, $\eta_3$
was determined by requiring that it have a pole of order $N$ at
$\eta_1=1$ (or $P_1$), a zero of order $N$ at $\eta_1=\infty$ (or $P_2$),
and equals 1 at $\eta_1=0$ (or $P_3$).

This curve has all the expected properties.  For example,  $\eta_1^N\eta_2\eta_3=(-1)^N$.  Likewise, under $\eta_1\to \eta_1^{-1}$,
$\eta_2 $ and $\eta_3$ are mapped to $\eta_3^{-1}$ and $\eta_2^{-1}$.
This is the expected discrete symmetry \olpo.  Finally, under
$\eta_1\to\bar\eta_1$, we have also $\eta_i\to\bar\eta_i$ for $i>1$,
which is the antiholomorphic symmetry \tolpo.

\subsec{Extension To $D_n$ Groups}

Our next task is to generalize this result to groups of type $D$ or $E$.
We must find additional semiclassical limits contributing zeroes and poles of $\eta_2$ and $\eta_3$,
since $h\not= N$ for these groups.
It will turn out that $\eta_1$ does not have additional zeroes or poles.

\nref\landlop{K. Landsteiner and E. Lopez, ``New Curves From
Branes,'' Nucl. Phys. {\bf B516} (1998) 273,
hep-th/9708118.}%
\nref\wittold{E. Witten, ``Toroidal Compactification Without Vector
Structure,' JHEP {\bf 9802:006} 1998, hep-th/9712028.}%
\nref\mikhailov{A. Mikhailov, ``Momentum Lattice For CHL String,''
Nucl. Phys. {\bf B534} (1998) 612. hep-th/9806030.}%
We consider first the $D_n$ groups because in this case the extra
semiclassical limit has a relatively familiar origin. In
considering the $D_n$ singularities and associated groups, we can
assume that $n\geq 4$, since $D_3$, for example, coincides with
the group $A_3$. There are in fact for $n\geq 4$ two different
types of $D_n$ singularity in $M$-theory
\refs{\landlop-\mikhailov}.  The more familiar one gives a gauge
group $D_n=SO(2n)$, and the less familiar $D_n$ singularity gives
a gauge group $Sp(n-4)$.  The familiar $D_n$ singularity can be
deformed away, a process that corresponds to Higgsing of the
$SO(2n)$ gauge symmetry down to an abelian group. For the exotic
$D_n$ singularity, the Higgsing process, or equivalently the
deformation of the singularity, can proceed as long as there is
nonabelian gauge symmetry,\foot{This assertion can be deduced
from supersymmetry.  Supersymmetry in seven dimensions relates
gauge fields to scalar fields in the adjoint representation. The
scalars can receive expectation values, breaking the group
(generically) to an abelian subgroup.} that is as long as
$n-4\geq 1$.  But when we reach $n=4$, the gauge symmetry
$Sp(n-4)$ becomes trivial; we remain with a ``frozen'' $D_4$
singularity that cannot be deformed or resolved.

\def\O{{\bf O}}
The existence of a second type of $D_n$ singularity has a simple
explanation from the point of view of Type IIA superstring theory.
In that theory, we can get $D_n$ gauge symmetry in
dimension seven by considering an $\O 6^-$ orientifold plane with
$n$ pairs of $D6$-branes.  The $D6$-brane charge, including
the charge $-2$ of the orientifold plane, is $n-2$.  This object
lifts in $M$-theory to a spacetime with a $D_n$ singularity.

We can make another Type IIA configuration that looks the same at
infinity and in particular has the same
$D6$-brane charge by taking an $\O 6^+$ orientifold plane (whose charge
is $+2$) with $n-4$
pairs of $D6$-branes.  In this case the gauge group is $C_{n-4}=Sp(n-4)$.
Since this Type IIA spacetime looks the same except for microscopic
details near the singularity, it lifts in $M$-theory to an object
that has the same $D_n$ singularity, though the microscopic details
of the physics near the singularity are different.  This is the $D_n$
singularity in $M$-theory that has gauge group $Sp(n-4)$.

Now we want to find a new semiclassical limit for $M$-theory on a
manifold asymptotic to a cone on $Y_\Gamma$, where $\Gamma$ is of
type $D$. Our basic idea is that the new limit comes from the
manifold $X_{1,\Gamma}=\S^3\times \R^4/\Gamma$, but now with the
singularity $\R^4/\Gamma$ giving $Sp(n-4)$ gauge symmetry rather
than $D_n$. Thus, $\N_\Gamma$ will interpolate for $D_n$ between
four different classical limits: $X_{1,\Gamma}$ with the two
different kinds of $D_n$ singularity and also $X_{2,\Gamma}$ and
$X_{3,\Gamma}$.

To see just where the new version of $X_{1,\Gamma}$ should come
in, it helps to use the following fact about orientifolds. In
Type IIA, there is an NS two-form field $B$. $B$ is odd under a
reversal of orientation of the string worldsheet. So in an
orientifold spacetime, $B$ is actually twisted by the orientation
bundle of spacetime and can be integrated over an unorientable
cycle $P\cong{\RP}^2$ that encloses the orientifold plane. We
have \ref\witbar{E. Witten, ``Baryons And Branes In Anti de Sitter
Space,'' JHEP {\bf 9807:006} 1998, hep-th/9805112.}
\eqn\kilobb{\eqalign{\int_PB&= 0~{\rm for}~\O 6^-\cr
                     \int_PB&= \pi~{\rm for}~\O 6^+.\cr}}
In going to $M$-theory, $P$ lifts to a three-cycle $\S^3/\Gamma\subset\R^4/\Gamma$
that encloses the singularity, and $B$ lifts to the $M$-theory three-form
field $C$.
So we find that
\eqn\rigob{\int_{\S^3/\Gamma}C=0}
for a $D_n$ singularity that gives $D_n$ gauge symmetry, but
\eqn\libop{\int_{\S^3/\Gamma}C=\pi}
for a $D_n$ singularity that gives $C_{n-4}$ gauge symmetry.

If we apply this to $M$-theory on $X_{1,\Gamma}=\S^3\times \R^4/\Gamma$,
an $\S^3/\Gamma$ that encloses the singularity is our friend
 $D_1'$.  So the argument of $\eta_1$, namely $\alpha_1'=\int_{D_1'}C$, should
be 0 or $\pi$ for the two cases in the  semiclassical limit on
$X_{1,\Gamma}$.  The modulus of $\eta_1$ is 1 in either case
since the reasoning of section 4 based on the behavior of the
volume defects is applicable.
  This means that $\eta_1$ must
be 1 in the large volume limit on $X_{1,\Gamma}$ if the gauge
symmetry is $D_n$, but it must be $-1$ if the gauge symmetry is
$C_{n-4}$.

The last statement, in particular, means that $\eta_1$ does not
have a zero or pole in the new semiclassical limit.
That is just as well, since we had no problem with the zeroes or poles
of $\eta_1$.  But we must investigate the behavior of $\eta_2 $  and
$\eta_3$ at the new limit point, which we now identify as $\eta_1=-1$.

Let again $u$ be the  amplitude for a membrane wrapped on the
supersymmetric cycle $Q_1\cong\S^3\subset X_{1,\Gamma}$. One would
expect at first sight that $u$ would be the instanton amplitude
of the $Sp(n-4)$ gauge theory.  If so, as in section 6.2, a local
parameter near $\eta_1=-1$ would be $u^{1/h'}$, where $h'$ is the
dual Coxeter number of $Sp(n-4)$ and the $h'$ root results from
chiral symmetry breaking. Actually, as we explain presently, at
$\eta_1=-1$, the wrapped membrane is equivalent to {\it two}
$Sp(n-4)$ instantons.  So the amplitude for one instanton is
$u^{1/2}$ and (taking the $h'$ root) the local parameter on a
curve $\N_\Gamma$ that parameterizes quantum vacua is
$u^{1/2h'}$.  Since $\eta_2$ and $\eta_3$ are proportional to
$u^{\pm 1}$ in a classical limit of $X_{1,\Gamma}$ (the derivation
of this statement in section 4 just involved the behavior of the
metric parameters $f_i$ and not the microscopic properties of the
singularity), it follows that at $\eta_1=-1$, $\eta_2$ has a zero
of order $2h'$, and $\eta_3$ has a pole of the same order.

So far we know for $\eta_2$ a pole of order $N$ and zeroes of order
$h$ and $2h'$, while for $\eta_3$, the orders of zeroes and poles
are exchanged.  A happy fact now is that for $D_n$
\eqn\pokoo{N=h+2h'.}
In fact, $N=4n-8$ for $D_n$, while $h=2n-2$, and for $Sp(n-4)$,
$h'=n-3$.  So we can assume that
the four semiclassical limits that we have identified give the
complete story.

On this basis, we can determine $\eta_2$ and $\eta_3$.  Requiring
them to have zeroes and poles of the specified order at $\eta_1=0,1,-1$
and $\infty$,  and also requiring $\eta_2=1$ at $\eta_1=\infty$
and $\eta_3=1$ at $\eta_1=0$, we get
\eqn\mokoo{\eqalign{\eta_2 & = {\eta_1}^{-N}{(\eta_1-1)^h(\eta_1+1)^{2h'}}\cr
\eta_3 & = { (\eta_1-1)^{-h}(\eta_1+1)^{-2h'}}.\cr}}
All the expected properties are satisfied.  For example,
$\eta_1^N\eta_2\eta_3=1$ (note that $(-1)^N=1$ for $D_n$, and also
for the $E$ groups considered in section 6.4), and the global symmetries
\olpo\ and \tolpo\ are clear.

It remains to explain why for $Sp(n-4)$ a membrane is equivalent to
two instantons.  Consider in Type IIA superstring theory
a system of $s$ $D6$-branes away from
an $\O 6^\pm$ orientifold plane.  The gauge group is $U(s)$.  The RR
fields couple to the $U(s)$ gauge fields by
\eqn\ikopo{\int C\wedge {\rm ch}(F)}
where ${\rm ch}(F)$ is the Chern character of the $U(s)$ gauge
fields (represented as a polynomial in the curvature $F$).
Because of this coupling, a $U(s)$ instanton has a membrane charge
of one.

Now, let the $D6$-branes approach an $\O 6^-$ plane. The gauge group
is enhanced from $U(s)$ to $D_s=SO(2s)$.  An instanton number one
gauge field in $U(s)$ has instanton number one when embedded in $SO(2s)$.
So a membrane corresponds to a single instanton of $D_s$; that is why,
for $D_s$, the membrane amplitude $u$ is the same as the instanton amplitude.

What happens if instead the $D6$-branes approach an $\O 6^+$ plane?
The gauge group is enhanced from $U(s)$ to $Sp(s)$.  A $U(s)$ instanton
has instanton number {\it two} when embedded in $Sp(s)$.\foot{When
we embed $U(s)$ in $Sp(s)$, the $2s$-dimensional representation $V$
of $Sp(s)$ transforms as ${\bf s}\oplus \bar{\bf s}$ of $U(s)$.
So under a minimal $SU(2)$ subgroup of $U(s)$, in which we can embed
a field of $U(s)$ instanton number one, $V$ contains
{\it two} copies of the ${\bf 2}$ of $SU(2)$, plus singlets.  That is
why its instanton number is two in $Sp(s)$.  An instanton number
one field in $Sp(s)$ can be made by embedding instanton number
one in a subgroup $SU(2)\cong Sp(1)\subset Sp(s)$, such that $V$ decomposes
as {\it one} copy of the ${\bf 2}$ of $SU(2)$, plus singlets.}
So a gauge field that has membrane charge one has $Sp(s)$ instanton
number two, as we claimed.  That is why, if the membrane amplitude
is $u$, the amplitude for an $Sp(s)$ instanton is $u^{1/2}$.

\bigskip\noindent{\it A Comment}

In Type IIA language, we have learned that an $\O 6^+$ plane wrapped on $\S^3$
(perhaps with some $D6$-branes) can be continuously deformed to
an $\O 6^-$ plane wrapped on $\S^3$ (with four more $D6$-branes).

One might ask whether a similar deformation can be made before wrapping
on $\S^3$.  The answer is ``no.''  An $\O6^-$ plane of worldvolume
$\R^7$ cannot be deformed smoothly to an $\O6^+$ plane of the same worldvolume,
because they give different gauge symmetries.  Gauge theory in seven
dimensions is infrared-free, and the gauge group can be determined
simply by detecting the massless particles and measuring their low
energy interactions.  Similarly, an $\O6^-$ plane wrapped on
a manifold $M$ of dimension $n<3$ (and so compactified down to $7-n>4$
dimensions) cannot be smoothly deformed to an $\O6^+$ plane.
Three is the smallest number of compact dimensions for which something
like this can happen to an $\O6$ plane,
because four is the largest dimension in which
gauge interactions are non-trivial in the infrared.

\subsec{The $E$ Series}

For any finite subgroup $\Gamma$ of $SU(2)$, let $k_i$ be the dimensions
of the irreducible representations of $\Gamma$.  By a well known fact
about finite groups, the order $N$ of $\Gamma$ is
\eqn\popol{N=\sum_ik_i^2.}
The $k_i$ are
the Dynkin indices of the $A$, $D$, or $E$ group associated with $\Gamma$.
(They are the  usual labels on the nodes of the extended Dynkin diagram
of $G$; alternatively, they
are 1 and the coefficients of the highest root of $G$ when expanded
in simple positive roots.)  It can be shown that the dual Coxeter number
of $G$ is
\eqn\nopol{h=\sum_ik_i.}
Comparing the two formulas, we see we will have $N=h$ if and only if
\eqn\jopol{\sum_ik_i=\sum_ik_i^2.}
that is, if and only if the $k_i$ are all one.  Among the $A$, $D$, and $E$
groups that is true only for the $A$ series, which is why we need
new semiclassical limits for the $D$ and $E$ groups.

A similar problem appeared in studying $\Tr\,(-1)^F$ in the minimal
supersymmetric gauge theory in four dimensions with gauge group $G$
(the same theory that also arises in our present problem).
The index was computed \ref\witold{E. Witten,
``Constraints On Supersymmetry Breaking,'' Nucl. Phys.
{\bf B202} (1982) 253.} by compactifying the spatial
directions to a three-torus $\T^3$, and then performing some quantum mechanics
on the space of flat connections.  A flat connection on $\T^3$
has holonomies
that are three commuting elements $(g_1,g_2,g_3)\in G$; we refer to
a triple of such elements as a commuting triple.
By assuming that the $g_i$ could be conjugated to the maximal torus
in $G$ and then quantizing, it seemed that $\Tr\,(-1)^F$ it should
equal $r+1$, where $r$ is the rank of $G$.  From the point of view of
chiral symmetry breaking, one expects $\Tr\,(-1)^F=h$.  Since
$r+1$ is the same as the number of nodes on the extended Dynkin diagram,
the formula one needs here is
\eqn\hopol{\sum_i 1 =\sum_ik_i.}
This formula, which has an obvious resemblance to \jopol, is again
valid if and only if the $k_i$ are all one.
This contradiction was eventually resolved
\nref\witres{E. Witten, ``Supersymmetric Index In Four-Dimensional
Gauge Theories,'' hep-th/0006010.}%
\refs{\wittold,\witres} by observing that the moduli space of commuting
triples has several components.  Each component has a rank $r_a$
(the rank of the subgroup of $G$ that commutes with the triple) and
the formula for $\Tr\,(-1)^F$ is $\sum_a(r_a+1)$, generalizing
the naive $r+1$ that arises if one assumes that every commuting
triple can be conjugated to the maximal torus.
So one expects
\eqn\tumigo{\sum_ik_i=\sum_a(r_a+1).}
By now
there is an extensive literature  on commuting triples
\nref\keur{A. Keurentjes, A. Rosly, and A. V. Smilga,
``Isolated Vacua In Supersymmetric Yang-Mills Theories,''
hep-th/9805183.}%
\nref\newkeur{A. Keurentjes, ``Non-Trivial Flat Connections On The
3-Torus I: $G_2$ And The Orthogonal Groups,'' hep-th/9901154;
``Non-Trivial Flat Connections On The 3-Torus, II,'' hep-th/9902186.}%
\nref\kac{V. G. Kac and A. V. Smilga, ``Vacuum Structure In Supersymmetric
Yang-Mills Theory With Any Gauge Group,'' hep-th/9902029.}%
\nref\bfm{A. Borel, R. Freedman, and J. Morgan, ``Almost
Commuting Elements In Compact Lie Groups,'' math.gr/9907007.}%
\refs{\keur -\bfm} with proofs of this assertion.

Commuting triples are related to  $A-D-E$ singularities in $M$-theory with
reduced rank gauge group
for the following reason \sethietal.
Consider the $E_8\times E_8$
heterotic string on  $\T^3$.
A commuting triple in $E_8\times E_8$ (with a condition on the Chern-Simons
invariant) gives a heterotic string vacuum with unbroken supersymmetry.
If the triple cannot be conjugated to the maximal torus, the rank of the
unbroken gauge group
and the dimension of the moduli space are reduced relative to the usual value.
In terms of the duality between the heterotic string on $\T^3$
and $M$-theory on K3, the reduction in dimension of the moduli space
is achieved by constraining the K3 surface to have $A$, $D$, or $E$
singularities that generate a gauge group of reduced rank.

{}From this point of view, for every component of the moduli space
of commuting triples in a group of type $A$, $D$, or $E$, there is
a corresponding $A$, $D$, or $E$ singularity in $M$-theory. For
example, for $A_n$, every commuting triple can be conjugated to
the maximal torus, so there is only one kind of $A_n$ singularity
in $M$-theory.  However, for $D_n$, there are two components of
the moduli space of commuting triples \wittold, so there are two
types of $D_n$ singularity in $M$-theory.  These two possibilities
were used in section 6.3.  For $E_6$, $E_7$, and $E_8$, the number
of components of the moduli space of commuting triples is
respectively $4,$ $6$, or 12; these are also the numbers of
different kinds of $M$-theory singularities of type $E_6$, $E_7$,
and $E_8$.

For the $E$-groups, the singular manifold $X_{1,\Gamma}$ thus represents
4, 6, or 12 possible semiclassical limits
 in $M$-theory.  Our proposal is that
the curve $\N_\Gamma$ interpolates between all of these possibilities
as well as two more classical limits corresponding to $X_{2,\Gamma}$
and $X_{3,\Gamma}$.  To implement this proposal, we need to recall
(from \bfm\ and \sethietal)
some facts about commuting triples.

For every simple Lie group $G$, every component of the moduli space
of commuting triples is associated with an integer $t$ that divides
some of the $k_i$.  (One way to characterize $t$ is that it is the least
order of any one of the commuting elements $(g_1,g_2,g_3)$ in a given
component of commuting triples.)  For $A_n$, $t$ must be 1
(as all $k_i=1$ for $A_n$), for $D_n$, $t\leq 2$ (as the $k_i$
are all 1 or 2), and  for $E_6,E_7, $ or $E_8$, the possible
values are $t\leq 3$, $\leq 4$, and $\leq 6$, respectively.
  We will call $t$ the index
of the component or the associated singularity.
For every $t$ with $2\leq t\leq 6$,
there is a unique $A$, $D$, or $E$  group of lowest rank for which $t$ divides
some of the $k_i$.  It is $D_4$ for $t=2$, $E_6$ for $t=3$,
$E_7$ for $t=4$, and $E_8 $ for $t=5,6$.

For these values of $t$ and
these groups, there is a ``frozen''
$M$-theory singularity that produces no gauge symmetry at all.
Thus, frozen singularities appear for $D_4$, $E_6$, $E_7$, and $E_8$.
By adjusting the moduli of a K3 surface that contains a frozen singularity,
one can enhance the singularity to one of higher rank.
If the rank
of the singularity
is increased by $s$, one gets gauge symmetry of rank $s$.
(As we explain below, a frozen singularity is not completely
classified by $t$ if $t>2$.  However, the gauge symmetry that develops
upon further enhancement of the singularity depends only on $t$.)
For example, a frozen $D_4$ singularity that is enhanced to $D_{4+s}$
gives $C_s=Sp(s)$ gauge symmetry (this case appeared in section 6.3),
while if it is enhanced to $E_6$, $E_7$, or $E_8$, the resulting
gauge symmetry is $C_2=Sp(2)$, $B_3=SO(7)$, or $F_4$.  These
last statements can be derived using a generalization of the Narain lattice
that is adapted to this problem \mikhailov\ and
have also been derived in $F$-theory \sethietal.
A frozen $E_6$ singularity that is enhanced to $E_7$ or $E_8$ gives
$SU(2)$ or $G_2$ gauge symmetry, respectively,
and a frozen $E_7$ singularity that
is enhanced to $E_8$ gives $SU(2)$ gauge symmetry \sethietal.

In each case, the gauge symmetry associated with a given $A$, $D$, or
$E$ singularity is strongly constrained by the following fact.
Let $G$ be a group of type $A$, $D$, or $E$.  Let $t$ be a positive
integer that divides some of the Dynkin indices $k_i$
of $G$. Let $J_t$ be the set of values of $i$
such that $k_i$ is divisible by $t$, and let $\# J_t$ be the number of
its elements.
Let  $K_t$ be the gauge group
in  $M$-theory
at a $G$-singularity of index $t$.
Then the Dynkin indices of $K_t$ are $k_i/t$, for all $i\in J_t$.
The dual Coxeter number $h_t$ of $K_t$ is the sum of the Dynkin indices
so
\eqn\jiho{h_t={1\over t}\sum_{i\in J_t}k_i.}
The statement that $k_i/t$ are the Dynkin indices of $K_t$ implies,
in particular, that the rank $r_t$ of $K_t$ obeys
\eqn\iho{r_t+1=\# J_t.}
 As an example,
take $G=E_8$ and $t=3$.  The $k_i$ for $E_8$ are $1,2,2,3,3,4,4,5,6$,
so the $k_i$ that are divisible by 3 are
 $3,3,6$ and after dividing them by 3, we get $1,1,2$ for the Dynkin indices
of $K_3$.  The only Lie group with precisely
these Dynkin indices is  $G_2$, so
an $E_8$ singularity of index $t=3$ gives $G_2$ gauge symmetry.
A frozen $E_6$ singularity has $t=3$, so an equivalent statement is that if
a frozen $E_6$ singularity is enhanced to $E_8$, the gauge symmetry
that is generated is $G_2$.

A component of the moduli space of commuting triples in $G$, or equivalently
an $M$-theory singularity of type $G$, is not completely classified by the value
of $t$ (though the group $K_t$ does depend only on $t$).  The remaining information one needs is the Chern-Simons invariant of the flat bundle on $\T^3$
that is determined by the commuting triple.  As  proved in \bfm\ (in accord with
conjectures in \witres), the values of the Chern-Simons invariants for flat bundles with a given  $t$ are
\eqn\hinin{{2\pi \mu\over t},}
where if $t\geq 2$,
$\mu$ runs over all positive integers less than $t$ that are prime to $t$,
 and $\mu=0$ if $t=1$.
We let $\phi(t)$ be the number of such $\mu$.
$t$ and $\mu$ give a complete classification of the components of the moduli
space of commuting triples.   With this information and \iho, \tumigo\ can be
deduced from the elementary number theory formula
\eqn\pinin{k=\sum_{t|k}\phi(t),}
for all positive integers $k$.  One simply writes
\eqn\ilpino{\sum_ik_i=\sum_i\sum_{t|k_i}\phi(t)=\sum_t\sum_{i\in J_t}\phi(t)
=\sum_{t,\mu}\sum_{i\in J_t}1=\sum_{t,\mu}\# J_t=\sum_{t,\mu}(r_{t}+1).}

According to \sethietal, for the $M$-theory singularity associated
with a commuting triple, the Chern-Simons invariant has the following
interpretation.  The period of the $C$-field over a copy of $\S^3/\Gamma$
that encloses the $\R^4/\Gamma$ singularity is
\eqn\polpol{\int_{\S^3/\Gamma}C={2\pi\mu\over t}.}
In our application, we can interpret this
as the period of the $C$-field on the cycle $D_1'\cong\S^3/\Gamma$
in $X_{1,\Gamma}=\S^3\times\R^4/\Gamma$.  In other words, it is the
argument of $\eta_1$ in the classical limit.  The modulus
of $\eta_1$ is 1 in the classical limit because of the behavior
of the volume defects as studied in section 4.
So in the construction of the curve $\N_\Gamma$, the only value of $\eta_1$
at which a classical limit $X_{1,\Gamma}$ may appear with
 given $\mu$ and $t$  is
\eqn\ikolo{\eta_1=\exp(2\pi i\mu/t).}
We will assume that all the possible classical limits corresponding
to all values of $t$ and $\mu$ (for given $\Gamma$)  do appear.

To determine the curve $\N_\Gamma$, we
still need to know how $\eta_2$ and $\eta_3$
behave at the distinguished values of $\eta_1$ corresponding
to the new classical limits.  Recalling that $K_t$ is the gauge
group in a classical limit with a given $t$,   we will assume
that a membrane wrapped on $\S^3\subset \S^3\times \R^4/\Gamma=X_{1,\Gamma}$
is equivalent to $t$ instantons in $K_t$.  This was demonstrated for
$t=1,2$ at the end of section 6.3; we will have to assume that it is also
true for $t>2$.  On this assumption, by the same reasoning as in section
6.3, $\eta_2$ and $\eta_3$ have respectively a zero and a pole of order
$th_t$
at $\eta_1=\exp(2\pi i\mu/t)$.

Let us now add up the orders of the zeroes of $\eta_2$ (or poles of $\eta_3$)
to show that we have the right number.  Summing over $t$ and $\mu$,
the orders of the zeroes add up to
\eqn\olol{N'=\sum_{t,\mu}th_t=\sum_t th_t\phi(t).}
Using \jiho, this is equivalent to
\eqn\pololo{N'=\sum_t\sum_{i\in J_t}k_i\phi(t)=\sum_ik_i\sum_{t|k_i}\phi(t)
=\sum_ik_i^2=N.} We have used the same formula \pinin\ that has been used
in verifying \tumigo.
This equality $N'=N$ is what we need, since $N$ is the total
multiplicity of known poles of  $\eta_2$ (the only pole we know of being
a pole of order $N$ at $\eta_1=0$).

Having reconciled the numbers of zeroes and poles of $\eta_2$ and $\eta_3$,
we can finally express these observables
as functions of $\eta_1$.  Given the locations and orders of the zeroes
and poles, and the fact that $\eta_2(\infty)=\eta_3(0)=1$, we must have
\eqn\pkok{\eqalign{\eta_2 & =
{\eta_1}^{-N}{\prod_{t,\mu}\left(\eta_1-\exp(2\pi i\mu/t)
\right)^{th_t}}\cr
\eta_3 & = \prod_{t,\mu}\left(1-\exp(-2\pi i\mu/t)\eta_1\right)^{-th_t}
.\cr}}
These formulas obey all the requisite symmetries and conditions,
and this is our proposal for $\N_\Gamma$.

\bigskip
We would like to thank J. Gomis, S. Gukov, K. Intriligator, S.
Sethi, and N. Warner for discussions.  This work was initiated
while the first author was a Moore Scholar at Caltech. Work of the
second author has been supported in part by NSF Grant PHY-0070928
and the Caltech Discovery Fund.

\listrefs
\end